\documentclass[11pt]{article}
\usepackage{amssymb,amsmath,times,graphicx}
\newcommand{\rbr}[1]{\left(#1\right)}
\newcommand{\sbr}[1]{\left[#1\right]}

\newcommand{\lop}[1]{]#1]}
\newcommand{\rop}[1]{[#1[}
\newcommand{\cbr}[1]{\left\{#1\right\}}
\newcommand{\mo}[1]{\left|#1\right|}
\newcommand{\mmo}[1]{\left|\left|#1\right|\right|}
\newcommand{\eval}[2]{\left .#1\right|_{#2}}
\newcommand{\spr}[2]{\langle #1 \,, #2\rangle} % Scalar product
\newcommand{\av}[1]{\prec #1 \succ}
\newcommand{\beq}{\begin{eqnarray}}
\newcommand{\eeq}{\end{eqnarray}}
\newcommand{\sbeq}{\begin{subequations}\begin{equation}}
\newcommand{\smeq}{\end{equation}\begin{equation}}
\newcommand{\seeq}{\end{equation}\end{subequations}}
\newcommand{\nn}{\nonumber\\}

\newcommand{\eq}[1]{(\ref{#1})}
\newcommand{\fou}[1]{\check{#1}}                  % Fourier transform
\newcommand{\fvol}[1]{\frac{d^d #1}{(2\pi)^d}}    % Volume in Fourier space
\newcommand{\de}{\partial}                        % Partial derivative
\newcommand{\der}[2]{\frac{d #1}{d #2}}           % Total derivative
\newcommand{\fset}[2]{\chi_{#1}\left(#2\right)}   % Characteristic function of a set
\newcommand{\dirac}[2]{\delta^{(#1)}(#2)}         % Delta Dirac
\newcommand{\flu}[1]{\delta #1}                   % Variation
\newcommand{\vv}[1]{\mathbf{#1}}                  % Boldface vector
\newcommand{\bj}{\bar{\jmath}}
\newcommand{\btheta}{\bar{\theta}}

\newcommand{\fbtheta}{\check{\bar{\theta}}}
\newcommand{\diff}{\varkappa}
\newcommand{\eddy}{\varkappa_{\star}}
\newcommand{\scale}{{\bar{m}}}
\newcommand{\mf}{m_{F}}

\newcommand{\dd}[1]{\eta_{#1}}
\newcommand{\ivec}{\mathcal{T}}
\newcommand{\ac}{\mathcal{A}}
\newcommand{\partfun}{\mathcal{Z}}
\newcommand{\mes}{\mathcal{D}}
\newcommand{\stf}[1]{\mathcal{S}_{#1}}
\newcommand{\mstf}[1]{\tilde{\mathcal{S}}_{#1}}
\newcommand{\stfv}{\mathcal{V}}
\newcommand{\sffv}{\mathfrak{V}}
\newcommand{\cf}{\mathcal{C}}
\newcommand{\rv}{\mathfrak{R}}
\newcommand{\zemo}{\mathfrak{Z}}
\newcommand{\inop}[1]{{\mathcal{M}_{#1}^{\star}}}

\newcommand{\exop}[2]{e^{\mathcal{M}_{#1}(#2)}}
\newcommand{\inopker}{M}
\newcommand{\oper}{\mathcal{O}}

\newcommand{\re}[1]{\mathcal{R}_{#1}}
\newcommand{\relin}[1]{\mathcal{L}_{#1}}

\newcommand{\gf}{\mathcal{G}}

\newcommand{\gfv}{\mathcal{U}}
\newcommand{\ff}{\mathcal{F}}
\newcommand{\matgrap}[3]{\parbox{#2}{\includegraphics[width=#1]{#3}}} % new graphicx
\title{Scaling, renormalization and statistical conservation laws in the 
Kraichnan model of turbulent advection}
\date{}
\author{A.~Kupiainen and P.~Muratore-Ginanneschi\\
\small{Department of Mathematics, University of Helsinki, 
P.O. Box 4, 00014 Helsinki, Finland}\\
\small{emails:}\\
\small {antti.kupiainen@helsinki.fi}\\
\small{paolo.muratore-ginanneschi@helsinki.fi}} 
\begin{document}
\setlength{\unitlength}{1mm}
\maketitle
\begin{abstract}
We present a systematic way to compute the scaling exponents of the structure 
functions of the Kraichnan model of turbulent advection in a series of powers 
of $\xi$, adimensional coupling constant measuring the degree of roughness of 
the advecting velocity field.
We also investigate the relation between standard and renormalization group 
improved perturbation theory. The aim is to shed light on the relation between 
renormalization group methods and the statistical conservation laws of the 
Kraichnan model, also known as zero modes. 
\end{abstract}
\thispagestyle{empty}
\newpage
\addtolength\textheight{2mm}
\pagenumbering{roman}
\tableofcontents
\goodbreak
\addtolength\textheight{-2mm}
\newpage
\pagenumbering{arabic}
\setcounter{page}{1}

\section{Introduction}
\label{sec:intro}
Fully developed turbulence seems to have properties that are
familiar from an another branch of physics, the theory
of critical phenomena. Thus certain observables are scale
invariant, i.e. they exhibit power-law dependence on length scale.
This power-law seems to be reasonably well captured
by dimensional analysis, with however systematic discrepancies
occurring that don't seem to have simple structure.
This resembles the phenomena of second order phase transitions
where dimensional arguments  (mean field theory) do a reasonable
job, but don't fully account for the true scaling exponents.
Furthermore, in both cases the observed scaling exponents
seem to exhibit universality, i.e. a relative independence
on many details of the system: in the case of critical phenomena
details of the microscopic Hamiltonian are unimportant, only
symmetries matter, in the case of turbulence, microscopic
details of the forcing mechanism that maintains the turbulent
state seem irrelevant.

Important differences occur too. On the practical level,
in the theory of critical phenomena a rather explicit
starting point for calculating correlation functions exists
in the form of Gibbs measure given in terms of
an explicit Hamiltonian. This is directly accessible
numerically and analytically one can gain qualitative 
understanding by perturbative study around the
upper  (and sometimes lower) critical dimension. 
In case of turbulence, while the dynamical equations
governing fluid motion have been known for long,
the analog of Gibbs distribution is not. The nature of the stationary
state describing temporal averages of measurements
is a dynamical problem that is unsolved. Also, there doesn't
seem to be a parameter in the problem whose special value
would make the dimensional analysis exact (like the dimension
of space in critical phenomena) and which could then
provide a basis for a perturbative study of the 
problem of anomalous scaling.

On the more fundamental level there also are differences.
The modern renormalization group (RG) theory of critical phenomena is
based on locality in position space. The effective theory of any given scale
is given by a Gibbs state defined by a local Hamiltonian. The RG relates these
different effective theories to each other. In turbulence the stationary state 
is characterized by fluxes ("cascades") of conserved quantities (energy and in 2d
enstrophy). This cascade process is believed to be local in wave number space
and it is not clear what the right RG description is. Both {\it direct} 
\cite{FoNeSt76,FoNeSt77,DM79,Ronis87} and {\it inverse}  
(see \cite{EG94} for review and also \cite{AdAnVa} for criticism) RG's have 
been proposed in the past. 
The former is analogously to the theory of critical phenomena based on 
locality and coarse graining in physical space whereas the latter coarse 
grains in wave vector space preserving locality there.

Theoretical progress in critical phenomena came in two ways: by exactly 
solvable models with nontrivial scaling (the 2d Ising model) and by perturbative 
RG analysis near a Gaussian theory ($\epsilon$-expansion). For Navier-Stokes 
turbulence these options are not available. The analog of the $\epsilon$-expansion 
is provided by considering random stirring where the power spectrum of the force 
concentrates at large wave numbers $k$, being proportional to $k^{4-d+2\,\epsilon}$. 
Unfortunately for small $\epsilon$ this is very different from the turbulent 
situation where the force is concentrated in low wave numbers. 
Nevertheless interesting lessons can be learned also from such small
$\epsilon$-expansions as witnessed by the renormalization group studies 
carried out by the St. Petersburg school (see for example \cite{AdAnVa,Va} 
and references therein).

A different approach which can be pursued in the investigation of
fully developed turbulence is to consider phenomenological models able 
to capture some of the properties of turbulent fluid motion. 
Among such models, the Kraichnan model of passive advection 
\cite{Kr68,Kr94} has permitted in the last years to shed some light the 
mechanisms underlying the genesis of intermittency
\cite{GK95,CFKL95,BGK96,CFKL96,BGK98,FMV98}. An exhaustive review of the 
results derived from the Kraichnan model can be found in \cite{FGV01}.

The simplifying assumption which defines the Kraichnan model is the 
replacement of the Navier-Stokes velocity field advecting the scalar 
observable with a random field, Gaussian and $\delta$-correlated in time. 
This latter assumption is crucial as it ensures that the correlation 
functions of the scalar field satisfy closed Hopf equations that allow to relate the 
$n$-point functions to $n-2$-point functions via Green functions of 
differential operators built out of the spatial part of the velocity covariance. 
The latter involves a parameter $\xi$ describing the smoothness of the velocity 
field: its realizations are Holder continuous with exponent less than $\xi/ 2$. 
 
The properties of the theory versus $\xi$ are particularly interesting.
At $\xi$ equal zero, the effect of the advection is to ``renormalize'' 
the microscopic molecular diffusivity to a macroscopic eddy-diffusivity.
The resulting theory is Gaussian and provides a natural starting point
for a perturbative investigation of the system in powers of $\xi$. 
Although it can be argued that the value of $\xi$ ideally corresponding 
to a turbulent flow is equal to $4/3$, the scalar field tends already 
for small values of $\xi$ to a steady state where an inertial range sets 
in. Thus, the turbulent regime it is accessible in perturbation theory.
This is at variance with what happens for the $\epsilon$-expansion 
of the Navier-Stokes equation where the perturbative expansion has its
starting point in a model with vanishing inertial range. 

The scalar correlation functions were seen to have the  zero molecular 
diffusivity limit order by order of the $\xi$-expansion. 
This result was subsequently proved rigorously for all $\xi$  in  \cite{Ha03}.
However, the main result of the $\xi$-perturbation theory has been
the derivation of corrections to the naive
scaling prediction of scaling exponents of the structure functions of the scalar
\cite{GK95,BGK96,Ga97,BGK98,FMV98,FMNV99} a result that also was obtained in
a perturbative expansion of
the structure functions in inverse powers of the spatial dimension
\cite{CFKL95,CFKL96,MMG01}. 

Both perturbative approaches were based on the study of the Hopf 
equations satisfied by the equal time correlations of the scalar. 
In the inertial range where forcing and dissipative effects 
are negligible, these equations reduce in the stationary state to the
annihilation
of the correlation function of order $n$ by a linear operator 
$\inop{n}$. For each $n$, the operators $\inop{n}$ admit
scaling zero modes that can be computed perturbatively in $\xi$
(or in $1/d$ ) and can be shown to determine the leading scaling behavior
of the structure functions. 

The zero modes are statistically conserved quantities of the scalar field.
The presence of such conservation laws provides a mechanism underlying the 
phenomenon of intermittency. Indeed, they were observed numerically both in
passive scalar advection by a two-dimensional Navier-Stokes velocity 
field \cite{CV01} and in shell models \cite{WB96,AMG99,ABCPV01} (see also  
\cite{JPV92,BoJePaVu98}).

The concept of zero mode has therefore proved fruitful both to shed light
on properties of more realistic models of turbulence and in the 
analysis of the Kraichnan model. However, an issue which was 
left open is how to compare the small $\xi$ expansion of \cite{GK95,BGK96}
with the field-theoretic methods based on perturbative expansions of the 
Martin-Siggia-Rose functional that describes the time space time correlation 
functions of the theory. The interest of such question is threefold.

First, the Martin-Siggia-Rose formalism allows to study all the
correlation functions described by the theory without restriction to equal
times. Hence it permits to inquire what the standard perturbation theory 
in $\xi$ can say about more general observables. 

The second reason
of interest is that the Martin-Siggia-Rose provides a natural framework
to define an algorithm to compute, at least in principle, all higher
orders of the perturbative expansion. In the zero mode approach, scaling exponents
have been computed using scaling Ans\"atze, which have provided the first
order in $\xi$ term (or in some special cases exact results, see for 
example \cite{Ve96}). 
On the contrary, using the Martin-Siggia-Rose formalism, 
ultra-violet renormalization and short distance expansion, 
expressions up to $O(\xi^3)$ of the scaling exponents were derived 
in \cite{AAV98,AABKV01a,AABKV01b}. 

The fact that ultra-violet renormalization
proves useful in the context of the Kraichnan model might appear at first 
glance surprising. In general, renormalization is used to make sense of
perturbative expansions affected by ultra-violet divergences. On the other
hand the Kraichnan model, at least as it should be defined whenever the small
$\xi$ expansion is used, exhibits a well defined small scale 
behavior. It is therefore interesting to establish a precise connection
between the zero-mode and the ultra-violet renormalization methods.
Zero modes provide the leading inertial range asymptotics of equal time
correlations. It was pointed out already in \cite{BGK98} that taking the 
Mellin transform yields a well defined way to identify the 
zero mode contribution given the full, all scales, expression of
correlation functions. In the present paper this idea is developed
in order to show how it can be implemented in principle to all orders
in perturbation theory. As an example, explicit expressions of
isotropic and anisotropic scaling exponents of the structure functions 
are obtained to second order in $\xi$. The result does not 
just recover the result found with the renormalization group in 
\cite{AAV98} but establishes the relation between the two methods and the
reason why the predictions must coincide to all orders in perturbation 
theory. \\

The third reason of interest is the relationship between the different 
renormalization groups, direct and inverse, that have been proposed for 
turbulence. In particular the concept of inverse or
infra-red renormalization has been proposed as an alternative 
tool to solve scaling in the Kraichnan model \cite{Ga96,Ga97}. 
 The special features of the 
Kraichnan model are probably not suited to settle the issue which of the approaches
is more natural.  
However it provides a simple case study of  what infra-red 
renormalization is about.

The scope of the present paper is to address in a comprehensive 
way the these three issues.

The paper is organized as follows. In the first section~\ref{sec:ps} 
the passive scalar equation is introduced and the 
definition of the Kraichnan ensemble is given. In section~\ref{sec:velasympt}
the asymptotic expression of the velocity field in the physically relevant 
ranges are derived. 
The analysis of the correlation of the velocity field
shows that the roughness parameter $\xi$ lend itself as a
parameter for a perturbative construction of the solution of the
Kraichnan model. 

In section~\ref{sec:hopf} the Hopf equations governing
the dynamics of equal time correlation functions are recalled.   
Section~\ref{sec:anomalous} delves into the relation between the Mellin transform
of the solution and statistical conservation laws. Statistical conservation laws
are specified by homogeneous solutions of the Hopf equations admissible in the
inertial range but not matching the boundary conditions at large spatial scales.
Such solutions, referred as zero modes can be interpreted as the residues of
the first poles of the Mellin transform of the full solution of the Hopf equation. 

In section~\ref{sec:ansatz} we recall the analysis of the zero modes in
\cite{BGK98} which allows to predict 
the inertial range asymptotics of the structure functions of the scalar field.
The prediction is encoded in a scaling Ansatz which can be tested in a perturbative
expansion in powers of the H\"older exponent of the velocity field.

The perturbative expansion is couched in the field theoretic formalism
through the introduction of a Martin-Siggia-Rose generating function 
(section~\ref{sec:msr}). The limit of zero H\"older exponent  is shown to 
provide a Gaussian limit around which it is possible to develop a perturbative
expansion according to standard Feynman rules.

In section~\ref{sec:smallxi} the explicit expression of the 
scaling exponent is given up to second order in $\xi$. The 
calculation is based on the use of Mellin transform technique described
in section~\ref{sec:ansatz}. Technical details of the evaluation
of the integrals are deferred to appendices~\ref{ap:firstorder}, 
~\ref{ap:secondorder} and \cite{website}.

A consequence of the scaling Ansatz of section~\ref{sec:ansatz}
is  that the scaling exponent of the structure functions also
govern the blow up rate of scalar gradients versus the dissipative scale.
This is recalled in section~\ref{sec:gradients} where the exponents
are again evaluated to the second order  in $\xi$, the evaluation
being considerably easier than via the structure functions.

In section~\ref{sec:rg:general} the Wilson's formulation of the renormalization 
group is recalled first in the traditional direct form and then in
the inverse setup. In section~\ref{sec:ir} the infrared scaling fields exhibiting
the anomalous scaling of the structure functions are derived and
in section~\ref{sec:uv} the ultraviolet scaling fields responsible to the scaling
of the scalar gradients.

The last section is devoted to conclusions. 
In an appendix some details of computations are collected.

\section{Passive advection by a turbulent fluid and the Kraichnan model}
\label{sec:ps}

The passive scalar equation describes a scalar quantity $\theta(\vv{x},t)$ 
which is advected by a fluid moving with velocity $\vv{v}(\vv{x},t)$ and diffuses 
with molecular diffusivity $\kappa$: 
\begin{eqnarray}
\de_t \theta+\vv{v}\cdot \de_{\vv{x}} \theta-\frac{\kappa}{2}\,\de^{2}\theta=f
\label{ps:ps}
\end{eqnarray}
The molecular diffusivity $\kappa$ describes microscopic dissipative effects. 
The role of the force term $f$ is to provide a source for the scalar in order 
to sustain the system which otherwise would decay under 
the effect of dissipation. 

A turbulent steady state with large inertial range 
sets in if the forcing acts over spatial scales much larger than those
where dissipation becomes relevant. The details of the forcing
are supposed to be irrelevant for inertial range scaling. It is therefore 
convenient to chose the forcing as a Gaussian field of zero average and 
covariance
\begin{eqnarray}
\av{ f(\vv{x}_1,t_1) f(\vv{x}_2,t_2)}=\delta(t_2-t_1)\,
F\left(\frac{\vv{x}_1-\vv{x}_2}{L_F}\right)
\label{ps:forcing}
\end{eqnarray}
The spatial part of the covariance $F$ is a  smooth function
decaying rapidly at infinity and satisfying 
\begin{eqnarray}
F\left(0\right)=F_{\star}>0
\label{ps:forceasympt}
\end{eqnarray}
Hence the constant $L_F$ specifies the characteristic scale
(the integral scale or the correlation length) of the forcing.
The delta correlation in time in (\ref{ps:forcing}) means
that (\ref{ps:ps}) is a stochastic (partial) differential equation.

In realistic models of turbulence the velocity field is specified by
a solution of the incompressible Navier-Stokes equation. One would
like to understand the typical behavior of the scalar given
a typical realization of $\vv{v}$ from an ensemble of such solutions. 
In the Kraichnan model \cite{Kr94} this ensemble is replaced
by an ensemble of Gaussian random velocity fields which is chosen
so as to mimic some properties that are thought to be crucial
of real turbulent velocity ensembles.

One considers random velocity fields with Gaussian statistics
having  zero average and covariance 
\begin{eqnarray}
\av{ \vv{v}^{\alpha}(\vv{x}_1,t_1) \vv{v}^{\beta}(\vv{x}_2,t_2)}
=\delta(t_2-t_1)\,D^{\alpha\,\beta}\left(\vv{x}_1-\vv{x}_2;m,M\right).
\label{ps:velcor}
\end{eqnarray} 
Real turbulent velocities have two characteristic length
scales, the short, dissipative scale $M^{-1}$ and the long
integral scale  $m^{-1}$. These are modeled in the Kraichnan
ensemble by an ultraviolet cutoff $M$ and infrared cutoff $m$ 
in wave numbers entering the spatial part of the covariance in
(\ref{ps:velcor}). Moreover, realistic turbulent velocities are approximately
self similar for scales between the dissipative and
the integral scales.

The simplest choice for the spatial part of the covariance 
having such properties is
\beq
D^{\alpha\,\beta}\left(\vv{x};m,M\right)
=D_{0}\xi \int\fvol{q}\,\frac{e^{\imath \vv{q} \cdot \vv{x}}}{q^{d+\xi}}\,
\Pi^{\alpha \beta}(\hat{\vv{q}})\fset{[m,M]}{q}
\label{ps:velspatial}
\eeq
where
\beq
\Pi^{\alpha \beta}(\hat{\vv{q}}):=
\delta^{\alpha\,\beta}-\hat{\vv{q}}^\alpha\,\hat{\vv{q}}^\beta
\eeq
with $\hat{\vv{q}}$ denoting the unit vector $\vv{q}/q$, $q=|\vv{q}|$ and
$\fset{[m,M]}{q}$ is the characteristic function of the interval $[m,M]$.

We consider in this paper only incompressible velocities $\de\cdot\vv{v}=0$
which is guaranteed by the tensor $\Pi^{\alpha\beta}$.
It is convenient to introduce the covariance of velocity differences:
\begin{eqnarray}
\av{ [\vv{v}^{\alpha}(0,t')-\vv{v}^{\alpha}(\vv{x},t')] 
[\vv{v}^{\beta}(0,t)-\vv{v}^{\beta}(\vv{x},t)] }=2\,\delta(t-t')
d^{\alpha\beta}(\vv{x};m,M)
\end{eqnarray}
where we have introduced the spatial velocity difference covariance
\begin{eqnarray}
d^{\alpha\beta}(\vv{x};m,M)=D^{\alpha\,\beta}\left(0;m,M\right)-
D^{\alpha\,\beta}\left(\vv{x};m,M\right)=
D_{0}\xi \int\fvol{q}\,
\frac{1-e^{\imath \vv{q} \cdot \vv{x}}}{q^{d+\xi}}\,
\Pi^{\alpha \beta}(\hat{\vv{q}})\fset{[m,M]}{q}
\label{ps:structure}
\end{eqnarray}
As seen in detail later, $d^{\alpha\beta}$ is approximatively
scale invariant in the inertial range: 
\begin{eqnarray}
d^{\alpha \beta}(\lambda\vv{x};m,M)
\sim \lambda^{\xi}\,d^{\alpha \beta}(\vv{x};m,M)
\end{eqnarray}
for 
\begin{eqnarray}
M^{-1}\,\ll\,|\vv{x}|\,\,,\lambda\,|\vv{x}| \,,\qquad 
\lambda\,|\vv{x}|\,\,,|\vv{x}|\,\ll\,m^{-1}
\end{eqnarray}
The constant 
\begin{eqnarray}
D^{\alpha\,\beta}\left(0;m,M\right)=
D \,(m^{-\xi}-M^{-\xi})\,\delta^{\alpha\,\beta}
\label{ps:eddy}
\end{eqnarray}
where 
\begin{eqnarray}
D:=D_{0}\,\frac{d-1}{d}\,\frac{\Omega_d}{(2\,\pi)^d}\,, 
\qquad
\Omega_{d}=\frac{2\,\pi^{d/2}}{\Gamma(d/2)}
\label{ps:constant}
\end{eqnarray}
describes the mean square velocity field which blows up as the integral scale
$m^{-1}$ tends to infinity.

Finally, the delta correlation in time of the velocity fields
guarantees the statistical invariance of the velocity differences under
Galilean transformations $\vv{v}'(\vv{x},t)=\vv{v}(\vv{x}+\vv{u} t,t)-\vv{u}$,
an important property of the Navier-Stokes equation. 
More important, it leads to a relatively explicit solution of
the scalar statistics.

Equation (\ref{ps:ps}) together with (\ref{ps:velcor}) defines an 
infinite dimensional stochastic differential equation with 
multiplicative noise. 
In order for this object to be well defined according to the general rules
of stochastic calculus \cite{Ok,ME81}, it is necessary to specify how 
the product between the velocity and the scalar field in (\ref{ps:ps})
is defined as the continuum limit of a finite difference stochastic
equation. The Kraichnan model is supposed to be the limit
of a physical system with finite time correlations. 
Hence it is natural to regard (\ref{ps:ps}),
(\ref{ps:velcor}) as infinite dimensional stochastic differential equation
in the Stratonovich sense.

\section{Asymptotics expressions of the velocity ensemble}
\label{sec:velasympt}

Far from being the only possible, the choice (\ref{ps:velspatial})
suits the derivation of explicit asymptotic expressions of the 
velocity field. 
Of physical relevance is the behavior of the velocity
field in the dissipative and inertial range. 

\subsection{Dissipative range asymptotics}

The dissipative range is defined by the inequalities
\begin{eqnarray}
m x\, \ll\, M x\,\ll\,1
\label{velasympt:dissipativerange}
\end{eqnarray}
Under this assumption the Fourier exponential in (\ref{ps:velspatial}) 
can be expanded in Taylor series.
Up to leading order, the expansion yields
\begin{eqnarray}
D^{\alpha\,\beta}\left(\vv{x};m,M\right)\sim D^{\alpha\,\beta}\left(0;m,M\right)
-\frac{D\,\xi\, M^{-\xi}_{v}\, (M x)^2\,\ivec^{\alpha\,\beta}(\hat{\vv{x}},2)}
{(2-\xi)(d-1)(d+2)}
\label{velasympt:dissipative}
\end{eqnarray}
having introduced the rank-two real-space tensor
\begin{eqnarray}
\ivec^{\alpha\,\beta}(\hat{\vv{x}},z)=
\delta^{\alpha\,\beta}-\frac{z}{d-1+z}\,
\hat{\vv{x}}^{\alpha}\hat{\vv{x}}^{\beta}
\label{velasympt:realranktwo}
\end{eqnarray}
Whenever
\begin{eqnarray}
\frac{m}{M}<<1
\end{eqnarray}
the {\em eddy diffusivity}
\begin{eqnarray}
\eddy:=D\,m^{-\xi}
\label{velasympt:eddy}
\end{eqnarray}
dominates the velocity field in the dissipative range. 
The leading correction to the 
constant mode of the velocity field is smooth and vanishing 
as $M^{-\xi}$ tends to zero.

\subsection{Inertial range asymptotics}

The integral (\ref{ps:structure})
is convergent both if the ultra-violet cut-off $M$ is set to infinity
and the infra-red cutoff $m$ to zero. Thus, the statistics of the
velocity differences exists in such limit. It can be determined by
considering the Mellin transform of the spatial correlation of the
velocity field
\begin{eqnarray}
\tilde{D}^{\alpha\,\beta}(\vv{x};m,z)=D_{0}\,\xi\,
\int_{0}^{\infty}\frac{dw}{w}\frac{1}{w^{z}}
\underset{q>m}{\int}\fvol{q}  \frac{e^{\imath w\,\vv{q}\cdot \vv{x}}}
{q^{d+\xi}}\,\Pi^{\alpha\,\beta}(\hat{\vv{q}})
\label{velasympt:Mellin}
\end{eqnarray}
The integral is convergent for  $\Re z<0$ 
and can be performed explicitly. 
The result is
(see appendix~\ref{ap:inertial} for details): 
\begin{eqnarray}
\tilde{D}^{\alpha\,\beta}(\vv{x};m,z)=
D\,\xi\,c(z)\frac{m^{z-\xi}\,x^z}{z\,(z-\xi)}
\,\ivec^{\alpha\,\beta}(\hat{\vv{x}},z)
\label{velasympt:real}
\end{eqnarray}
with $D$ and 
$\ivec^{\alpha\,\beta}$ respectively defined by (\ref{ps:constant}) 
and (\ref{velasympt:realranktwo}) whilst
\begin{eqnarray}
c(z):=\frac{(d-1+z)}{(d-1)}
\frac{\Gamma\left(\frac{d+2}{2}\right)\Gamma\left(1-\frac{z}{2}\right)}
{2^{z}\,\Gamma\left(\frac{d+2+z}{2}\right)}\,,\qquad\qquad c(0)=1
\label{velasympt:g}
\end{eqnarray}
is a function with simple poles for $z$ a positive even
integer .

The small scale asymptotics is derived by evaluating the inverse 
Mellin transform involving an integral over $z$ along $\Re z={\rm const}<0$
by pushing the contour to the right and picking residues from the poles.
This gives
\begin{eqnarray}
D^{\alpha\,\beta}(\vv{x};m)= D\,m^{-\xi}\delta^{\alpha\,\beta}-
d^{\star\,\alpha\,\beta}(\vv{x})+o(m^{2-\xi}x^{2})
\label{velasympt:inertial}
\end{eqnarray}
Thus, the residue of the pole at zero corresponds to the eddy diffusivity.
The pole at $\xi$ specifies instead the inertial range asymptotic of the
structure tensor of the velocity field:
\begin{eqnarray}
d^{\star\,\alpha\,\beta}(\vv{x}):=D\,c(\xi)\,
x^{\xi}\,\ivec^{\alpha\,\beta}(\hat{\vv{x}},\xi)
\label{velasympt:structure}
\end{eqnarray}

Note that the blowup of $c(\xi)$ at $\xi=2$ is 
related to the fact that if $\xi$ is equal two 
the pole at $z$ equal $\xi$ in (\ref{velasympt:real})  turns from 
simple to double. 
This indicates  the existence of logarithmic corrections to 
the analytic behavior, proportional to $x^2$, of the velocity field 
structure function. Logarithmic corrections are suppressed for 
example by redefining
\beq
D_{0}\rightarrow \frac{D_{0}}{\Gamma\rbr{1-\frac{\xi}{2}}}
\eeq
The rescaling does not affect universal quantities in the 
small $\xi$ limit and will be neglected in the present paper.

Let us finally discuss the behavior of the velocity
covariance around $\xi=0$.
At fixed ultra-violet cut-off it vanishes linearly
with $\xi$. On the other hand, the removal of the ultra-violet cut-off
reduces the dissipative range to the the single point  $x$ equal zero.
There the velocity field coincides with the large scale constant mode.
In the inertial range, by (\ref{velasympt:inertial}) the velocity field
is vanishing with $\xi$ for any nonzero $m$ and $x$ also after sending 
$M$ to infinity. We will see that the expansion around $\xi$ equal zero 
provides a viable analytic tool for the investigation of 
universal properties of advection.  

\section{Hopf's equations and statistical conservation laws} 
\label{sec:hopf}

Let $\theta(\vv{x},t)$ be the solution of the stochastic
differential equation (\ref{ps:ps}) with suitable initial
condition. The equal time  correlation functions:
\begin{eqnarray}
\cf_{2\,n}:=\av{ \prod_{i=1}^{2 n} \theta(\vv{x}_i,t)}
\end{eqnarray}
satisfy in the Kraichnan model a solvable hierarchy of Hopf equations
which are simplest to derive in the Ito representation \cite{Ok,ME81} 
of the equation. Letting $\vv{v} \cdot\de_{\vv{x}}\theta$ be defined
with the Ito convention the equation (\ref{ps:ps}) becomes
\begin{eqnarray}
\de_t \theta + \vv{v} \cdot\de_{\vv{x}}\theta-\frac{1}{2}
\left[\kappa \,\delta^{\alpha\,\beta}+D^{\alpha\,\beta}(0;m,M) \right]
\de_{\alpha}\de_{\beta}\theta=f
\label{hopf:equation}
\end{eqnarray}
By (\ref{ps:eddy}), in the Ito representation the molecular 
viscosity is renormalized by the velocity field to
\begin{eqnarray}
\diff=\kappa+D\,(m^{-\xi}-M^{-\xi})
\label{hopf:viscosity}
\end{eqnarray}
In the inertial range where dissipative effects are negligible the diffusion
felt by the scalar field reduces to the eddy diffusivity $\eddy$ introduced in
equation \eq{velasympt:eddy}.

A direct application of the Ito formula then yields
\begin{eqnarray}
\de_{t}\cf_{2\,n}-\frac{\diff}{2}\,
\sum_i\,\de^{2}_{\vv{x}_i}\cf_{2\,n}-
\sum_{i<j}D^{\alpha\,\beta}(\vv{x}_i-\vv{x}_j;m,M)
\de_{\vv{x}_i^{\alpha}}\de_{\vv{x}_j^{\beta}}
\cf_{2\,n}=\ff_{2n}
\label{hopf:general}
\end{eqnarray}
with
\sbeq
\ff_{2n}:=\cf_{(2\,n-2)}\otimes F\equiv
\sum_{i<j}\,
\cf_{(2\,n-2)}(\vv{x}_1,\underset{\hat{i}\hat{j}}{...},\vv{x}_{2 n},t)\,
F\left(\frac{\vv{x}_{i}-\vv{x}_{j}}{L_{F}}\right)
\label{hopf:forcing}
\smeq
\ff_{2}:= F\left(\frac{\vv{x}_{i}-\vv{x}_{j}}{L_{F}}\right)
\label{hopf:forcing2p}
\seeq
Odd order correlation functions will be ignored since they vanish in 
the steady state due to parity invariance.
 
For a translation invariant initial condition for $\theta$ (say 0)
$\mathcal{C}_{2\,n}$ is translation invariant i.e.
\begin{eqnarray}
\sum_{i}\de_{\vv{x}_i^{\alpha}} \cf_{2\,n}=0
\end{eqnarray}
the Hopf equations reduce to the final form
\begin{eqnarray}
\de_{t}\cf_{2\,n}-\frac{\kappa}{2}\,
\sum_i\,\de^{2}_{\vv{x}_i}\cf_{2\,n}+
\sum_{i<j}d^{\alpha\,\beta}(\vv{x}_i-\vv{x}_j;m,M)
\de_{\vv{x}_i^{\alpha}}\de_{\vv{x}_j^{\beta}}
\cf_{2\,n}=\ff_{2n}
\label{hopf:hopf}
\end{eqnarray}
Note that owing to translational invariance the $2n$-point correlation 
in $d$ dimensions is a function of $d_{n}=(2n-1)d$ variables. 

Since (\ref{hopf:hopf}) depends on the velocity statistics only
through the structure function it has coefficients that have limits as 
the ultra-violet and infra-red cut-offs of the velocity field are removed. 
In that limit we may define the elliptic negative definite operators
\begin{eqnarray}
\inop{2\,n}:=\sum_{i<j}
d^{\star\,\alpha\,\beta}(\vv{x}_i-\vv{x}_j)\de_{\vv{x}_i^{\alpha}}
\de_{\vv{x}_j^{\beta}}
\label{hopf:operator}
\end{eqnarray}
Since the coefficients $d_{\alpha\,\beta}^{\star}$ are
homogeneous functions of degree $\xi$ these differential
operators are self similar, namely homogeneous of degree
$\xi-2$. The stationary state correlation functions
of the scalar satisfy the equations
\begin{eqnarray}
-\,\inop{2\,n}\cf_{2\,n}=\frac{\kappa}{2}\,
\sum_{i=1}^{2\,n}\,\de^{2}_{\vv{x}_i}\cf_{2\,n}+
\ff_{2\,n}
\label{hopf:steadystate}
\end{eqnarray} 
The terms collected on the right hand side of (\ref{hopf:steadystate}) 
are related to the non-universal dependencies of the dynamics. Self-similarity 
breaking from the large scale can occur through the integral scale of the 
forcing $L_{F}$. Comparing the $\kappa$ and the $\inop{2\,n}$ terms 
hints at the existence of a scalar field dissipative scale
\begin{eqnarray}
\ell=\left(\frac{2\,\kappa}{D}\right)^{\frac{1}{\xi}}
\label{hopf:dissipative}
\end{eqnarray}

The existence and uniqueness of the solutions 
of (\ref{hopf:steadystate}) has been proved at zero molecular dissipation
for all values of the H\"older exponent $\xi<2$ of the velocity field \cite{Ha03}.
They are given as
\begin{eqnarray}
\cf_{2\,n}(\mathbf{X};L_{F})=-\,
\int d\mathbf{Y} \inopker_{2n}^{-1}(\mathbf{X},\mathbf{Y})
\ff_{2n-2}(\mathbf{Y};L_{F})
\label{hopf:solution}
\end{eqnarray}
for $\mathbf{X}\,,\mathbf{Y}\in \mathbb{R}^{d_{n}}$. The kernel 
$\inopker_{2n}^{-1}$ of the operator $\inop{2n}^{-1}$ was proved to be 
locally integrable and the integrals converge absolutely as long as 
the forcing scale $L_F$ is finite. By the homogeneity of $\inop{2n}^{-1}$ we get 
immediately that
\begin{eqnarray}
\mathcal{C}_{2\,n}(\mathbf{X}, L_F)=L_F^{n\,(2-\xi)}
\mathcal{C}_{2\,n}\left(\frac{\mathbf{X}}{L_F}, 1\right)
\label{hopf:dimensional}
\end{eqnarray}
i.e. the canonical dimension of $\mathcal{C}_{2\,n}$ is $n(2-\xi)$.

\section{Zero modes and short distance asymptotics}
\label{sec:anomalous}
 
From (\ref{hopf:dimensional}) we see that the large $L_F$
behavior of the correlation functions $\mathcal{C}_{2\,n}$ is 
dominated by the large scale velocity: they blow up like $L_F^{n\,(2-\xi)}
\mathcal{C}_{2\,n}(0, 1)$. Sub-leading terms in the inertial range can be 
extracted by considering the Mellin transforms
\begin{eqnarray}
\tilde{\cf}_{2n}(\mathbf{X},L_{F};z):=
\int_{0}^{\infty}\frac{dw}{w}\frac{\cf_{2\,n}(w \mathbf{X},L_{F})}{w^{z}}=
L_{F}^{n(2-\xi)}\frac{X^{z}}{L_{F}^{z}}\,\int_{0}^{\infty}\frac{dw}{w}
\frac{\cf_{2\,n}(w \hat{\mathbf{X}},1)}{w^{z}}
\label{anomalous:Mellindef}
\end{eqnarray}
These integrals converge for the real part of $z$ small
enough and are expected to extend to meromorphic functions of $z$
at least for generic $\xi$. 

The Mellin transform of the Hopf equation \eq{hopf:steadystate} is
\begin{eqnarray}
-\inop{2n}X^{z+2-\xi}\tilde{\cf}_{2n}(\hat{\mathbf{X}},L_{F};z+2-\xi)=
\frac{\kappa}{2}\de_{X}^2X^{z+2}\tilde{\cf}_{2n}(\hat{\mathbf{X}},L_{F};z+2)+
X^{z}\tilde{\ff}_{2n}(\hat{\mathbf{X}},L_{F};z)
\label{anomalous:Mellinequation}
\end{eqnarray}
It was observed in \cite{BGK98} that poles of $\tilde{\cf}_{2n}$ can occur 
either for values of $z$ for which $\tilde{\ff}_{2n}$ has a pole
or for $z$ such that the operator  $\inop{2n}$ has a zero mode 
\begin{eqnarray}
\inop{2n} \zemo=0
\label{anomalous:zeromode}
\end{eqnarray}
which is a homogeneous function of degree $z$.
The poles of $\tilde{\ff}_{2n}$ are in view of (\ref{hopf:forcing})
determined by solving the Hopf equations lower in the hierarchy.
One then ends up with an asymptotic short distance expansion
for
$\mathcal{C}_{2\,n}$ 
in terms of homogeneous functions of the coordinates
\begin{eqnarray}
\mathcal{C}_{2\,n}(\mathbf{X},L_{F})=
\sum_{j} X^{\zeta_{n,j}}\,L_{F}^{n\,(2-\xi)-\zeta_{n,j}}\,A_j(\hat{\mathbf{X}})
\label{anomalous:expan}
\end{eqnarray}
Since the forcing covariance
is smooth and has a Taylor expansion $F(x)=\sum f_n |x|^n$ we conclude
that the scaling exponents $\zeta_{nj}$ that may enter in (\ref{anomalous:expan})
are either the homogeneity degrees of the zero modes of $\inop{2n} $
or they are determined in terms of the previous ones $\zeta_{n-1,j}$.

The general situation is illustrated by the case of the two-point 
function $\cf_{2}$ which may be computed explicitly by quadrature
if the forcing is isotropic \cite{BGK98}:
\begin{eqnarray}
\cf_{2}(x;L_{F})&=&(d+\xi)\,\int_{x}^{\infty}\!\!\!dx_1
\frac{\int_{0}^{x_1}dx_2\,F(x_2/L_{F})\,x_2^{d-1}}{x_1^{d-1}\,[
d^{\star \alpha}_{\quad\alpha}(x_1)+(d+\xi)\,\kappa]}
\label{anomalous:second}
\end{eqnarray}
with $d^{\star \alpha\,\beta}$ specified by (\ref{velasympt:structure}).
In such a case, at zero molecular viscosity, 
equation \eq{anomalous:Mellinequation} reduces to
\begin{eqnarray}
-d^{\star\,\alpha\,\beta}(\vv{x})\de_{\alpha}\de_{\beta} x^{z+2-\xi} 
\tilde{\cf}_{2}(1,L_{F};z+2-\xi)=x^{z}\,\tilde{F}(1,L_{F};z)
\end{eqnarray}
with solution
\begin{eqnarray}
\tilde{\cf}_{2}(x,L_{F};z)=-\,L_{F}^{2-\xi}\frac{x^{z}}{L_{F}^{z}}\,
\frac{(d+\xi-1)\,\tilde{F}(1,1;z-2+\xi)}{c(\xi)\,(d-1)(d+z-2+\xi)\,z\,D}
\label{anomalous:twopoints}
\end{eqnarray}
Zero modes correspond to the poles in $z$ equal zero and $2-d-\xi$ respectively
associated to short and large distance asymptotics \cite{CFKL96,FGV01}.
The only zero mode contributing in short distances is the constant,
the poles of $\tilde{F}$ are at $z=2-\xi+n$ for nonnegative integer $n$.

\section{Anomalous scaling of the structure functions}
\label{sec:ansatz}

Correlation functions that probe the sub-leading terms in the
short distance expansion (\ref{anomalous:expan}) are provided
by the {\it structure functions } of the scalar
namely
\begin{eqnarray}
\mathcal{S}_{2\,n}(\vv{x},L_F,\ell):=
\av{[\theta(\vv{x})-\theta(0)]^{2\,n}}
\label{anomalous:structure}
\end{eqnarray}
where we denoted explicitly the dependence on the forcing scale
and $\kappa$ via (\ref{hopf:dissipative}). Let us consider for
simplicity isotropic forcing so that $\mathcal{S}_{2\,n}$ is only
a function of  $x=|\vv{x}|$. We may scale it out as
\begin{eqnarray}
\mathcal{S}_{2\,n}(x,L_F,\ell)=x^{(2-\xi)n}
\mathcal{S}_{2\,n}\rbr{1, \frac{L_F}{x},\frac{\ell}{x}}
\label{anomalous:structure1}
\end{eqnarray}
Suppose $\mathcal{S}_{2\,n}(1,L,\ell)$ had a limit as $\ell$ 
tends to zero and $L$ tends
to infinity. Then we would conclude 
\begin{eqnarray}
\mathcal{S}_{2\,n}(x,L_F,\ell)=
A_n x^{(2-\xi)n}\rbr{1+o\rbr{\frac{x}{L_F},\frac{\ell}{x}}},
\label{anomalous:structure2}
\end{eqnarray}
i.e. the structure
function scaling exponent would be $n(2-\xi)$. This
is the prediction of the Obukhov-Corssin \cite{Ob49,Cor51} theory, analogous to the 
Kolmogorov 41 theory \cite{Ko41a,Ko41b,MoYa,Fr} for Navier-Stokes turbulence 
that predicts there a scaling exponent $n/3$ for the velocity n-point
structure function.

To discuss the validity of the Obukhov-Corssin theory, we note first
that the limit $\ell\to 0$ exists for the correlation
functions and thus for the structure functions by the results
of \cite{Ha03}. The large $L_F$ behavior depends on 
the nature of the terms
entering  the expansion (\ref{anomalous:expan}), i.e. the zero modes
of the inertial operators. 
Structure functions are obtained from correlation functions by applying
the finite increment operator $\mathcal{I}_{\vv{x}}$ 
\begin{eqnarray}
\stf{2\,n}(\vv{x},t)=
\mathcal{I}_{\vv{x}}\mathcal{C}_{2\,n}(\vv{x}_1,\dots,\vv{x}_{2n},t)
\label{ansatz:increment}
\end{eqnarray}
The operator $\mathcal{I}_{\vv{x}}=\prod_i\iota_{\vv{x}}^{(i)}$ 
with $i$ counting the number of fields in the correlation function 
generates finite field increments according to the rule
$\iota_{\vv{x}}f(\vv{y})=f(\vv{x})-f(0)$.
From (\ref{ansatz:increment}) it follows immediately that the only zero
modes of $\cf_{2n}$ contributing to $\stf{2n}$ can be
the {\it irreducible} ones, i.e. those depending on
all the independent variables $\vv{x}_1-\vv{x}_i$, $i=2\dots n$.

The existence and the properties of zero modes were thoroughly 
investigated in \cite{GK95,CFKL95,BGK96,CFKL96,BGK98,FGV01}. 
It was shown in \cite{GK95,BGK96} for small $\xi$ and in \cite{CFKL95}
for large $d$ that for each $n>1$ there is a unique irreducible 
zero mode $\zemo_{2n}$. 
The irreducible zero mode mode has scaling dimension 
$\zeta_{2n}= (2-\xi)n-\rho_{2n}$ with $\rho_{2n}>0$. 
In \cite{BGK98} further arguments were presented
for the conclusion that this zero mode enters the expansion
(\ref{anomalous:expan}) and is the dominant term.
This means that we may write
\begin{eqnarray}
\mathcal{S}_{2\,n}(x, L_F,\ell)=
x^{(2-\xi)n}\,\rbr{\frac{L_{F}}{x}}^{\rho_{2n}}s_{2\,n}
\rbr{\frac{x}{L_F},\frac{\ell}{x},\xi}\rbr{\frac{F_{\star}}{D}}^{n}
\label{anomalous:ansatz}
\end{eqnarray}
where  the function $s_{2n}$ has a nonzero limit as $\ell\to 0$
and $L_{F}\to\infty$:
\beq
\lim_{L_{F}\to\infty}\lim_{\ell\to 0}
s_{2\,n}\rbr{\frac{x}{L_{F}},\frac{\ell}{x},\xi}:=
s_{2n}(\xi) >0.
\eeq
The ratio between the constants  $F_{\star}$ and $D$, defined respectively in
(\ref{ps:forceasympt}) and (\ref{ps:eddy}),  is introduced
for later convenience.

It was also argued in \cite{GK95,BGK96} that the sub-leading terms
for the asymptotics of $\mathcal{S}_{2\,n}$ have exponents
well separated from the leading one for small $\xi$, namely
\beq
s_{2\,n}\rbr{\frac{x}{L_F},0,\xi}:=
s_{2n}(\xi) +O\rbr{\rbr{\frac{x}{L_{F}}}^{2-O(\xi)}}
\label{ansatz:zeromode}
\eeq
We may call this the {\it scaling Ansatz} for the 
Kraichnan model.

It is useful to express this result in terms of the Mellin transform. 
Let us shift for convenience $z$ by $2 n$ which is the $\xi$ equal zero
theory scaling exponent:  
\begin{eqnarray}
\mstf{2n}(x,L;z+2\,n):=\int_{0}^{\infty}\frac{dw}{w}
\frac{\stf{2n}(w x,L)}{w^{z+2\,n}}
\label{ansatz:Mellindef}
\end{eqnarray} 
which converges for $\Re z$ small enough (actually $\Re z<-O(\xi)$). 
No specific assumption is made on the origin of the integral scale $L$
appearing in \eq{ansatz:Mellindef}. 
The scaling Ansatz (\ref{ansatz:zeromode}) then yields
\sbeq
\mstf{2n}(x,L;z):=-\,
\frac{A(z,\xi)\,L^{-n\,\xi-z}\,x^{2\,n+z}}{z-\sigma_{2\,n}(\xi)}
\label{ansatz:Mellin}
\smeq
\sigma_{2\,n}(\xi):=-n\,\xi-\rho_{2\,n}(\xi)
\label{ansatz:Mellinpole}
\seeq
with the function $A$ having poles for values of $z$ differing from 
$\sigma_{2\,n}$ by terms at least of the order $O(\xi^{0})$. Furthermore,
$A$ takes on the first pole of the Mellin transform the value 
\begin{eqnarray}
A(\sigma_{2\,n}(\xi),\xi)=s_{2\,n}(\xi) \left(\frac{F_{\star}}{D}\right)^{n}
\label{ansatz:amplitude}
\end{eqnarray} 

In the following sections structure functions will be computed using 
perturbation theory around $\xi$ equal zero. It is useful to observe that
the small $\xi$ expansion of the Mellin transform generates a Laurent series
in $z$. The residues of the expansion have simple relations with the expansion
in powers of $\xi$ of the scaling exponent:
\begin{eqnarray}
\lefteqn{\frac{d\,}{d\xi}\ln \cbr{\mstf{2n}(x,L;z+2\,n)L^{n\xi}}
=}
\nn
&&\frac{\sigma_{2\,n}^{'}(0)}{z}
+O(z^0,\xi^0)+\xi\left[\frac{(\sigma_{2\,n}^{'})^{2}(0)}{z^2}+
\frac{\sigma_{2\,n}^{''}(0)}{z}+O(z^0)\right]+O(\xi^2)
\label{ansatz:logseries}
\end{eqnarray}
Analogous relations hold true to all orders in $\xi$ and also for
anisotropic forcing. Thus, the Mellin transform permits
to extract, systematically at any order in $\xi$, zero mode contributions. 
Furthermore, as it will be seen below, taking the Mellin transform of the 
Feynman diagrams generated by the Martin-Siggia-Rose formulation \cite{MaSiRo73} 
of the Kraichnan model greatly simplifies the explicit evaluation of the 
corresponding integrals. 
The fact is well known in field theory as the Mellin transform has 
the effect to map a theory with massive propagators into a massless one
\cite{Zi,OO00}.

\section{Martin-Siggia-Rose formalism}
\label{sec:msr}

The perturbative analysis of the zero modes of the inertial operators $\inop{2n}$
in the parameter $\xi$ is based on the fact that the velocity
structure function (\ref{velasympt:structure}) becomes
constant as $\xi$ tends to zero and the operators become Laplacians. It can then
be checked that the solutions of the Hopf equations are correlation functions
of a Gaussian field $\theta$.
For small $\xi$ the distribution of $\theta$ should be given by a perturbation of 
this Gaussian. It is straightforward to derive a perturbation expansion from the
Ito stochastic differential equation (\ref{hopf:equation}).

For this, let us observe that the solution of \eq{hopf:equation}
can be written as
\beq
\theta(\vv{x},t)=\int d^dy\, \rv(\vv{x},t|\vv{y},t_0)\theta(\vv{y},t_0)
+\int_{t_0}^tds\int d^{d}y\,\rv(\vv{x},t| \vv{y},s)f(\vv{y},s)
\label{msr:theta}
\eeq
with $\rv$ solution of the stochastic differential equation
\sbeq
\rbr{\de_{t}-\frac{\varkappa}{2}\de^2_{\vv{x}}}\rv= -\vv{v}\cdot \de_{\vv{x}} \rv
\label{msr:stochres}
\seeq
with $\rv(\vv{x},t| \vv{y},t)=\dirac{d}{\vv{x}-\vv{y}}$
and  the product is defined with Ito convention. This equation can be solved
as a series in multiple stochastic integrals
\beq
\rv(t|s)=\sum_0^\infty\int_s^{t_1}\dots\int_{t_n}^{t}R(t|t_n) 
\vv{v}(t_n)dt_n\cdot \partial R(t_{n}|t_{n-1})
\dots\vv{v}(t_1)dt_1\cdot\partial R(t_{1}|s)
\label{msr:series}
\eeq
which converges in $L^2$ of the probability space of $\vv{v}$  \cite{LeJan}. Here
$R$ is the fundamental solution to the heat equation
\beq
R(\vv{x},t|\vv{x}^{\prime},t^{\prime})=
\int\frac{d^dp}{(2\pi)^d}\,e^{\imath \vv{p}\cdot(\vv{x}-\vv{x}^{\prime})
-\frac{\varkappa}{2}\,p^2\,(t-t^{\prime})}
\label{msr:freepropagator}
\eeq
Inserting  \eq{msr:series} to  \eq{msr:theta} and averaging over $\vv{v}$ 
and $f$ leads
to an expansion of the correlation functions in terms of 
integrals involving $R$ and the covariances of 
$\vv{v}$ and $f$. 

A convenient way to generate this series is 
provided by the so called  Martin-Siggia-Rose (MSR) formalism 
 (see \cite{Ca99,Va,Zi} and references  therein). 
It provides a graphical representation of the 
perturbation theory analogous to the Feynman rules of 
quantum field theory.  

The MSR formalism is derived
by introducing the generating function 
\beq
\partfun(\jmath,\bj)=\av{\int \mes[\theta]\,e^{\spr{\jmath}{\theta}}
\delta (\de_{t}\theta+\vv{v}\cdot\de_{\vv{x}}\theta
-\frac{\varkappa}{2}\de^2_{\vv{x}}\theta -f-\imath \bj)}_{v,f}
\label{msr:partfundelta}
\eeq
where
\begin{eqnarray} 
\spr{\jmath}{\theta}=\int_{-\infty}^{\infty}dt\int d^dx\,\jmath \,\theta 
\end{eqnarray} 
and $\av{\bullet}_{v,f}$ denotes the average with respect to 
the velocity and forcing fields. 
By  introducing of an auxiliary {\em ``ghost''}
field $\btheta$ the MSR functional becomes 
\beq
\partfun(\jmath,\bj)=\av{\int \mes[\theta \btheta]\,
e^{\spr{\jmath}{\theta}+\spr{\btheta}{\bj}-\ac}}_{f,v}
\label{msr:velpartfun}
\eeq
with
\beq
\ac=-\imath 
\spr{\btheta}{\rbr{\de_{t}+\vv{v}\cdot\de-\frac{\varkappa\,\de^2}{2}}\theta}
-\imath \spr{\btheta}{f}
\label{msr:action}
\eeq
Inverting the order of integration and averaging over the velocity and
forcing fields leads to 
\beq
\mathcal{Z}(\jmath,\bj)=\av{
e^{-\ac_1(\theta\,,\btheta)-\frac{1}{2}\spr{\btheta}{F\btheta}
+\spr{\jmath}{\theta}+\spr{\btheta}{\bj}}}_{G}
\label{msr:msr}
\eeq
where 
\begin{eqnarray}
\ac_1(\theta\,,\btheta)=
\frac{1}{2}\spr{\btheta\de_{\alpha}\theta}{D^{\alpha\,\beta}
\btheta\de_{\alpha}\theta}
\label{msr:interaction}
\end{eqnarray}
and the average is with respect the Gaussian ``measure'' with
covariance given by
\begin{eqnarray}
\av{\theta(\vv{x},t)\theta(\vv{x}^{\prime},t^{\prime})}_{G}=
\int\frac{d^dp}{(2\pi)^d}\,\frac{e^{\imath \vv{p}\cdot(\vv{x}-\vv{x}^{\prime})
-\diff\,p^2\,|t-t^{\prime}|}}{\diff p^2} F(\vv{p}).
\label{msr:freecor}
\end{eqnarray}
and
\beq
\av{\theta(\vv{x},t)\btheta(\vv{x}^{\prime},t^{\prime})}_{G}=\imath
H_{0}(t-t^{\prime})R(\vv{x},t|\vv{x}^{\prime},t^{\prime})
\label{msr:freeres}
\eeq 
The free response function  \eq{msr:freeres} involves the "Ito-Heaviside function"
\begin{eqnarray} 
H_{0}(t)=\left\{ 
\begin{array}{cc} 
1 & t\,>\,0\\ 
0 & t\,\leq\,0  
\end{array}\right. 
\label{msr:Heaviside} 
\end{eqnarray} 
which insures that equal time contractions in  \eq{msr:interaction} will not occur 
as follows from the Ito convention used for the stochastic integrals
in  \eq{msr:series} . 
The generating function satisfies  the normalization
\beq
\partfun(0,0)=1.
\eeq

The perturbation expansion for $\mathcal{Z}(\jmath,\bj)$ is obtained 
by expanding $e^{-\ac_1}$ in powers of $\ac_1$ and 
expressing the resulting expectations of $\theta\, ,\, \btheta$ 
in graphical terms as explained in Section 8. 
Before going to that let us make two comments concerning
the infrared cutoffs and the small parameter in the expansion.

The  free response function  \eq{msr:freeres} depends on the eddy diffusivity 
\eq{velasympt:eddy} which diverges as the infrared cutoff $m$ of 
the velocity field tends to zero. Thus this limit cannot be taken directly 
in \eq{msr:msr}. In section~\ref{sec:hopf} the properties of 
equal time scalar correlations were discussed in the  case when
$m=0$. This limit can be taken in the MSR formalism 
only for equal time correlation and structure functions. 
In particular it can be shown \cite{MG06} that order by order in perturbation 
theory, if the integral scale of the forcing $L_{F}$ is kept fixed, the limit 
for $m$ tending to zero exists and leads to the scaling predictions 
of \cite{BGK96}. 

On the other hand, in the MSR formalism it is natural 
to study the opposite limit when the 
integral scale of the velocity field $m^{-1}$ is smaller than that of the forcing. 
As argued in section~\ref{sec:ansatz}, anomalous scaling of the 
structure functions is a consequence of the existence of 
homogeneous zero modes of equation \eq{anomalous:zeromode}
where the infrared cutoffs don't occur. 
Thus, it is plausible that the scaling exponents
don't depend on the order in which the infrared cutoffs are removed.
It would be interesting to spell this out more explicitly. In what follows
we will keep $m$ fixed and study the $m_F\to 0$ limit.

Finally let us comment on the small parameter of the expansion.
In section~\ref{sec:velasympt} it was shown that the
spatial part of the velocity covariance vanishes almost everywhere 
(i.e. at $x\neq 0$) as $\xi\to 0$ whilst being bounded from above by the eddy
diffusivity $\eddy$.  Thus apart from the constant mode the \eq{msr:interaction} 
can be viewed as a small perturbation in the limit $\xi$ tending to zero.

\section{Small $\xi$ expansion}
\label{sec:smallxi}

The general features of the small $\xi$ expansion can be summarized as follows. 
The coefficients of the expansion are determined by integrals symbolically 
represented by Feynman diagrams. 
The basic ingredients of the Feynman diagrams are the free scalar
correlation (\ref{msr:freecor}) and response functions  (\ref{msr:freeres})
together with the velocity correlation \eq{ps:velcor}. They have the graphical
representation
%%%%%%%%%%%%%%%%%%%%%%%%%%%%%%%%%%%%%%%%%%%%%%%%%%%%%%%%%%%%%%%%%%%%%%%%%%
%%%%%%%%%%%%%%%%%%%%%%%%%%%%%%%%%%%%%%%%%%%%%%%%%%%%%%%%%%%%%%%%%%%%%%%%%%
\sbeq
\av{ \theta(\vv{x},t)\theta(\vv{x}^{\prime},t^{\prime})}_{G}\,=
\matgrap{1.8cm}{1.0cm}{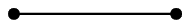}
\label{fig:correlation}
\smeq
\av{ \theta(\vv{x},t)\btheta(\vv{x}^{\prime},t^{\prime})}_{G}\,=
\matgrap{1.8cm}{1.0cm}{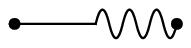}
\label{fig:response}
\smeq
\av{ v^{\alpha}(\vv{x},t)v^{\beta}(\vv{x}^{\prime},t^{\prime})}\,=
\matgrap{1.8cm}{1.0cm}{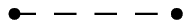}
\label{fig:velocity}
\seeq
%%%%%%%%%%%%%%%%%%%%%%%%%%%%%%%%%%%%%%%%%%%%%%%%%%%%%%%%%%%%%%%%%%%%%%%%%%
%%%%%%%%%%%%%%%%%%%%%%%%%%%%%%%%%%%%%%%%%%%%%%%%%%%%%%%%%%%%%%%%%%%%%%%%%%
where end-line dots represent external points. In order to exhibit the
expansion parameter, Feynman diagrams are constructed by connecting 
free response and correlations lines through the $O(\xi^{0})$ part of the 
interaction vertex
\beq
\ac_{I}=\frac{1}{\xi}\,\ac_{1}
=\matgrap{1.4cm}{1.0cm}{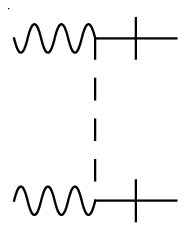}
\label{fig:vertex}
\eeq
%%%%%%%%%%%%%%%%%%%%%%%%%%%%%%%%%%%%%%%%%%%%%%%%%%%%%%%%%%%%%%%%%%%%%%%%%%
%%%%%%%%%%%%%%%%%%%%%%%%%%%%%%%%%%%%%%%%%%%%%%%%%%%%%%%%%%%%%%%%%%%%%%%%%%
The bars transversal to scalar lines in \eq{fig:vertex} 
represent spatial derivatives. Finally, all contributions are reordered
in powers of the H\"older exponent of the velocity field by expanding
the residual $\xi$ dependence of the interaction. 

It is worth stressing that this perturbation series is {\it ultraviolet finite}
i.e. the resulting integrals have a well defined limit as the ultra-violet 
cut-off $M$ tends to infinity. 
Singular behavior is instead exhibited in the limit of infinite integral scale, 
$m$ tending to zero.

Structure functions are obtained by applying the finite increment operator
\eq{ansatz:increment} to the perturbative expressions of equal time scalar 
correlation functions. 
This operation removes order by order in perturbation theory all terms 
proportional to powers of the integral scale otherwise present for dimensional 
reasons in the correlation functions. The resulting perturbative 
expansion of structure functions contains only 
expressions exhibiting logarithmic divergences in the infra-red.

In section~\ref{sec:ansatz} it was argued that $\stf{2n}$ is in the 
inertial range a homogeneous function of the spatial separation 
of degree $\zeta_{2n}$  determined by the unique irreducible zero-mode of the Hopf 
equation of order $2\,n$. According to \eq{ansatz:logseries} $\zeta_{2n}$ 
can be straightforwardly evaluated order by order in $\xi$ by taking
the Mellin transform of the perturbative expression of $\stf{2n}$. 
Working under these assumptions, in the following two subsections the 
calculation of $\zeta_{2n}$ is outlined for the first two orders in $\xi$.
For simplicity we set $M=\infty$ in the calculation. 
More details are deferred to appendices~\ref{ap:firstorder}~and~\ref{ap:secondorder}.

\subsection{Zeroth order approximation}
\label{sec:smallxi:zerorder}

The zeroth order of the perturbative expansion corresponds to neglecting
the interaction term \eq{msr:interaction} in \eq{msr:msr}. In order to simplify
the notation we keep the explicit non perturbative $\xi$
dependence \eq{velasympt:eddy} in the eddy diffusivity $\eddy$.
This trivial $\xi$ dependence can be always expanded a posteriori to check
of the decoupling of the scaling exponents into the part predicted by 
canonical dimensional analysis and a part associated to self-similarity 
breaking in the inertial range.

The second order structure function of the Gaussian theory is obtained from 
\eq{msr:freecor} and  
\eq{ps:forcing}. In the limit $L_F\to\infty$
we get 
\beq
\stf{2}^{(0)}(\vv{x})= \lim_{L_{F}\uparrow \infty}\lim _{\kappa\downarrow 0}\,2\,
\int \fvol{p}\frac{1-e^{\imath \vv{p}\cdot\vv{x}}}{D\,p^2}
\,L_{F}^{d}\fou{F}(L_{F} p)
=\frac{2}{D}\,
\int \fvol{p}(\hat{\vv{p}}\cdot\vv{x})^2
\,\fou{F}(p)
\label{smallxi:zero:two}
\eeq
In particular, if the forcing correlation is 
isotropic we get
\beq
\stf{2}^{(0)}(\vv{x})=\frac{x^2\,F_{\star}}{d\,D}.
\label{smallxi:zero:two1}
\eeq

Structure functions of higher order are given in the Gaussian limit as
\begin{eqnarray}
\stf{2\,n}^{(0)}(\vv{x})=\frac{(2\,n)!}{2^{n}\,n!}
[\stf{2}^{(0)}(\vv{x})]^{n}
\label{smallxi:zero:free}
\end{eqnarray}
If the forcing is anisotropic, scaling properties of the structure 
functions are identified by expanding them in a functional basis 
invariant under rotations. In $d$-dimensions, this scope is achieved 
by resorting to an expansion in hyperspherical harmonics \cite{WenAvery}. 
Under rather general conditions, the generic Gaussian structure function
takes the form
\begin{eqnarray}
\stf{2\,n}^{(0)}(\vv{x})=\sbr{\frac{x^{2}F_{\star}}{d\,D}}^{n}
\,\sum_{j} K_{j}\,Y_{j\,0}(\hat{\vv{x}})+O(L_{f}^{-2})
\label{smallxi:zero:decomposition}
\end{eqnarray}
with $K_{j}$ some non-universal constants depending on the forcing.

\subsection{First order approximation}
\label{sec:smallxi:firstorder}

To first order in $\xi$ the {\em structure functions} 
require the evaluation of the diagrams
%%%%%%%%%%%%%%%%%%%%%%%%%%%%%%%%%%%%%%%%%%%%%%%%%%%%%%%%%%%%%%%%%%%%%%%%%%
%%%%%%%%%%%%%%%%%%%%%%%%%%%%%%%%%%%%%%%%%%%%%%%%%%%%%%%%%%%%%%%%%%%%%%%%%%
\beq
\stfv_{(1;2)}\,=\,\mathcal{I}_{\vv{x}}
\matgrap{2.4cm}{1.0cm}{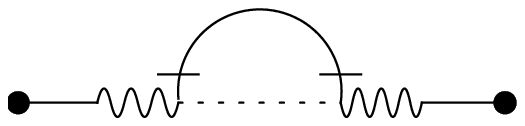}
\label{fig:twopoints}
\eeq 
and
\beq
\stfv_{(1;4)}\,=\,\mathcal{I}_{\vv{x}}
\matgrap{1.8cm}{1.0cm}{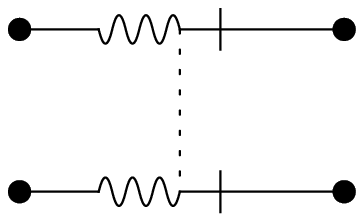}
\label{fig:fourpoints}
\eeq
%%%%%%%%%%%%%%%%%%%%%%%%%%%%%%%%%%%%%%%%%%%%%%%%%%%%%%%%%%%%%%%%%%%%%%%%%%
%%%%%%%%%%%%%%%%%%%%%%%%%%%%%%%%%%%%%%%%%%%%%%%%%%%%%%%%%%%%%%%%%%%%%%%%%%
In \eq{fig:twopoints}, \eq{fig:fourpoints} the increment operator 
is given by eq. \eq{ansatz:increment}.
As $L_{F}\to\infty$ the momentum transfer
along free correlation lines is suppressed. In such a limit the two diagrams 
can be recast in the form
\sbeq
\stfv_{(1;2)}^{(0)}=\stfv_{(1;4)\,\alpha}^{(0)\,\alpha}\de^2 \stf{2}^{(0)}
\label{smallxi:firstorder:dia12}
\smeq
\stfv_{(1;4)}^{(0)}=\stfv_{(1;4)}^{(0)\,\alpha\,\beta}
(\de_{\alpha} \stf{2}^{(0)})(\de_{\beta} \stf{2}^{(0)})
\label{smallxi:firstorder:dia14}
\seeq
using the general notation
\beq
\stfv_{\bullet}^{(n)\,\bullet}:=\left.\frac{d^{n}\,}{d\xi^{n}}\right|_{\xi=0}
\stfv_{\bullet}^{\bullet}
\label{smallxi:notation}
\eeq
to count the number of derivatives with respect to $\xi$ of a diagram 
and
%%%%%%%%%%%%%%%%%%%%%%%%%%%%%%%%%%%%%%%%%%%%%%%%%%%%%%%%%%%%%%%%%%%%%%%%%%
%%%%%%%%%%%%%%%%%%%%%%%%%%%%%%%%%%%%%%%%%%%%%%%%%%%%%%%%%%%%%%%%%%%%%%%%%%
\beq
\stfv_{(1;4)}^{\alpha\,\beta}=
\mathcal{I}_{\vv{x}} 
\matgrap{1.4cm}{1.0cm}{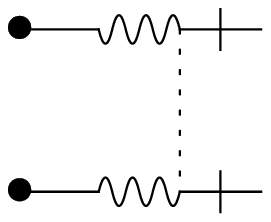}
\setlength{\unitlength}{0.1cm}
\begin{picture}(0,0) 
\put(6.0,5.0){\mbox{\small$\alpha$}} 
\put(6.0,-4.0){\mbox{\small$\beta$}} 
\end{picture}
\hspace{1.0cm}
=\frac{2\,D_{0}\,m^{\xi}}{D}\int
\frac{d^dp}{(2\pi)^d}\frac{1-e^{\imath\vv{p}\cdot\vv{x}}}{p^2}
\frac{\Pi^{\alpha\,\beta}(\hat{\vv{p}})}{p^{d+\xi}}\fset{\lop{m,\infty}}{p}
\label{smallxi:firstorder:v14}
\end{eqnarray}
%%%%%%%%%%%%%%%%%%%%%%%%%%%%%%%%%%%%%%%%%%%%%%%%%%%%%%%%%%%%%%%%%%%%%%%%%%
%%%%%%%%%%%%%%%%%%%%%%%%%%%%%%%%%%%%%%%%%%%%%%%%%%%%%%%%%%%%%%%%%%%%%%%%%%
In \eq{smallxi:firstorder:v14} the convention is adopted to denote with 
truncated scalar correlation lines in the diagrammatic representation the
absence of momentum transfer along those lines.
The factor two in \eq{smallxi:firstorder:v14} stems from the action of the
finite increment operator $\mathcal{I}_{\vv{x}}$ on the two response 
functions in the integrand. From \eq{smallxi:firstorder:v14} it is readily
verified the expected  the ultra-violet convergence of 
$\stfv_{(1;4)}^{(0)\,\alpha\,\beta}$ as well as its logarithmic divergence
in the infra-red.

The sum of \eq{smallxi:firstorder:dia12} and 
\eq{smallxi:firstorder:dia14} weighted by combinatorial factors yields 
the inertial range asymptotics of structure functions 
\begin{eqnarray}
\lefteqn{\stf{2\,n}(\vv{x},m)=\frac{(2\,n)!}{2^{n}\,n!}\left\{
\sbr{\stf2^{(0)}(\vv{x})}^{n}+\frac{n\,\xi}{2}\,
\sbr{\stf2^{(0)}(\vv{x})}^{n-1}\, 
\stfv^{(0)\,\alpha\,\beta}_{(1;4)}
\de_\alpha\de_\beta\,\stf2^{(0)}(\vv{x})\right.}
\nonumber\\
&&\left.+\frac{\xi\,n(n-1)}{2}\,[\stf2^{(0)}(\vv{x})]^{n-2}\,
\stfv^{(0)\,\alpha\,\beta}_{(1;4)}[\de_\alpha \stf2^{(0)}(\vv{x})] 
[\de_\beta\stf2^{(0)}(\vv{x})] \right\}+O(\xi^2,L_{F}^{-2})
\label{smallxi:firstorder:structure}
\end{eqnarray}
Some straightforward algebra permits to recast the result in a more 
compact form
\begin{eqnarray}
\stf{2\,n}(\vv{x};m)=\left\{ 1+ \frac{\xi}{2}\,
\stfv^{(0)\,\alpha\,\beta}_{(1;4)}
\de_\alpha\de_\beta\,\right\}\,\stf{2\,n}^{(0)}(\vv{x})
+O(\xi^2,L_{F}^{-2})
\label{smallxi:firstorder:final}
\end{eqnarray}
The Mellin transform defined as in (\ref{ansatz:Mellindef}) 
acts on the perturbative expression of structure function as
\begin{eqnarray}
\widetilde{[\stf{2\,n}-\stf{2\,n}^{(0)}]}(\vv{x},m;z+2\,n)=\frac{\xi}{2}\,
\stfv^{(0)\,\alpha\,\beta}_{(1;4)}(z+2)\,
\de_\alpha\de_\beta\,\stf{2\,n}^{(0)}(\vv{x})+
O(\xi^2,L_{F}^{-2})
\label{smallxi:firstorder:Mellinfinal}
\end{eqnarray}
where
\begin{eqnarray}
\stfv_{(1;4)}^{(0)\,\alpha\,\beta}(z+2)
=\frac{2\,D_{0}}{D}\int_{0}^{\infty}\frac{dw}{w}\frac{1}{w^{z+2}}
\underset{p\geq m}\int\frac{d^dp}{(2\pi)^d}
\frac{1-e^{\imath\,w\,\vv{p}\cdot\vv{x}}}{p^2}
\frac{\Pi^{\alpha\,\beta}(\hat{\vv{p}})}{p^{d}}
\label{smallxi:firstorder:mv14}
\end{eqnarray}
The explicit evaluation of this integral is performed in 
appendix~\ref{ap:firstorder}. 

The relations
\begin{eqnarray}
&&x^{2}\,\de^2\,(x^{2\,n}\,Y_{j\,l})= 
[2\,n\,(2\,n+d-2)-j\,(j+d-2)]\,x^{2\,n}\,Y_{j\,l}
\nonumber\\
&&\vv{x}^{\beta}\cdot \de_{\beta}\,(x^{2\,n}\,Y_{j\,l})=
2\,n\,\,x^{2\,n}\,Y_{j\,l}
\label{smallxi:firstorder:angularprojection}
\end{eqnarray}
together with (\ref{ansatz:logseries}) yield the first 
order result for the leading scaling exponents for each angular 
component of the structure functions: 
\begin{eqnarray}
\zeta_{2\,n,j}=2\,n-\left[\frac{n\,(d+2\,n)}{d+2}
-\frac{(d+1)\, j\, ( d + j-2)}{2\,(d+2)(d-1)}\right]\,\xi+O(\xi^2)
\label{smallxi:firstorder:exponent}
\end{eqnarray}
The anomalous part of the scaling exponent is identified
by expanding the $\xi$ dependence of the free structure function 
upon the eddy diffusivity $\eddy$:
\begin{eqnarray}
\rho_{2\,n,j}=\left[\frac{2\,n\,(n-1)}{d+2}
-\frac{(d+1)\, j\, ( d + j-2)}{2\,(d+2)(d-1)}\right]\,\xi+O(\xi^2)
\label{smallxi:firstorder:anomalous}
\end{eqnarray}
The result is in agreement with those given in 
\cite{BGK96,AAV98,A98,ALPP00}. Note that the isotropic exponent $j=0$ dominates 
inertial range scaling.

\subsection{Second order approximation}
\label{sec:smallxi:secondorder}

To second order in $\xi$ and as
 $L_{F}\to\infty$ we obtain the
representation
\begin{eqnarray}
\lefteqn{\stf{2\,n}(\vv{x};m)=\left\{1+ \frac{\xi}{2}\,
\stfv^{(0)\,\alpha\,\beta}_{(1;4)}
\de_\alpha\de_\beta\,+\frac{\xi^2}{2}\,\left[
\stfv^{(1)\,\alpha\,\beta}_{(1;4)}\de_\alpha\de_\beta\,
+\stfv^{(0)\,\alpha\,\beta}_{(2;4)}\de_\alpha\de_\beta\,
\right]\right\}\stf{2\,n}^{(0)}(\vv{x})}
\nonumber\\
&&
+\xi^2
\left\{\stfv^{(0)\,\alpha\,\beta\,\mu}_{(2;6)}
\de_\alpha\de_\beta\,\de_{\mu}+\frac{1}{8}\,
\stfv_{(2;8)}^{(0)\,\alpha\,\beta\,\mu\,\nu}
\de_\alpha\de_\beta\,
\de_{\mu}\de_{\nu}\right\}\,\stf{2\,n}^{(0)}(\vv{x})+
O(\xi^3,L_{F}^{-2})
\label{smallxi:secondorder:final}
\end{eqnarray}
The 
expansion coefficients in \eq{smallxi:secondorder:final} require 
the evaluation of two new diagrams 
%%%%%%%%%%%%%%%%%%%%%%%%%%%%%%%%%%%%%%%%%%%%%%%%%%%%%%%%%%%%%%%%%%%%%%%%%%
%%%%%%%%%%%%%%%%%%%%%%%%%%%%%%%%%%%%%%%%%%%%%%%%%%%%%%%%%%%%%%%%%%%%%%%%%%
\beq
\stfv_{(2;4)}^{\alpha\,\beta}&=&\mathcal{I}_{\vv{x}}
\setlength{\unitlength}{0.1cm}
\begin{picture}(0,8) 
\put(0.0,0.0){\mbox{$\matgrap{2.4cm}{1.0cm}{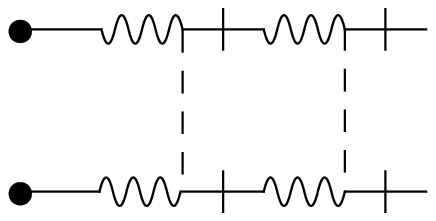}$}}
\put(26.0,5.0){\mbox{\small$\alpha$}} 
\put(26.0,-4.0){\mbox{\small$\beta$}} 
\end{picture}
\nn[0.5cm]
&=&\frac{2\,D_{0}^{2}}{D^{2}}
\underset{\substack{q\geq m\\ p \geq m}}{\int}
\frac{d^dp\,d^dq}{(2\,\pi)^{2\,d}}\,
\frac{1-\cos [(\vv{p}+\vv{q})\cdot \vv{x}]}{(\vv{q}+\vv{p})^2}\,
\frac{\vv{q}_{\mu}\,\vv{q}_{\nu}}{q^2}
\frac{\Pi^{\mu\,\nu}(\hat{\vv{p}})\,\Pi^{\alpha\,\beta}(\hat{\vv{q}})\,
}{p^{d+\xi}\,q^{d+\xi}}
\label{smallxi:secondorder:v24}
\\
\stfv_{(2;6)}^{\alpha\,\beta\,;\mu}&=&\mathcal{I}_{\vv{x}}
\setlength{\unitlength}{0.1cm}
\begin{picture}(0,14) 
\put(0.0,0.0){\mbox{$\matgrap{2.8cm}{1.0cm}{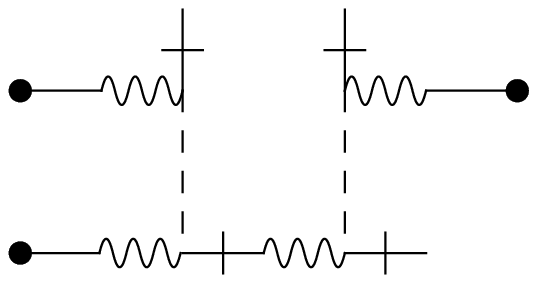}$}}
\put(9.0,9.0){\mbox{\small$\mu$}} 
\put(18.0,9.0){\mbox{\small$\alpha$}} 
\put(24.0,-6.0){\mbox{\small$\beta$}} 
\end{picture}
\nn[0.8cm]
&=&\frac{-\,\imath\,D_{0}^2}{D^2}
\underset{\substack{q\geq m\\ p \geq m}}{\int} 
\frac{d^dq\,d^dp}{(2\,\pi)^{2\,d}}\,
\frac{(e^{\imath \vv{q}\cdot \vv{x}}-1)(e^{-\imath (\vv{q}+\vv{p})\cdot \vv{x}}-1)
(e^{\imath \vv{p}\cdot \vv{x}}-1)}{(q^2+\vv{q}\cdot\vv{p}+p^2)\,q^2}
\frac{\vv{q}_{\nu}\,\Pi^{\mu\,\nu}(\hat{\vv{p}})\, 
\Pi^{\alpha\,\beta}(\hat{\vv{q}})}{p^{d+\xi}\,q^{d+\xi}}
\label{smallxi:secondorder:v26}
\eeq
Note that in 
\eq{smallxi:secondorder:final} tensor indices are symmetrically contracted.
Consequently only the {\em index-symmetric} part of the above diagrams contributes 
to the structure function whence the {\em omission of the semicolon} between 
indices in \eq{smallxi:secondorder:final}. 

The third new coefficient appearing in  \eq{smallxi:secondorder:final}
for $L_{F}$ tending to infinity is
%%%%%%%%%%%%%%%%%%%%%%%%%%%%%%%%%%%%%%%%%%%%%%%%%%%%%%%%%%%%%%%%%%%%%%%%%%%%%%%
%%%%%%%%%%%%%%%%%%%%%%%%%%%%%%%%%%%%%%%%%%%%%%%%%%%%%%%%%%%%%%%%%%%%%%%%%%%%%%%
\newline
%%%%%%%%%%%%%%%%%%%%%%%%%%%%%%%%%%%%%%%%%%%%%%%%%%%%%%%%%%%%%%%%%%%%%%%%%%%%%%%
%% the commented diagram seems to have problems when compiled with pdflatex   %
%%%%%%%%%%%%%%%%%%%%%%%%%%%%%%%%%%%%%%%%%%%%%%%%%%%%%%%%%%%%%%%%%%%%%%%%%%%%%%%
%\beq
%\stfv_{(2;8)}^{\alpha\,\beta\,;\mu\,\nu}\,
%=\mathcal{I}_{\vv{x}}
%\setlength{\unitlength}{0.1cm}
%\begin{picture}(0,0) 
%\put(36.0,6.2){\mbox{$\matgrap{-1.4cm}{1.0cm}{EPSFIGURES/vertex14.eps}$}}
%\put(16.0,5.0){\mbox{\small$\alpha$}} 
%\put(16.0,-4.0){\mbox{\small$\beta$}} 
%\put(0.0,0.0){\mbox{$\matgrap{1.4cm}{1.0cm}{EPSFIGURES/vertex14.eps}$}}
%\put(20.0,5.0){\mbox{\small$\mu$}} 
%\put(20.0,-4.0){\mbox{\small$\nu$}} 
%\end{picture}
%\hspace{3.8cm}= \stfv_{(1;4)}^{\alpha\,\beta}\,
%\stfv_{(1;4)}^{\mu\,\nu}
%\label{smallxi:secondorder:v28}
%\eeq
\beq
\stfv_{(2;8)}^{\alpha\,\beta\,;\mu\,\nu}\,
=\mathcal{I}_{\vv{x}}
\setlength{\unitlength}{0.1cm}
\begin{picture}(0,0) 
\put(37.3,6.2){\mbox{$\matgrap{-3.8cm}{1.0cm}{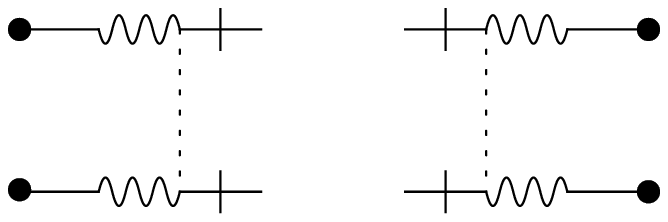}$}}
\put(16.0,5.0){\mbox{\small$\alpha$}} 
\put(16.0,-4.0){\mbox{\small$\beta$}} 
\put(20.0,5.0){\mbox{\small$\mu$}} 
\put(20.0,-4.0){\mbox{\small$\nu$}} 
\end{picture}
\hspace{3.8cm}= \stfv_{(1;4)}^{\alpha\,\beta}\,
\stfv_{(1;4)}^{\mu\,\nu}
\label{smallxi:secondorder:v28}
\eeq
\newline
%%%%%%%%%%%%%%%%%%%%%%%%%%%%%%%%%%%%%%%%%%%%%%%%%%%%%%%%%%%%%%%%%%%%%%%%%%
%%%%%%%%%%%%%%%%%%%%%%%%%%%%%%%%%%%%%%%%%%%%%%%%%%%%%%%%%%%%%%%%%%%%%%%%%%
Thus, the absence of momentum flow
across inner scalar correlation lines reduces the evaluation 
of $\stfv_{(2;8)}^{(0)\,\alpha\,\beta\,;\mu\,\nu}$ and
$\stfv_{(1;4)}^{(1)\,\alpha\,\beta}$ to the first order integral computed 
in appendix~\ref{ap:firstorder}. 

As expected by dimensional considerations, the integrals are convergent 
for large momentum values and logarithmic for small momenta.  

The Mellin transform (\ref{ansatz:Mellindef}) of 
(\ref{smallxi:secondorder:final}) acts on the individual
Feynman diagrams as
\begin{eqnarray}
\lefteqn{\widetilde{[\stf{2\,n}
-\stf{2\,n}^{(0)}]}(\vv{x},m;z+2\,n)=}
\nonumber\\
&&\frac{1}{2}
\left\{\xi\,\tilde{\stfv}^{(0)\,\alpha\,\beta}_{(1;4)}(z+2)
+\xi^2\left[\tilde{\stfv}^{(1)\,\alpha\,\beta}_{(1;4)}(z+2)+
\tilde{\stfv}^{(0)\,\alpha\,\beta}_{(2;4)}(z+2)\right]\right\}
\,\de_\alpha\de_\beta\,\stf{2\,n}^{(0)}(\vv{x})
\nonumber\\
&&
+\xi^2\,\left\{\tilde{\stfv}^{(0)\,\alpha\,\beta\,\mu}_{(2;6)}(z+3)
\de_\alpha\de_\beta\,\de_{\mu}+\frac{1}{8}\,
\widetilde{(\stfv^{(0)\,\alpha\,\beta}_{(1;4)}
\stfv^{(0)\,\mu\,\nu}_{(1;4)})}(z+4)\,\de_\alpha\de_\beta\,
\de_{\mu}\de_{\nu}\right\}\,\stf{2\,n}^{(0)}(\vv{x})
\nn&&+O(\xi^3,L_{F}^{-2})
\label{smallxi:secondorder:Mellinfinal}
\end{eqnarray}
where
\begin{eqnarray}
&&\!\!\!\!\!\!\!\!\!\!
\tilde{\stfv}_{(2;4)}^{(0)\,\alpha\,\beta}(z+2)\,=
\int_{0}^{\infty}\frac{dw}{w}\frac{2\,D_{0}^2}{D^{2}\,w^{z+2}}
\underset{\substack{q\geq m\\ p \geq m}}{\int}
\frac{d^dp\,d^dq}{(2\,\pi)^{2\,d}}\,
\frac{1-e^{\imath\,w(\vv{p}+\vv{q})\cdot \vv{x}}}{(\vv{q}+\vv{p})^2}\,
\frac{\vv{q}_{\mu}\,\vv{q}_{\nu}}{q^2}
\frac{\Pi^{\mu\,\nu}(\hat{\vv{p}})\,\Pi^{\alpha\,\beta}(\hat{\vv{q}})\,
}{p^{d}\,q^d}
\label{smallxi:secondorder:mv24}
\\
&&\!\!\!\!\!\!\!\!\!\!
\lefteqn{\tilde{\stfv}_{(2;6)}^{(0)\,\alpha\,\beta\,;\mu}(z+3)\,
=\int_{0}^{\infty}\frac{dw}{w}\frac{-\,\imath\,D_{0}^{2}}{D^{2}\,w^{z+3}}}
\nonumber\\
&&\times
\underset{\substack{q\geq m\\ p \geq m}}{\int} 
\frac{d^dq\,d^dp}{(2\,\pi)^{2\,d}}\,
\frac{(e^{\imath w\,\vv{q}\cdot \vv{x}}-1)
(e^{-\imath w\,(\vv{q}+\vv{p})\cdot \vv{x}}-1)
(e^{\imath w\,\vv{p}\cdot \vv{x}}-1)}{(q^2+\vv{q}\cdot\vv{p}+p^2)\,q^2}
\frac{\vv{q}_{\nu}\,\Pi^{\mu\,\nu}(\hat{\vv{p}})\, 
\Pi^{\alpha\,\beta}(\hat{\vv{q}})}{p^d\,q^d}
\label{smallxi:secondorder:mv26}
\end{eqnarray}
Guidelines for the evaluation and explicit expressions of the Mellin
transform of the diagrams are deferred to appendix~\ref{ap:secondorder}.

Representing \eq{smallxi:secondorder:final} in terms of hyperspherical 
harmonics \cite{WenAvery} yields the  the scaling
exponents of the structure function for each value of the angular momentum $j$:
\begin{eqnarray}
\lefteqn{\zeta_{2\,n,j}=2\,n-\left[\frac{n\,(d+2\,n)}{d+2}
-\frac{(d+1)\, j\, ( d + j-2)}{2\,(d+2)(d-1)}\right]\,\xi
+(d+1)\,\left\{\frac{4\,(16 - 5\, d)\, n^2}{(d-1)\,(d+2)^3\,(d+4)}
\right.}
\nonumber\\
&&+\frac{4\,[4-(j-2)\,j+d^2\,(5+2\, j)+d\,j\,(2\, j-5)-9\,d]\,n
-48\,(d-1)\, n^3-9\, d\,\,j\,(d+j-2)}{(d-1)^2\,(2 + d)^3\,(4 + d)}
\nonumber\\
&&+(n-1)\,\mathrm{Hyp}_{21}\rbr{1, 1,2 + \frac{d}{2}, \frac{1}{4}}
\left[\frac{3\,d\,(6+d)\,n}{(2+d)^3\,(4+d)\,(1+d)}
+6\,n\,\frac{n\,(4 + d\ (7 + d))-4}{(d^2-1)\,(2 + d)^3\,(4 + d)}
\right.
\nonumber\\
&&-\left.\left.\frac{3\,(d^3+6\,d^2+d-4)\,j\,(d+j-2)}
{2\,(d-1)^2\,(2+d)^3\,(4+d)\,(d+1)}
\,\right]\,\right\}\,\xi^2+O(\xi^3)
\label{smallxi:secondorder:exponent}
\end{eqnarray}
with $\mathrm{Hyp}_{21}$ denotes the Gauss hypergeometric series 
\cite{AS}:
\begin{eqnarray}
\mathrm{Hyp}_{21}(a,b,c,x)=\frac{\Gamma(c)}{\Gamma(a)\,\Gamma(b)}\,
\sum_{n=0}^{\infty}\frac{\Gamma(a+n)\,\Gamma(b+n)\,x^n}
{\Gamma(c+n)\,\Gamma(n+1)}
\end{eqnarray}
The result is in agreement with those of ref.'s 
\cite{AAV98,AABKV01a,AABKV01a} derived using the ultra-violet 
renormalization group. From the computational point of view the Mellin 
transform applied here to the perturbative expansion does not
provide the simplest scheme to derive the scaling exponents. It is
conceptually important because it shows how to relate zero modes
of the Hopf equations to the more general diagrammatic expansion.  

For completeness sake, it is worth noticing that gluing together 
truncated scalar correlation lines of 
\eq{smallxi:secondorder:v24}, \eq{smallxi:secondorder:v26} and 
\eq{smallxi:secondorder:v28} permits for any given structure
function to reconstruct the diagrammatic expression of
the non-universal contributions vanishing in the limit $L_{F}$ tending 
to infinity. For example, the full diagrammatic expression of the fourth 
order structure function is
%%%%%%%%%%%%%%%%%%%%%%%%%%%%%%%%%%%%%%%%%%%%%%%%%%%%%%%%%%%%%%%%%%%%%%%%%%
%%%%%%%%%%%%%%%%%%%%%%%%%%%%%%%%%%%%%%%%%%%%%%%%%%%%%%%%%%%%%%%%%%%%%%%%%%
\beq
\lefteqn{\stf{4}(\vv{x};m)=3\,\mathcal{I}_{\vv{x}}
\begin{array}{c} \matgrap{2.0cm}{1.0cm}{EPSFIGURES/correlation.eps} 
\\ \matgrap{2.0cm}{1.0cm}{EPSFIGURES/correlation} \end{array}
\hspace{1.0cm}
+6\,\rbr{\xi\,\eval{\mathcal{I}_{\vv{x}}}{\xi=0}+
\xi^2\eval{\mathcal{I}_{\vv{x}}\der{}{\xi}}{\xi=0}}
\begin{array}{c} \matgrap{2.0cm}{1.0cm}{EPSFIGURES/correlation.eps} 
\\ \matgrap{2.0cm}{1.0cm}{EPSFIGURES/twopoints.eps} \end{array}}
\nn&&
+3\,\xi^2\eval{\mathcal{I}_{\vv{x}}}{\xi=0}
\begin{array}{c} \matgrap{2.2cm}{1.0cm}{EPSFIGURES/twopoints.eps} 
\\ \matgrap{2.2cm}{1.0cm}{EPSFIGURES/twopoints.eps} \end{array}
\hspace{1.5cm}
+3\,\xi^2\eval{\mathcal{I}_{\vv{x}}}{\xi=0}
\begin{array}{c} \matgrap{2.0cm}{1.0cm}{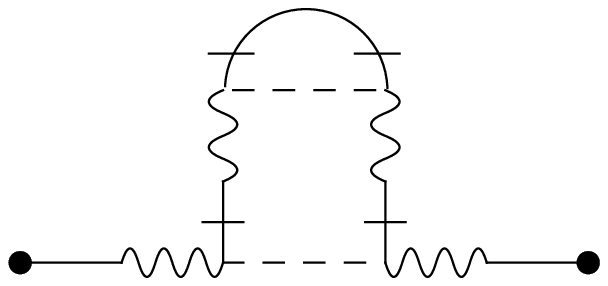} 
\\ \matgrap{2.0cm}{1.0cm}{EPSFIGURES/correlation.eps} \end{array}
\nn&&
+12\,\rbr{\xi\,\eval{\mathcal{I}_{\vv{x}}}{\xi=0}+
\xi^2\eval{\mathcal{I}_{\vv{x}}\der{}{\xi}}{\xi=0}} 
\matgrap{1.6cm}{1.0cm}{EPSFIGURES/fourpoints.eps}
\hspace{1.0cm}
+6\,\xi^2\,\eval{\mathcal{I}_{\vv{x}}}{\xi=0}
\matgrap{2.5cm}{1.0cm}{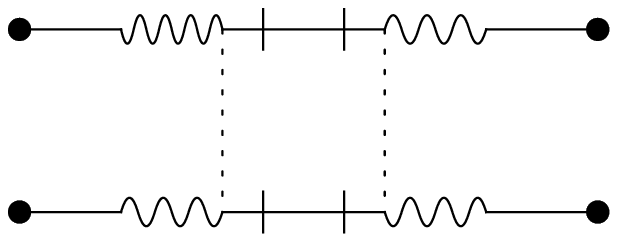}
\nn[0.1cm]&&
+12\,\xi^2\,\eval{\mathcal{I}_{\vv{x}}}{\xi=0}
\matgrap{2.5cm}{1.0cm}{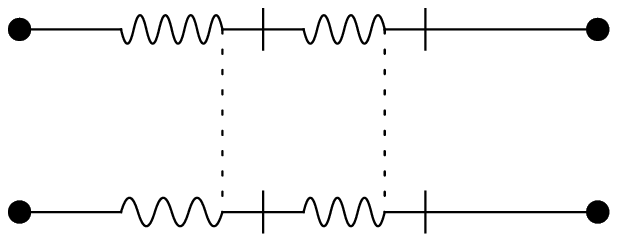}
\hspace{1.8cm}
+24\,\xi^2\,\eval{\mathcal{I}_{\vv{x}}}{\xi=0}
\matgrap{2.5cm}{1.0cm}{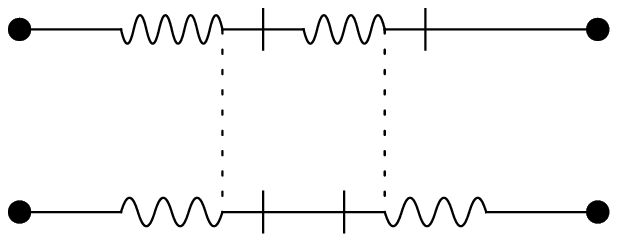}
\nn[0.1cm]&&
+24\,\xi^2\,\eval{\mathcal{I}_{\vv{x}}}{\xi=0}
\matgrap{2.5cm}{1.0cm}{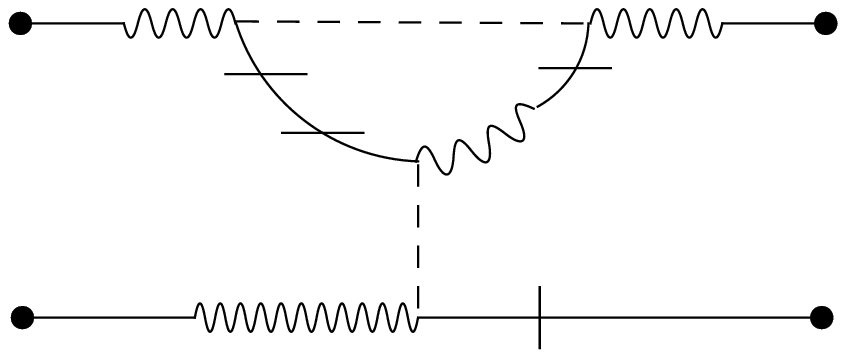}
\hspace{1.8cm}
+24\,\xi^2\,\eval{\mathcal{I}_{\vv{x}}}{\xi=0}
\matgrap{2.5cm}{1.0cm}{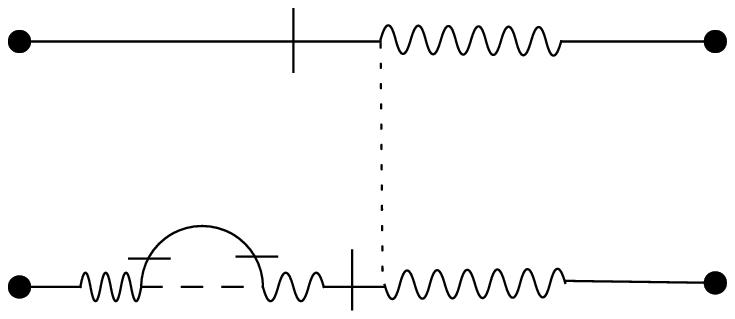}
\hspace{1.8cm}+O(\xi^3)
\label{smallxi:secondorder:fourpoints}
\eeq
%%%%%%%%%%%%%%%%%%%%%%%%%%%%%%%%%%%%%%%%%%%%%%%%%%%%%%%%%%%%%%%%%%%%%%%%%%
%%%%%%%%%%%%%%%%%%%%%%%%%%%%%%%%%%%%%%%%%%%%%%%%%%%%%%%%%%%%%%%%%%%%%%%%%%

\section{Gradient correlations and the role of the dissipative scale}
\label{sec:gradients}

From the short distance behavior of the structure
functions of the scalar for zero molecular diffusivity 
we infer that the the field $\theta(\vv{x})$ will not be differentiable:
it is only H\"older continuous with exponent $1-{\cal O}(\xi)$.
Hence correlation functions of the gradients of $\theta$ should
blow up as the dissipative scale is taken to zero. 
    
The scaling Ansatz (\ref{anomalous:ansatz}) 
imposes the existence of a precise relation between the rate of 
these divergences and the scaling exponents of the structure functions. 
It is convenient to illustrate the argument in the fully isotropic case, 
the generalization to anisotropy being straightforward. Using the hypothesis of 
inertial range universality, $L$ and $\ell$ will respectively the integral and 
dissipative scale of the scalar field disregarding of the mechanism responsible 
for their onset.

Radial gradient correlations at equal points and structure 
functions are related by the incremental ratio
\begin{eqnarray}
\gf_{2\,n}(L,\ell)&:=&\av{\left[\frac{\vv{x}^{\alpha}}{x}
\de_{\alpha}\theta(\vv{x})\right]^{2\,n}}=
 \lim_{x\,\downarrow\,0}
\av{\left[\frac{\theta(\vv{x})-\theta(0)}{x}\right]^{2\,n}}
\nonumber\\
&=& \lim_{x\,\downarrow\,0} x^{-2\,n}
\,S_{2\,n}\left({x},\ell,L\right)
\label{gradients:radial}
\end{eqnarray} 
By scaling
\begin{eqnarray}
S_{2\,n}\left({x},\ell,L\right)=x^{2\,n}\ell^{-\xi\,n}c_n(L/\ell).
\nonumber
\end{eqnarray}
Supposing 
\beq
c_n(L/\ell)\sim (L/\ell)^{\alpha_n}
\eeq
and matching at $x=\ell$ with the scaling Ansatz (\ref{anomalous:ansatz})
we infer
\begin{eqnarray}
\gf_{2\,n}(L,\ell)\propto \ell^{-\,n\,\xi}
\,\left(\frac{L}{\ell}\right)^{\rho_{2\,n}}
\label{gradients:connection}
\end{eqnarray}
As the dissipative scale tends to zero gradient correlations are 
seen to blow up at equal points as a power law with exponent 
determined by the anomalous scaling exponent of the structure 
functions. Thus (\ref{gradients:connection}) relates the existence of 
anomalous scaling and the {\em dissipative anomaly} in the energy flux of 
the scalar field. Taking the angular average of (\ref{gradients:connection})
relates it to the dissipative anomaly
\begin{eqnarray}
\gf_{2n}(L,\ell)=
\av{[(\de^{\alpha} \theta\de_{\alpha} \theta)(0,t)]^{n} }
\int \frac{d \Omega_{d}}{\Omega_{d}}\cos_{\angle}^{2n}(\vv{x}\cdot\de\theta)
\label{gradients:dissipative}
\end{eqnarray}
where the angular average is given by
\begin{eqnarray}
\int \frac{d \Omega_{d}}{\Omega_{d}}\cos_{\angle}^{2n}(\vv{x}\cdot\de\theta)=
\frac{\Gamma\left(2  n + 1\right) 
\Gamma\left(\frac{d}{2}\right)}
{4^n \Gamma\left(n + 1\right) \Gamma\left(\frac{d}{2} + n\right)}.
\label{gradients:angular-average}
\end{eqnarray}

\subsection{Perturbative expansion for radial gradient correlations}
\label{sec:gradients:uv}

Eq. (\ref{gradients:connection}) suggests that it should 
be possible to determine the scaling exponents from a perturbative
expansion of radial gradient correlations rather than of the structure
functions \cite{ALPP00}. By dimensional analysis a similar expansion is seen to 
generate Feynman diagrams logarithmic at all momentum scales. The 
evaluation of the associated integrals is therefore greatly simplified.

The identity
\beq
\av{(\vv{x}^{\alpha}\de_{\alpha}\theta)^{2n}}
=x^{2n}\av{(\hat{\vv{x}}^{\alpha}\de_{\alpha}\theta)^{2n}}
=x^{2n}\gf_{2\,n}
\eeq
permits to derive the perturbative expansion of radial gradients from
\begin{eqnarray}
\lefteqn{\av{(\vv{x}^{\alpha}\de_{\alpha}\theta)^{2n}}}
\nonumber\\
&&=
\left\{ 1+ \frac{\xi}{2}\,
\gfv^{(0)\,\alpha\,\beta}_{(1;4)}
\de_\alpha\de_\beta\,+\frac{\xi^2}{2}\,\left[\,
\gfv^{(1)\,\alpha\,\beta}_{(1;4)}\de_\alpha\de_\beta\,
+\gfv^{(0)\,\alpha\,\beta}_{(2;4)}\de_\alpha\de_\beta\,
\right]\right\}\av{(\vv{x}^{\alpha}\de_{\alpha}\theta)^{2n}}_{G}
\nn&&
+\xi^2\,
\left\{\gfv^{(0)\,\alpha\,\beta\,\mu}_{(2;6)}
\de_\alpha\de_\beta\,\de_{\mu}+\frac{1}{8}\,
\gfv^{(0)\,\alpha\,\beta}_{(1;4)}
\gfv^{(0)\,\mu\,\nu}_{(1;4)}\de_\alpha\de_\beta\,
\de_{\mu}\de_{\nu}\right\}\,\av{(\vv{x}^{\alpha}\de_{\alpha}\theta)^{2n}}_{G}
+O(\xi^3)
\label{gradients:secondorder:analytic}
\end{eqnarray}
where $\av{\bullet}_{G}$ denotes averaging with respect to the
Gaussian measure.
The argument of the logarithms appearing in the expansion is now
the ratio between the integral and dissipative scales of the
velocity field. As in section~\ref{sec:smallxi} the result is
written by keeping fixed the eddy diffusivity \eq{velasympt:eddy} and in
the limit of integral scale of the forcing $L_{F}$ tending to infinity.

The $\gfv$-coefficients in \eq{gradients:secondorder:analytic} are related
through the general definition
\beq
\gfv^{(n)\, \bullet}_{\bullet}=
\eval{\der{^{n}}{\xi^{n}}}{\xi=0}\gfv^{\bullet}_{\bullet}
\eeq
to the evaluation of the Feynman diagrams
\begin{eqnarray}
\gfv_{(1;4)}^{\alpha\,\beta}&=&\matgrap{1.4cm}{1.0cm}{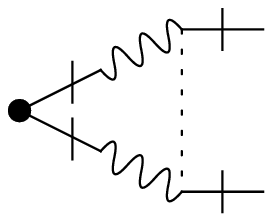}
\setlength{\unitlength}{0.1cm}
\begin{picture}(0,0) 
\put(6.0,5.0){\mbox{\small$\alpha$}} 
\put(6.0,-4.0){\mbox{\small$\beta$}} 
\end{picture}\hspace{1.0cm}
=\frac{D_{0}\,m^{\xi}}{D}
\underset{M\geq p\geq m}\int
\frac{d^dp}{(2\pi)^d}\frac{(\vv{p}\cdot \vv{x})^2}
{p^2}\frac{\Pi^{\alpha\,\beta}(\hat{\vv{p}})}{p^{d+\xi}}
\nonumber\\
&=&
\frac{(d+1)\,x^2}{(d-1)\,(d+2)\,\xi}\left[1-\left(\frac{M}{m}\right)^{-\xi}\right]
\ivec^{\alpha\,\beta}(\hat{\vv{x}},2)
\label{gradients:firstorder:v14}
\end{eqnarray}
and
\begin{eqnarray}
\gfv_{(2;4)}^{\alpha\,\beta}&=&
\matgrap{2.3cm}{1.0cm}{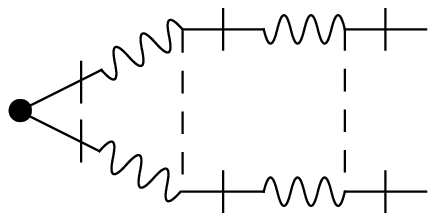} 
\setlength{\unitlength}{0.1cm}
\begin{picture}(0,0) 
\put(15.0,5.0){\mbox{\small$\alpha$}} 
\put(15.0,-4.0){\mbox{\small$\beta$}} 
\end{picture}
\nn[0.3cm]
&=&\frac{D_{0}^2}{D^2}
\underset{\substack{M\geq q \geq m \\ M \geq p \geq m}}
{\int}\frac{d^dp\,d^dq}{(2\,\pi)^{2\,d}}\,
\frac{[(\vv{p}+\vv{q})\cdot \vv{x}]^2}
{(\vv{q}+\vv{p})^2}\,\frac{\vv{q}_{\mu}\,\vv{q}_{\nu}}{q^2}
\frac{\Pi^{\mu\,\nu}(\hat{\vv{p}})\,\Pi^{\alpha\,\beta}(\hat{\vv{q}})}
{p^{d+\xi}\,q^{d+\xi}}
\label{gradients:secondorder:v24}
\\[0.3cm]
\gfv_{(2;6)}^{\alpha\,\beta\,;\mu}&=&
\matgrap{2.4cm}{1.0cm}{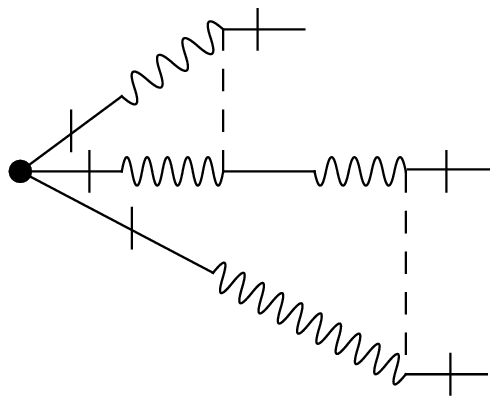} 
\setlength{\unitlength}{0.1cm}
\begin{picture}(0,0) 
\put(7.0,9.0){\mbox{\small$\alpha$}} 
\put(16.0,2.0){\mbox{\small$\mu$}} 
\put(16.0,-8.0){\mbox{\small$\beta$}} 
\end{picture}
\nn[0.4cm]&=&
\frac{D_{0}^2}{D^2}
\underset{\substack{M \geq q\geq m\\ M \geq p \geq m}}
{\int} \frac{d^dq\,d^dp}{(2\,\pi)^{2\,d}}\,
\frac{(\vv{q}\cdot \vv{x})\,[(\vv{q}+\vv{p})\cdot \vv{x}]\,( \vv{p}\cdot \vv{x})}
{(\vv{q}^2+\vv{q}\cdot\vv{p}+\vv{p}^2)\,q^2}
\frac{\vv{q}_{\nu}\,\Pi^{\mu\,\nu}(\hat{\vv{p}})\, 
\Pi^{\alpha\,\beta}(\hat{\vv{q}})}{p^{d+\xi}\,q^{d+\xi}}
\label{gradients:secondorder:v26}
\end{eqnarray}
%%%%%%%%%%%%%%%%%%%%%%%%%%%%%%%%%%%%%%%%%%%%%%%%%%%%%%%%%%%%%%%%%%%%%%%%%%
%%%%%%%%%%%%%%%%%%%%%%%%%%%%%%%%%%%%%%%%%%%%%%%%%%%%%%%%%%%%%%%%%%%%%%%%%%
The evaluation of the last coefficient appearing in 
\eq{gradients:secondorder:analytic} does not require the evaluation of
further integrals 
%%%%%%%%%%%%%%%%%%%%%%%%%%%%%%%%%%%%%%%%%%%%%%%%%%%%%%%%%%%%%%%%%%%%%%%%%%
%%%%%%%%%%%%%%%%%%%%%%%%%%%%%%%%%%%%%%%%%%%%%%%%%%%%%%%%%%%%%%%%%%%%%%%%%% 
\newline
%%%%%%%%%%%%%%%%%%%%%%%%%%%%%%%%%%%%%%%%%%%%%%%%%%%%%%%%%%%%%%%%%%%%%%%%%%%%%%%
%% the commented diagram seems to have problems when compiled with pdflatex   %
%%%%%%%%%%%%%%%%%%%%%%%%%%%%%%%%%%%%%%%%%%%%%%%%%%%%%%%%%%%%%%%%%%%%%%%%%%%%%%%
%\beq
%\gfv_{(2;8)}^{\alpha\,\beta\,;\mu\,\nu}\,=\,
%\setlength{\unitlength}{0.1cm}
%\begin{picture}(0,0) 
%\put(16,5.5){\mbox{$\matgrap{-1.4cm}{1.0cm}{EPSFIGURES/ope14.eps}$}} 
%\put(-1.0,5.0){\mbox{\small$\alpha$}} 
%\put(-1.0,-4.0){\mbox{\small$\beta$}} 
%\put(14,0.0){\mbox{$\matgrap{1.4cm}{1.0cm}{EPSFIGURES/ope14.eps}$}}  
%\put(30.0,5.0){\mbox{\small$\mu$}} 
%\put(30.0,-4.0){\mbox{\small$\nu$}} 
%\end{picture}\hspace{3.0cm}=
%\gfv_{(1;4)}^{\alpha\,\beta}
%\gfv_{(1;4)}^{\mu\,\nu}
%\label{fig:secondorder}
%\eeq
%
\beq
\gfv_{(2;8)}^{\alpha\,\beta\,;\mu\,\nu}\,=\,
\setlength{\unitlength}{0.1cm}
\begin{picture}(0,0) 
\put(2,0){\mbox{$\matgrap{2.7cm}{1.0cm}{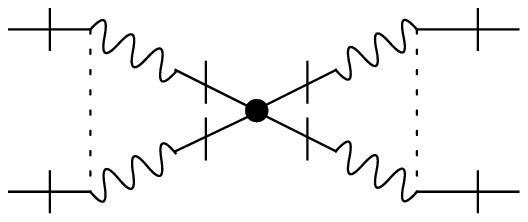}$}} 
\put(-1.0,5.0){\mbox{\small$\alpha$}} 
\put(-1.0,-4.0){\mbox{\small$\beta$}} 
\put(31.0,5.0){\mbox{\small$\mu$}} 
\put(31.0,-4.0){\mbox{\small$\nu$}} 
\end{picture}\hspace{3.2cm}=
\gfv_{(1;4)}^{\alpha\,\beta}
\gfv_{(1;4)}^{\mu\,\nu}
\label{fig:secondorder}
\eeq
\newline
%%%%%%%%%%%%%%%%%%%%%%%%%%%%%%%%%%%%%%%%%%%%%%%%%%%%%%%%%%%%%%%%%%%%%%%%%%
%%%%%%%%%%%%%%%%%%%%%%%%%%%%%%%%%%%%%%%%%%%%%%%%%%%%%%%%%%%%%%%%%%%%%%%%%% 
As in the case of structure function only the index-symmetric part of
the diagrams above matters in the evaluation of \eq{gradients:secondorder:analytic}
whence the omission of the semicolon between indices.

The Mellin transform proves again useful in extracting logarithms 
from \eq{gradients:secondorder:analytic}. The transform is well defined 
since the condition $M>m$ provides an infra-red cut-off:
\begin{eqnarray}
\tilde{\gf}_{2\,n}(m,M;z):=\int_{m}^{\infty}\frac{dw}{w}
\frac{\gf_{2\,n}(M w,m)}{w^z}\propto
\frac{m^{n\xi}}{z-n\,\xi-\rho_{2n}}
\left(\frac{M}{m}\right)^{z}
\label{gradients:Mellin}
\end{eqnarray} 
Evaluated at $m$ equal one or equivalently by keeping $\eddy$ constant in the
$\xi$-expansion, (\ref{gradients:Mellin}) permits to apply the formula 
(\ref{ansatz:logseries}) to the computation of the scaling exponent.
The explicit expressions of the Mellin transform of the $\gfv$-coefficients
together with some details about their computation are given 
appendix~\ref{ap:gradients}. Straightforward algebra then shows that from
(\ref{gradients:secondorder:analytic}) the $O(\xi^2)$ result 
(\ref{smallxi:secondorder:exponent}) for the anomalous scaling exponent
is recovered.
Finally it is readily verified that \eq{smallxi:secondorder:exponent} can
be related to the perturbative expansion of the structure function 
in the limit
\begin{eqnarray}
\gf_{2\,n}(m,M)=
\frac{\av{(\vv{x}^{\alpha}\de_{\alpha}\theta)^{2n }}}{x^{2n}}
=\lim_{x\downarrow 0}\,
\frac{\stf{2\,n}(x;m^{-1},M^{-1})}{x^{2n}}
\end{eqnarray}
Namely, the $\gfv$-coefficients in \eq{smallxi:secondorder:exponent} 
coincide with leading order of the Taylor expansion in the spatial 
increments of the  $\stfv$-coefficients in \eq{smallxi:secondorder:final}.

\section{Renormalization group methods and the validation of scaling Ans\"atze} 
\label{sec:rg:general}

The zero mode analysis leading to the scaling Ansatz (\ref{anomalous:ansatz}) 
justifies the exponentiation of the logarithms encountered in the perturbative 
expansion in powers of $\xi$.
However, the concept of a zero mode is based on the existence of a closed hierarchy 
of Hopf equations for equal time correlations.
This is a very special feature of the Kraichnan model which is not present
in more realistic models of fluid turbulence. It is therefore useful to look for a
validation of scaling directly in framework of the MSR
formalism.

Such a validation is well known in the statistical field theory
of dynamical and critical phenomena namely the Wilson formulation of the 
renormalization group (see for example \cite{WilKo74,Wil83}) and by its 
subsequent refinements (comprehensive presentations can be found for example 
in the monographes \cite{Ca,Co,LB,Ma,Va,Zi}). 

In the theory of critical phenomena
scaling of correlation functions occurs at large spatial scales.
They exhibit at the critical point a power law fall-off 
which is insensitive to the short distance details of the system.
The universality of the long distance or  infra-red behavior 
with respect to the ultra-violet is formalized by direct or ultra-violet 
renormalization. In the coming subsection the main idea will be recalled 
with the aim of applying it to the Kraichnan model where universality 
is expected to emerge from the renormalization of {\em infra-red} rather than 
{\em ultra-violet} degrees of freedom once the dissipative scale has been set 
to zero.

\subsection{Wilson direct (ultra-violet) renormalization group}
\label{sec:rg:general:uv}

We start by briefly recalling the Wilson renormalization group 
as applied to the study of correlation functions at large spatial scales.
A field theory or a spin system is specified by an action functional $\ac$
(or Hamiltonian) depending on fields, spins and the like here collectively  
denoted by $\phi$ where $\phi$ is a random field $\phi(\vv{x})$ with probability
distribution  formally given by
\begin{eqnarray}
\mathcal{P}[\phi]\propto e^{-\ac(\phi)}\mes[\phi]
\label{rg:general:uv:probability}
\end{eqnarray} 
Local observables $\oper(x)$ are  functions of $\phi$ and its 
derivatives at the   point $x$. In analogy to quantum field theory, 
they are often called  {\em operators}. 

At the critical point correlation functions of local operators exhibit
scaling
\beq
\av{\oper(\vv{x})\oper(\vv{y})}\overset{|\vv{x}-\vv{y}|\uparrow\infty}{\sim} 
|\vv{x}-\vv{y}|^{-2\dd{\oper}}
\eeq
with $\dd{\oper}$  the {\em scaling dimension} of the operator $\oper$. 
Scaling becomes exact in the  {\em scaling limit} i.e. for the random fields
\beq
\oper_{\star}\rbr{\vv{x}}=
\lim_{\lambda \downarrow 0} \lambda^{-\dd{\oper}}
\oper\rbr{\lambda^{-1}\vv{x}} 
\label{rg:general:uv:scaldim}
\eeq
whose two point function is 
\beq
\av{\oper_{\star}\rbr{\vv{x}}\oper_{\star}\rbr{\vv{y}}}\propto 
|\vv{x}-\vv{y}|^{-2\dd{\oper}}
\eeq
Thus, the scaling operator $ \oper_{\star}\rbr{\vv{x}}$
describes the long distance behavior of $ \oper\rbr{\vv{x}}$.

In statistical mechanics the fields $\phi$ and hence $\oper$ have an 
UV cutoff (e.g. the lattice spacing). $\oper_{\star}$ has no such cutoff. 
Wilson's idea was to combine the scaling limit with a coarse graining operation 
so that the fields retain a fixed UV cutoff and the operation, called 
Renormalization Group, can be viewed as a on mapping probability distributions 
(or actions). This leads to a theory of the scaling dimensions $\dd{\oper}$ .

In the simplest setup the fields $\phi$ have a cutoff in momentum space
e.g.  defined by having Fourier 
transform with support for radial values of the momentum  in $[0,M]$.  
Let $\varphi$ consist of the Fourier components of $\phi$ in the range
$[0,\lambda M]$  and  $\delta \phi$ the ones on $[\lambda M, M]$:
\begin{eqnarray}
\phi(\vv{x})=\varphi(\vv{x})+\flu{\phi}(\vv{x})
\label{rg:general:uv:decomposition1}
\end{eqnarray}
Define a rescaling operation
\begin{eqnarray}
\oper_{(\lambda)}(\vv{x}):=
\lambda^{-\dd{\oper}}\oper(\lambda^{-1}\vv{x})
\label{rg:general:uv:scaling}
\end{eqnarray}
and let 
\beq
\phi^{\prime}(x)=\varphi_{(\lambda)}(\vv{x})
\eeq
so $\phi^{\prime}$ has momenta on  $[0,M]$. Then we have the decomposition
\beq
\phi(\vv{x})=\phi^{\prime}_{(1/\lambda)}(\vv{x})+\flu{\phi}(\vv{x})
\label{rg:general:uv:decomposition}
\eeq
The expectation of  $\oper$ can be 
rewritten as
\begin{eqnarray}
\av{ \oper }=\frac{\int \mes[\phi]\,\oper(\phi)\,
e^{-\ac(\phi)}}{\int \mes[\phi]\,e^{-\ac(\phi)}}
=\frac{\int \mes[\varphi]\mes[\delta \phi]
\,\oper(\phi^{\prime}_{(1/\lambda)}+\delta \phi)\,
e^{-\ac(\phi^{\prime}_{(1/\lambda)}+\delta \phi)}}
{\int \mes[\varphi]
\mes[\delta \phi]\,e^{-\ac(\phi^{\prime}_{(1/\lambda)}+\delta \phi)}}
\label{rg:general:uv:average}
\end{eqnarray}
Integrating over $\delta \phi$,  this becomes
\begin{eqnarray}
\av{ \oper }=\frac{\int \mes[\phi^{\prime}]\,
(\relin{\lambda } \oper)(\phi^{\prime})\,
e^{-(\re{\lambda} \ac)(\phi^{\prime})}}
{\int \mes[\phi^{\prime}]\,e^{-(\re{\lambda} \ac)(\phi^{\prime})}}
\label{rg:general:uv:renormaverage}
\end{eqnarray}
where we defined the coarse-grained or renormalized operator 
\begin{eqnarray}
({\cal L}_{\lambda}\oper)(\phi^{\prime}):=\frac{\int
\mes[\delta \phi]\,\oper(\phi^{\prime}_{(1/\lambda)}+\delta \phi)\,
e^{-\ac(\phi^{\prime}_{(1/\lambda)}+\delta \phi)}}{\int
\mes[\delta \phi]\,e^{-\ac(\phi^{\prime}_{(1/\lambda)}+\delta \phi)}}
\label{rg:general:uv:definition}
\end{eqnarray}
and the renormalized action functional
\begin{eqnarray}
(\re{\lambda}\ac)(\phi^{\prime})=-\ln\,\int\mes[\delta \phi]\,
e^{-\ac(\phi^{\prime}_{(1/\lambda)}+\delta \phi)}.
\label{rg:general:uv:action}
\end{eqnarray}
These functionals depend on the field $\phi^{\prime}$ which has momentum support on
the original range $[0,M]$.  $\re{\lambda}$ is called the renormalization 
group transformation and clearly $\relin{\lambda }$ is the derivative 
$D\re{\lambda}$ of $\re{\lambda }$ at $\ac$ i.e. the linearized RG 
transformation.

The renormalization group transformation $\re{\lambda }$
satisfies the relation
\begin{eqnarray}
\re{\lambda_{1}}\re{\lambda_{2}}
=\re{\lambda_{1}\lambda_{2}}
\label{rg:general:uv:semigroup}
\end{eqnarray}
and defines a semi-group acting on the space of field 
functionals.

The physical interpretation of RG is that by {\em averaging} over the
ultra-violet degrees of freedom stored in $\flu{\phi}$ and {\em rescaling} 
the resulting theory coincides in the long distances with  the original theory
and differs only in irrelevant short distance properties. The limit 
$\lambda$ tending to zero should then describe universal long distance properties 
common to many action functionals $\ac$. It plays the role of the scaling limit 
in the Wilson formulation. The simplest limit is a fixed point
\begin{eqnarray}
\ac_{\star}=\re{\lambda }\ac_{\star}
\label{rg:general:uv:fixedpoint}
\end{eqnarray}
and given an action  $\ac$ there is at most one
choice of $\dd{\phi}$ of the scaling dimension of the basic fields such that the
renormalization group flow of the action $\ac$ converges to a fixed point.

{\em Scaling fields} in the Wilson theory are local eigenoperators of the 
linearized RG
$\relin{\lambda }^{\star}$ at the fixed point $\ac_{\star}$:
\begin{eqnarray}
\relin{\lambda }^{\star}\oper (\vv{x})=\lambda^{\dd{\oper}}
\oper (\lambda \vv{x}).
\label{rg:general:uv:sf}
\end{eqnarray}
These may often be found as follows. Suppose the limits
\begin{eqnarray}
\lim_{\lambda \downarrow 0}\re{\lambda }\ac=\ac_{\star}
\end{eqnarray}
and
\begin{eqnarray}
\lim_{\lambda \downarrow 0}
(\relin{\lambda }\oper)_{(\lambda)}=\oper_{\star}.
\label{rg:general:uv:sf1}
\end{eqnarray}
exist. Then $\oper_{\star}$ satisfies eq.~\eq{rg:general:uv:sf}.

We formulated the UV renormalization as the search of universal
long distance properties of a theory with fixed UV cutoff. In field theory
one is also interested in the continuum limit, i.e. the removal of the UV cutoff,
$M\to\infty$. Thus one considers a one parameter family of actions $\ac^M$
with UV cutoff M and possibly depending parametrically on $M$ (through
bare mass, coupling constant, wave function renormalization etc).
One then fixes some momentum scale $\bar m$ and splits the field
as
\begin{eqnarray}
\phi(\vv{x})=\varphi(\vv{x})+\flu{\phi}(\vv{x})
\label{rg-general:uv:decomposition1}
\end{eqnarray}
where $\varphi$ has UV cutoff $\bar m$ and the fluctuation part $\flu{\phi}$
momenta between $\bar m$ and $M$. Call the result after integrating
over $\flu{\phi}$ by $e^{-\ac^M_{\bar m}(\varphi)}$. The limit
\begin{eqnarray}
\lim_{M\to\infty}\ac^M_{\bar m}=\ac_{\bar m}
\label{rg-general:uv:effact}
\end{eqnarray}
is the effective action of scale $\bar m$.
This problem of continuum limit is obviously related to the previous one by 
trivial rescalings. For the limits to exist, $\ac^M$ 
(after rescaling to say unit cutoff)
has to approach a fixed point $\ac_\star$ of the Wilson RG 
as $M\to\infty$ and then $\ac_{\bar m}$
(after rescaling again) will lie on the unstable manifold of $\ac_\star$.

The Wilson idea can be applied with small changes to the time dependent
correlation functions of solutions of
stochastic (partial) differential equations. In that case the fields 
$\phi(\vv{x},t)$ depend also on time and $\mathcal{P}$ is given by the
MSR construction (and is not positive). The coarse graining takes place in space 
only whereas in time one scales. We write
\begin{eqnarray}
\oper_{(\lambda)}(\vv{x}, t):=
\lambda^{-\dd{\oper}}\oper(\lambda^{-1}\vv{x},\lambda^{\dd{t}}t)
\end{eqnarray}
where the time exponent $\dd{t}$ has to be determined. In the simplest
diffusion process $\dd{t}=-2$.

\subsection{Infra-red renormalization}
\label{sec:rg:general:ir}

In the theory of critical phenomena the concept of universality refers to 
the independence of the long distance properties of correlation functions 
on the short distance details of the Hamiltonian which in the RG language 
translates to the fact that all the Hamiltonians in the domain of attraction 
of a given fixed point have the same critical exponents.

In turbulence there is an inversion of scales.  There are 
many mechanisms which may stir the onset of turbulence at large scales
and universality refers to the independence of the inertial range scaling
on these long distance details of the forcing. It has
therefore been conjectured in the literature
(see \cite{EG94,Ga97} for a general 
discussion and further references)
 that it should be possible to prove the 
irrelevance of infra-red degrees of freedom for inertial range scaling through 
the use of an inverse renormalization group. The adjective ``inverse'' must be 
understood in the sense that the renormalization group
is constructed by averaging over fluctuations $\flu{\phi}$ with support in the
infra red. More precisely, Wilson's recursion scheme is implemented along the same
lines of the direct one encoded in formulae 
(\ref{rg:general:uv:decomposition})-(\ref{rg:general:uv:fixedpoint}) but with 
the following differences
\begin{itemize}
\item[i] The basic field $\phi$ is supported for radial values 
of the momentum in $\rop{m,\infty}$ with $m$ the infra-red cut-off. 
It is decomposed into a scaling field $\phi^{\prime}_{(1/\lambda)}$ 
with support in $\rop{\lambda\,m\,,\,\infty}$ and fluctuating field 
$\flu{\phi}$ with support in $\rop{m,\lambda\, m}$. 
In view of the inversion, the asymptotic regime is reached now for large 
values of $\lambda$.
\item[ii] If the (inversely) renormalized actions $\ac_{\lambda}$ converge
to a fixed point $\ac_{\star}$  as $\lambda$ tends to infinity then  
$\ac_{\star}$ describes the universal short distance properties of the theory.

\end{itemize}

In the following section the idea of inverse renormalization will be applied to the
Kraichnan model to provide a validation of scaling complementary to the zero
modes picture of section~\ref{sec:anomalous}.  

\section{Inverse (infra-red) renormalization group for the Kraichnan model}
\label{sec:ir}

The idea to investigate the Kraichnan model by implementing a Wilson's 
infra-red recursion scheme was put forward first in ref.'s \cite{Ga96,Ga97}. 
There, it was argued that the Kraichnan model with a
quasi-Lagrangian velocity field  \cite{BL87} had a inverse RG fixed point
and scaling fields with dimensions given by the exponents
found from the zero mode analysis of structure functions.

The purpose here it to carry out such an analysis in more detail
  to the canonical Eulerian representation of the 
Kraichnan model. 

\subsection{General settings for inverse renormalization}
\label{sec:ir:general}

It was observed in \cite{Ga96,Ga97} that one should consider the IRG
transformation in the space of MSR actions without the forcing which should
be treated as the correlation functions. The basic fields are 
\beq
\phi=(\theta\,,\btheta\,,\vv{v})
\eeq
The starting point is the measure
\begin{eqnarray}
e^{-\ac(\phi)}\mes\sbr{\phi}=
e^{-\ac_v(\theta\,,\btheta\,,\vv{v})}d\mu_R(\theta,\btheta)d\mu_D(\vv{v})
\label{ir:general:measure}
\end{eqnarray}
where
\begin{eqnarray}
\ac_{v}(\theta\,,\btheta\,,\vv{v})
:=-\imath\,\spr{\btheta}{v^{\alpha}\de_{\alpha}\theta}=
\matgrap{1.0cm}{1.0cm}{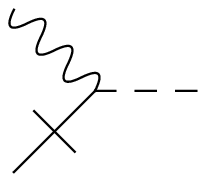}
\label{ir:general:vertex}
\end{eqnarray}
and $\mu_D$ the Gaussian measure with covariance $D$ and $\mu_R$ 
is the Gaussian "measure" with "covariance" given by the free 
response function \eq{msr:freepropagator}. To keep to the RG formalism
discussed in Section 10 we introduce to $R$ the same
infrared cutoff $m$ as in the velocity covariance $D$. This is
only for notational convenience as can easily be checked below.

The IRG transformation is defined by decomposing the fields as follows.
The scaling exponents of time and the velocity are tied together
by Galilean invariance, i.e. they are chosen so
as to preserve the material derivative
\beq
\nabla_{t}=\de_{t}+v^{\alpha}\de_{\alpha}
\eeq
under scaling. This means that
\beq
\dd{v }=-1-\dd{t} .
\label{ir:general:galilean}
\eeq
The velocity field is decomposed into a scaling and fluctuating part
\beq
\vv{v}=\vv{v}^{\prime}_{(1/\lambda)}+\flu{\vv{v}}.
\label{ir:general:velocitydec}
\eeq
It is readily seen that $\vv{v}^{\prime}$ is again Gaussian with covariance
\beq
\av{v^{\prime\alpha}(\vv{x},t)v^{\prime \beta}(\vv{y},s)}
=\lambda^{-2\,\dd{v}-\dd{t}-\xi}\delta(t-s)\,D^{\alpha\,\beta}(\vv{x},m)
\label{ir:general:velocityprime}
\eeq
whereas the covariance of the velocity fluctuation denoted by
\beq
\flu{v}^{\alpha}(\vv{x},t)\flu{v}^{\beta}(\vv{y},s)
\setlength{\unitlength}{0.1mm}
\begin{picture}(5,0.6)(0,0)
\put(-270,-20){\line(1,0){150}}
\put(-270,-20){\line(0,1){10}}
\put(-120,-20){\line(0,1){10}}
\end{picture}
:=\delta(t-s)\,\flu{D}^{\alpha\,\beta}(\vv{x}-\vv{y},m)
\eeq
is given as 
\beq 
\flu{D}^{\alpha\,\beta}(\vv{x},m)=D_{0}\,\xi\,\int\fvol{p}\,  
\frac{e^{\imath\,\vv{p}\cdot \vv{x}}}
{p^{d+\xi}}\,\Pi_{\alpha\,\beta}(\hat{\vv{p}})
\fset{[m,\lambda\,m]}{p}.
\label{ir:general:velcor}
\eeq
Similar decomposition of the fields $\theta$ and $\btheta$
\sbeq
\theta=\theta_{(1/\lambda)}^{\prime}+\flu{\theta}
\label{ir:general:scalardec}
\smeq
\btheta=\btheta_{(1/\lambda)}^{\prime}+\flu{\btheta}
\label{ir:general:ghostdec}
\seeq 
leads to 
\begin{eqnarray}
\av{\theta^{\prime}(\vv{x},t)\btheta^{\prime}(\vv{y},s)}=
\lambda^{-\eta_\theta-\eta_{\btheta} +d}
R(\vv{x}-\vv{y},t-s, \lambda^{2+\dd{t}}\eddy)
\equiv R^{\prime}(\vv{x}-\vv{x},t-s) .
\label{ir:general:response}
\end{eqnarray}
Thus we will fix
\beq
\dd{\theta}+\dd{\btheta}=d
\label{ir:general:fields}.
\eeq
The fluctuation covariance is given by
\beq
\flu{R}(\vv{x}-\vv{y},t-s)\equiv
\flu{\theta}(\vv{x},t)\flu{\btheta}(\vv{y},s)
\setlength{\unitlength}{0.1mm}
\begin{picture}(5,0.6)(0,0)
\put(-230,-20){\line(1,0){130}}
\put(-230,-20){\line(0,1){10}}
\put(-100,-20){\line(0,1){10}}
\end{picture}
=\imath H_{0}\rbr{t-s}
\int\fvol{p}\,e^{\imath \vv{p}\cdot (\vv{x}-\vv{y}) -
\eddy\,p^2\,(t-s)}\fset{[m,\lambda m]}{p}
\label{ir:general:covariance}
\eeq
The IRG transformation of the measure (\ref{ir:general:measure}) is given by
\begin{eqnarray}
e^{-\re{\lambda }\ac(\phi)}\mes\sbr{\phi^{\prime}}
=e^{-\ac^{\prime}_v(\theta^{\prime},\btheta^{\prime},\vv{v}^{\prime})}
d\mu_{R^{\prime}}(\theta^{\prime},\btheta^{\prime})
d\mu_{D^{\prime}}(\vv{v}^{\prime})
\label{ir:general:measureprime}
\end{eqnarray}
where
\begin{eqnarray}
\ac^{\prime}_v(\theta^{\prime},\btheta^{\prime},\vv{v}^{\prime})
=-\log\int e^{-\ac_v(\theta,\btheta,\vv{v})}
d\mu_{\delta R}(\delta\theta,\delta\btheta)d\mu_{\delta D}(\flu{\vv{v}} )
\label{ir:general:aprime}
\end{eqnarray}
and $\theta$, $\btheta$ and $\vv{v}$ have been decomposed according to 
\eq{ir:general:scalardec},\eq{ir:general:ghostdec} and \eq{ir:general:velocitydec}.

\subsection{Infra-red renormalization group flow of the action}
\label{sec:ir-action}

Let us do first the $\delta v$ integral in (\ref{ir:general:aprime}) -
\begin{eqnarray}
e^{-\ac^{\prime}_v(\phi^{\prime})}=\int e^{\imath \,\spr{\btheta}
{{v^{\prime}}^{\alpha}\de_{\alpha}\theta} -\frac{1}{2}
\spr{\btheta \de_{\alpha} \theta}
{\flu{D}^{\alpha\,\beta}\btheta\de_{\beta} \theta}}
d\mu_{\flu{R}}(\flu{\theta},\flu{\btheta})
\end{eqnarray} 
where 
\beq
\theta=\theta^{\prime}_{(1/\lambda)}+\flu{\theta} 
\eeq
The second
exponent on the RHS equals at $\flu{\theta}=0=\flu{\btheta}$
\beq
\spr{\btheta^{\prime}_{(1/\lambda)} \de_{\alpha} \theta^{\prime}_{(1/\lambda)}}
{\flu{D}^{\alpha\,\beta}\btheta^{\prime}_{(1/\lambda)}\de_{\beta} 
\theta^{\prime}_{(1/\lambda)}}
=\lambda^{2-\xi+\dd{t}}
\spr{\btheta^{\prime} \de_{\alpha} \theta^{\prime}}
{\flu{D}^{\alpha\,\beta}_{(\lambda)}\btheta^{\prime}\de_{\beta} 
\theta^{\prime}}.
\eeq
The large scale velocity is dominated by the constant mode of the
velocity field
\beq 
\flu{D}^{\alpha\,\beta}_{(\lambda)}
=\rbr{\lambda^{\xi}-1} \eddy\,\delta^{\alpha\,\beta}-
\flu{d}^{\alpha\,\beta}_{\star}(\vv{x},m)
+o\rbr{\frac{1}{\lambda}}
\label{ir:action:velcor}
\eeq
where
\beq
\flu{d}^{\alpha\,\beta}_{\star}(\vv{x},m)=D_{0}\,\xi\,
\int\fvol{p}\,  \frac{1-e^{\imath\,\vv{p}\cdot \vv{x}}}{p^{d+\xi}}\,
\Pi_{\alpha\,\beta}(\hat{\vv{p}})\fset{[0,m]}{p}
\label{ir:action:structure}
\eeq
Due to the unstable constant mode of the velocity field the only
scaling under which the IRG stabilizes as $\lambda$ tends to infinity 
is the Gaussian scaling
\beq
\dd{t}=-2.
\label{ir:action:time}
\eeq
From  \eq{ir:general:velocityprime} we see that  the covariance $D^{\prime}$
then tends to zero i.e. the velocity field disappears from the action.
In that limit then
\begin{eqnarray}
e^{-\ac^{\prime}_v(\phi^{\prime})}=\lim_{\lambda\to\infty}\,\int
e^{-\frac{\eddy}{2}\int dt \rbr{\int dx\, \btheta \de_{\alpha} \theta}^2}
d\mu_{\delta R}(\delta\theta,\delta\btheta).
\end{eqnarray} 
But 
\beq
\int dx\, \btheta \de_{\alpha} \theta=
\int dx\, \btheta^{\prime}_{(1/\lambda)} \de_{\alpha} 
\theta^{\prime}_{(1/\lambda)}+\int dx\, \delta\btheta \de_{\alpha} \delta\theta
\eeq
due to the disjoint supports of $\btheta^{\prime}_{(1/\lambda)}$
and $\delta\btheta$ in momentum space. Thus we end up
with the fixed point (up to a constant)
\beq
\ac_v^{\star}=\frac{\eddy}{2}\int dt \rbr{\int dx\, \btheta \de_{\alpha} \theta}^2 .
\label{ir:action:actionfp}
\eeq
The action  \eq{ir:action:actionfp} coincides with 
\beq
\ac_v^{\star}= -\ln
\av{ e^{\imath \spr{\btheta}{v^{\alpha}_{\star}\de_{\alpha}\theta}}}_{v^{\star}}
\eeq
where $\vv{v}_{\star}$ is a velocity field with constant covariance
\beq
\av{v_{\star}^{\alpha}(t) v_{\star}^{\beta}(s)}=
\delta(t-s) \eddy \delta^{\alpha\,\beta} 
\eeq
with $\eddy$ given by \eq{velasympt:eddy}. 
The only nonzero correlations of the fixed point theory are 
the multiple response functions
\beq
\av{\prod_{i=1}^{n}\theta(\vv{x}_i,t_i)\btheta(\vv{y}_i,s_i)}
=\eval{\prod_{i=1}^{n}\frac{\delta}{\flu{\jmath}(\vv{x}_i,t_i)}
\frac{\delta}{\flu{\bj}(\vv{y}_i,s_i)}}{\jmath=\bj=0}
\av{e^{\imath \spr{\jmath}{\rv H \bj}}}_{\vv{v}}
\label{ir:action:response}
\eeq
and thus they coincide with the velocity averages of the same response functions in
a random constant velocity field. 

Using  Hopf equations \eq{hopf:general}  the 
response functions of the Kraichnan model
with only two times are given as heat kernels
\beq
\av{\prod_{i=1}^{n}\theta(\vv{x}_i,t)\btheta(\vv{y}_i,s)}
=\exop{n}{t-s}(\vv{x}_1,\dots,\vv{x_n},\vv{y}_1,\dots,\vv{y}_n)
\label{ir:action:Green}
\eeq  
of the operators 
\beq
\mathcal{M}_{n}=\frac{\kappa}{2}\sum_{i}\de^2_{\vv{x}_i}+\frac{D\,m^{-\xi}}{2}
\rbr{\sum_{i}\de_{\vv{x}_i}}^2+\sum_{i<j}d^{\alpha\,\beta}
\de_{\vv{x}^\alpha_{i}}\de_{\vv{x}^\beta_{j}}
\label{ir:action:mn}
\eeq  
and the multiple time ones are simple combinations of these
with various $n$'s. Under the scaling of space and time where
$\eta_t=-2$ the $d^{\alpha\,\beta}$ will contract by $L^{-\xi}$
thus explaining the fixed point.

The reason for the trivial fixed point of the IRG lies, as stressed
in  \cite{Ga96,Ga97}, in the constant mode of the velocity field.
As we saw in Section 4 this mode decouples in the stationary
equal time correlators i.e. the second term drops out from the operators 
\eq{ir:action:mn} when acting on translationally invariant
functions. However, it doesn't decouple in the unequal time stationary
correlators. For example, the two-point function of the scalar is given by
\beq
\av{\theta(\vv{x},t)\theta(\vv{y},s)}=
\exop{1}{t-s} \av{\theta(\vv{x},s)\theta(\vv{y},s)}
\label{ir:int:example}
\eeq
and the time dependence is dominated by the eddy diffusivity $\eddy=D\,m^{-\xi}$.

It is possible to modify the velocity covariance so that the zero mode
decouples also from the time evolution while the stationary state
is left unchanged \cite{Ga96,Ga97}, an example being the quasi-Lagrangean
velocity field. Then one may choose scaling $\eta_t=-2+\xi$ and the
IRG fixed point is less trivial. However, in both cases 
the scaling properties of the fixed point action bear very little 
information about the structure functions $\stf{2 n}$.
To understand their scaling we need to find the relevant
scaling fields. In the next section we show how they appear
in our model of Eulerian velocities thus bypassing the
more involved quasi-Lagrangean formalism proposed in \cite{Ga96,Ga97}.
 
\section{Infra-red renormalization group analysis of the structure functions}
\label{sec:ir:structure}

By definition structure functions are the averages
\beq
\stf{2\,n}(\vv{x})=\av{\sbr{\theta(\vv{x})-\theta(0)}^{2n}
\frac{[\imath\,\spr{\btheta}{f}]^{2n}}{(2 n)!}}
\eeq
with respect to the  measure \eq{ir:general:measure} and the
Gaussian measure of the forcing field $f$.
For the renormalization group calculations it is convenient to take the forcing
covariance  \eq{ps:forcing} as 
\beq
\av{\fou{f}(\vv{p}_1,t_1) \fou{f}(\vv{p}_2,t_2)}
=F_{\star}\frac{\delta(p_1-\mf)}{\mf^{d-1}}\,
\dirac{d}{\vv{p}_1+\vv{p}_2}\,\delta(t-s)
\label{ir:structure:correlation}
\eeq
where $m_F=L_F^{-1}$.
The forcing is in such a case isotropic. An example of anisotropic
forcing is
\beq
\av{\fou{f}(\vv{p}_1,t_1) \fou{f}(\vv{p}_2,t_2)}
=F_{\star}\frac{\dirac{d}{\vv{p}_1-\vv{q}^{\star}}
+\dirac{d}{\vv{p}_1+\vv{q}^{\star}}}{2}\,
\dirac{d}{\vv{p}_1+\vv{p}_2}\,\delta(t-s)
\label{ir:structure:correlation2}
\eeq
with $\mo{q}^{\star}=\mf$. 

As in sections~\ref{sec:smallxi} and \ref{sec:gradients} we work 
under the assumption 
\beq
\frac{\mf}{m}\,\ll\,1.
\label{ir:structure:hypothesis}
\eeq
It should be stressed that this is for simplicity only and  the conclusions 
of the renormalization group analysis of the structure function hold true 
for an arbitrary value of the ratio of the integral scales provided the 
requirement of infra-red forcing is satisfied \cite{MG06}. In the language 
of renormalization group this latter requirement means that 
\beq
f=\flu{f}
\eeq
which obviously is true as soon as $m_F\in [0,\lambda m]$.
Thus the structure functions are given as the expectation value
\beq
S_{2n}(\vv{x};\phi)=\sbr{\mathcal{J}(\vv{x})}^{2\,n}\,
\frac{\imath^{2n}\,\spr{\flu{\btheta}}{F\,\flu{\btheta}}^{n}}{2^{n} n!}
\label{ir:structure:structure}
\eeq
where  $F$ is the spatial part of the forcing correlation and
using invariance under $\vv{x}\to -\vv{x}$ we replaced the
scalar increment by
\beq
\mathcal{J}(\vv{x})=
\frac{1}{2}\rbr{\theta(\vv{x},0)+\theta(-\vv{x},0)-2\,\theta(0,0)}
\label{ir:structure:increment}
\eeq

In the limit $\mf$ tending to zero, the structure function operator
involves of powers of
\beq
\lim_{\mf\downarrow 0} \spr{\flu{\btheta}}{F\,\flu{\btheta}}
\propto \int_{-\infty}^{\infty} dt\, \flu{\fbtheta}(0,t)\flu{\fbtheta}(0,t)
\label{ir:structure:forcing}
\eeq
which is a field functional {\em local in momentum space}.

\subsection{Scaling field for the structure functions}

At zeroth order in $\xi$ the isotropic component of the 
renormalized structure function has the scaling limit
\beq
\lim_{\lambda \uparrow \infty}\lambda^{\zeta_{2n,j}}\int d \Omega_{d} 
Y_{j0}^{*}(\hat{\vv{x}})
\,({\cal L}_{\lambda} S_{2n})
\rbr{\frac{\vv{x}}{\lambda};\phi}
=\int d \Omega_{d} Y_{j0}^{*}(\hat{\vv{x}})\stf{2\,n}^{(0)}(\vv{x})
\label{ir:structure:zero}
\eeq
for
\beq
\zeta_{2n,j}=2\,n
\eeq
 
\subsubsection{First order in $\xi$}

We will work out perturbative corrections to \eq{ir:structure:zero} 
in the limit $m_F\to 0$.
In this limit there is no momentum flow  along scalar correlation lines
even at finite $\lambda$ . In consequence, couplings generated by 
Wick contractions of scalar fields in the interaction vertex 
\eq{ir:general:vertex} with ghost fields in the forcing vertex 
\eq{ir:structure:forcing} factorize as
\beq
\lefteqn{
\spr{\btheta_{(1/\lambda)}}{v^{\alpha}_{(1/\lambda)}\de_{\alpha}\flu{\theta}}
\spr{\flu{\btheta}}{F\,\flu{\btheta}}\flu{\mathcal{J}}
\rbr{\frac{\vv{x}}{\lambda}}
\setlength{\unitlength}{0.1mm}
\begin{picture}(5,0.6)(0,0)
\put(-340,-25){\line(1,0){70}}
\put(-340,-25){\line(0,1){15}}
\put(-270,-25){\line(0,1){15}}
\put(-170,-20){\line(1,0){60}}
\put(-170,-20){\line(0,1){10}}
\put(-110,-20){\line(0,1){10}}
\end{picture}}
\nn&&
=\spr{\btheta_{(1/\lambda)}}{v^{\alpha}_{(1/\lambda)}}\int\fvol{p}\,
\imath p_{\alpha}\frac{2 \cos_{\angle}\rbr{\frac{\vv{x}}{\lambda}\cdot \vv{p}}-1}
{2\,p^2}\fou{F}(p)
\eeq
which vanishes because of parity (recall that $\fou{F}(p)\propto\delta(p)$
as  $m_F\to 0$). At leading order in $\xi$
the scaling field associated to the structure function operation
is given by    
\beq
\lefteqn{\lim_{\lambda\uparrow \infty}\lambda^{\zeta_{2n,j}}
\int d \Omega_{d} Y_{j0}^{*}(\hat{\vv{x}}) \,
({\cal L}_{\lambda}S_{2n})\rbr{\frac{\vv{x}}{\lambda};\phi}=}
\nn&&
\lim_{\lambda\uparrow \infty}
\int d \Omega_{d} Y_{j0}^{*}(\hat{\vv{x}}) 
\cbr{(1+\zeta_{2n,0}^{(1)}\ln\lambda)\,\stf{2n}(\vv{x})
+\xi\,\lambda^{\zeta_{2n}^{(0)}} \,\eval{\der{}{\xi}}{\xi=0}
\!\!\!
({\cal L}_{\lambda}S_{2n})\rbr{\frac{\vv{x}}{\lambda};\phi}}+O(\xi^2)
\label{ir:structure:limit}
\eeq
where
\beq
\lefteqn{\eval{\der{}{\xi}}{\xi=0}\!\!\!
({\cal L}_{\lambda}S_{2n})\rbr{\vv{x};\phi}}
\nn
&&=
\frac{1}{2}\,\rbr{\begin{array}{c} 2n \\ 2n-2 \end{array}}
\,\stf{2n-2}\rbr{\vv{x}}
\sbr{\flu{\stfv}_{(1;4)}^{(0)\,\alpha\,\beta}\rbr{\vv{x},\lambda}
+\sffv_{(1;4)}^{(0)\,\alpha\,\beta}(\vv{x},\phi_{(1/\lambda)})}
(\de_{\alpha}\de_{\beta}\stf{2})\rbr{\vv{x}}
\nn
&&+3\,\rbr{\begin{array}{c} 2n \\ 2n-4 \end{array}}
\stf{2n-4}\rbr{\vv{x}}
\,\sbr{\flu{\stfv}_{(1;4)}^{(0)\,\alpha\,\beta}\rbr{\vv{x},\lambda}
+\sffv_{(1;4)}^{(0)\,\alpha\,\beta}(\vv{x},\phi_{(1/\lambda)})}
(\de_{\alpha}\stf{2})\rbr{\vv{x}}
(\de_{\beta}\stf{2})\rbr{\vv{x}}
\eeq
The vertices are given by the small scale field functional
%%%%%%%%%%%%%%%%%%%%%%%%%%%%%%%%%%%%%%%%%%%%%%%%%%%%%%%%%%%%%%%%%%%%%%%
%%%%%%%%%%%%%%%%%%%%%%%%%%%%%%%%%%%%%%%%%%%%%%%%%%%%%%%%%%%%%%%%%%%%%%%
\beq
\sffv_{(1;4)}^{\alpha\,\beta}(\vv{x},\phi_{(1/\lambda)})
=\mathcal{J}^{2}_{(1/\lambda)}(\vv{x})
\spr{\btheta_{(1/\lambda)}}{v_{(1/\lambda)}^{\alpha}}
\spr{\btheta_{(1/\lambda)}}{v^{\beta}_{(1/\lambda)}}
:=\mathcal{I}_{\vv{x}} 
\matgrap{1.4cm}{1.0cm}{EPSFIGURES/vertex14.eps}
\setlength{\unitlength}{0.1cm}
\begin{picture}(0,0) 
\put(6.0,5.0){\mbox{\small$\alpha$}} 
\put(6.0,-4.0){\mbox{\small$\beta$}} 
\put(-6.0,4.6){\mbox{$\bigstar$}} 
\put(-2.4,0.0){\mbox{$\bigstar$}} 
\put(-6.0,-4.2){\mbox{$\bigstar$}} 
\end{picture}
\label{ir:structure:vertex}
\eeq
and the averaged value of its large scale (small momentum) counterpart 
\beq
\flu{\stfv}^{\alpha\,\beta}_{(1;4)}\rbr{\vv{x},\lambda}
=\frac{2\,D_{0}\,m^{\xi}}{D}\int
\fvol{p}\frac{1-e^{\imath \vv{p}\cdot \vv{x}}}{p^2}
\frac{\Pi^{\alpha\,\beta}(\hat{p})}{p^{d+\xi}}\fset{\lop{m,\lambda m}}{p}
=\mathfrak{P}_{\lambda}\mathcal{I}_{\vv{x}} 
\matgrap{1.4cm}{1.0cm}{EPSFIGURES/vertex14.eps}
\setlength{\unitlength}{0.1cm}
\begin{picture}(0,0) 
\put(6.0,5.0){\mbox{\small$\alpha$}} 
\put(6.0,-4.0){\mbox{\small$\beta$}} 
\end{picture}
\label{ir:structure:average}
\eeq
%%%%%%%%%%%%%%%%%%%%%%%%%%%%%%%%%%%%%%%%%%%%%%%%%%%%%%%%%%%%%%%%%%%%%%%
%%%%%%%%%%%%%%%%%%%%%%%%%%%%%%%%%%%%%%%%%%%%%%%%%%%%%%%%%%%%%%%%%%%%%%%
whilst the notation \eq{smallxi:notation} is adopted for derivatives 
with respect to the H\"older exponent $\xi$.
The symbol $\mathfrak{P}_{\lambda}$ in \eq{ir:structure:average} 
denotes the restriction of the momentum support of the velocity 
correlation. As $\lambda$ tends to infinity the diagram has the
limit
\beq
\lambda^2\,
\flu{\stfv}^{(0)\,\alpha\,\beta}_{(1;4)}\rbr{\frac{\vv{x}}{\lambda},\lambda}
\overset{\lambda\uparrow \infty}{=}\ln\lambda\,
\frac{(d+1)x^2}{(d-1)(d+2)}\ivec^{\alpha\,\beta}(\hat{\vv{x}},2)
+\flu{\stfv}^{(0)\,\alpha\,\beta}_{\star\,(1;4)}(\vv{x})
\eeq
Little algebra yields the final form of \eq{ir:structure:limit} 
\beq
\lefteqn{\lim_{\lambda\uparrow \infty}\lambda^{\zeta_{2n,j}}
\int d \Omega_{d} Y_{j0}^{*}(\hat{\vv{x}}) \,
({\cal L}_{\lambda}S_{2n})\rbr{\vv{x};\phi}
=\int d \Omega_{d} Y_{j0}^{*}(\hat{\vv{x}}) \stf{2n}(\vv{x})}
\nn&&
+\frac{\xi}{2}\,\int d \Omega_{d} Y_{j0}^{*}(\hat{\vv{x}})
\,\sbr{\flu{\stfv}^{(0)\,\alpha\,\beta}_{\star\,(1;4)}(\vv{x})
+\sffv_{(1;4)}^{(0)\,\alpha\,\beta}(\phi)}
\de_{\alpha}\de_{\beta}\stf{2n}(\vv{x})+O(\xi^2)
\label{ir:structure:scalingfield}
\eeq
provided $\zeta_{2n,j}$ is given by \eq{smallxi:firstorder:exponent}.

\subsubsection{Second order in $\xi$}

The calculation of the structure function  scaling field can be
inferred from the perturbation theory of section~\ref{sec:smallxi}.
The field independent part of the scaling field is given by
the structure function diagrams  in the presence of a fixed
arbitrary ultra-violet cutoff. The field dependent part 
 is obtained by pruning lines in these diagrams
 and replacing 
them with the corresponding pair of ultra-violet scaling fields.  
Accordingly, up to second order accuracy in $\xi$, 
the renormalized structure function takes the form
\begin{eqnarray}
\lefteqn{(\relin{\lambda}S_{2\,n})(\vv{x};\phi)}
\nn&&
=\cbr{1+ \frac{\xi}{2}\sum_{r=0}^{1}\xi^r\,\sbr{
\sffv_{(1;4)}^{(r)\,\alpha\,\beta}(\vv{x},\phi_{(1/\lambda)})+
\flu{\stfv}^{(r)\,\alpha\,\beta}_{(1;4)}(\vv{x},\lambda)
}}\de_\alpha\de_\beta\,\stf{2\,n}^{(0)}(\vv{x})
\nn&&
+\frac{\xi^2}{2}\,
\cbr{\sffv_{(2;4)}^{(0)\,\alpha\,\beta}(\vv{x},\phi_{(1/\lambda)})+
\flu{\stfv}^{(0)\,\alpha\,\beta}_{(2;4)}(\vv{x},\lambda)
}\de_\alpha\de_\beta\,\stf{2\,n}^{(0)}(\vv{x})
\nn&&
+\xi^2
\cbr{\sffv_{(2;6)}^{(0)\,\alpha\,\beta\,\mu}(\vv{x},\phi_{(1/\lambda)})
+\flu{\stfv}^{(0)\,\alpha\,\beta\,\mu}_{(2;6)}(\vv{x},\lambda)}
\de_\alpha\de_\beta\,\de_{\mu}\stf{2\,n}^{(0)}(\vv{x})
\nn&&
+\frac{\xi^2}{8}\,
\cbr{\sffv_{(2;8)}^{(0)\,\alpha\,\beta\,\mu\,\nu}(\vv{x},\phi_{(1/\lambda)})+
\flu{\stfv}_{(2;8)}^{(0)\,\alpha\,\beta\,\mu\,\nu}(\vv{x},\lambda)}
\de_\alpha\de_\beta\,\de_{\mu}\de_{\nu}\,\stf{2\,n}^{(0)}(\vv{x})
+O(\xi^3,L_{F}^{-2})
\label{ir:structure:secondorder}
\end{eqnarray}
The $\flu{\stfv}^{\bullet}_{\bullet}$'s denote the diagrams
\eq{smallxi:secondorder:v24},\eq{smallxi:secondorder:v26}, and  
\eq{smallxi:secondorder:v26} where velocity correlations have 
momentum support in $\lop{m\,,\lambda m}$. Infra-red logarithmic
behavior of the field dependent vertices is identified by considering 
the scaling limits
%%%%%%%%%%%%%%%%%%%%%%%%%%%%%%%%%%%%%%%%%%%%%%%%%%%%%%%%%%%%%%%%%%%%%%%%%%%%
%%%%%%%%%%%%%%%%%%%%%%%%%%%%%%%%%%%%%%%%%%%%%%%%%%%%%%%%%%%%%%%%%%%%%%%%%%%%
\beq
\lefteqn{\lim_{\lambda \uparrow \infty} \lambda^{2}\,\sffv_{(2;4)}^{\alpha\,\beta}
\rbr{\frac{\vv{x}}{\lambda},\phi_{(1/\lambda)}}=
\lim_{\lambda \uparrow \infty} \lambda^{2}\,
\mathcal{I}_{\vv{x}}\setlength{\unitlength}{0.1cm}
\begin{picture}(0,8) 
\put(0.0,0.0){\mbox{$\matgrap{2.4cm}{1.0cm}{EPSFIGURES/vertex24.eps}$}}
\put(26.0,5.0){\mbox{\small$\alpha$}} 
\put(26.0,-4.0){\mbox{\small$\beta$}} 
\put(4.0,4.6){\mbox{$\bigstar$}} 
\put(9.0,0.0){\mbox{$\bigstar$}} 
\put(4.0,-4.2){\mbox{$\bigstar$}} 
\end{picture}
\hspace{2.8cm}}
\nn[0.5cm]
&=&\frac{2\,(d+1)\ln\lambda}{(d-1)(d+2)}
\mathfrak{V}^{\mu\,\nu}_{(1;4)}\rbr{\vv{x},\phi}
\mathcal{Q}^{\alpha\,\beta}_{\mu\,\nu}
+O(\lambda^{0},\xi)
\label{structure:secondorder:v24}
\eeq
%%%%%%%%%%%%%%%%%%%%%%%%%%%%%%%%%%%%%%%%%%%%%%%%%%%%%%%%%%%%%%%%%%%%%%%%%%%%
%%%%%%%%%%%%%%%%%%%%%%%%%%%%%%%%%%%%%%%%%%%%%%%%%%%%%%%%%%%%%%%%%%%%%%%%%%%%
and
%%%%%%%%%%%%%%%%%%%%%%%%%%%%%%%%%%%%%%%%%%%%%%%%%%%%%%%%%%%%%%%%%%%%%%%%%%%%
%%%%%%%%%%%%%%%%%%%%%%%%%%%%%%%%%%%%%%%%%%%%%%%%%%%%%%%%%%%%%%%%%%%%%%%%%%%%
\beq
\lefteqn{
\lim_{\lambda \uparrow \infty} \lambda^{3}\,\sffv^{\alpha\,\beta\,;\mu}_{(2;6)}
\rbr{\frac{\vv{x}}{\lambda},\phi_{(1/\lambda)}}=
\lim_{\lambda \uparrow \infty} \lambda^{3}\,\mathcal{I}_{\vv{x}}
\setlength{\unitlength}{0.1cm}
\begin{picture}(0,14) 
\put(0.0,0.0){\mbox{$\matgrap{2.8cm}{1.0cm}{EPSFIGURES/vertex26.eps}$}}
\put(9.0,9.0){\mbox{\small$\mu$}} 
\put(18.0,9.0){\mbox{\small$\alpha$}} 
\put(24.0,-6.0){\mbox{\small$\beta$}} 
\put(4.0,3.0){\mbox{$\bigstar$}} 
\put(8.5,-1.2){\mbox{$\bigstar$}} 
\put(4.0,-5.8){\mbox{$\bigstar$}} 
\end{picture}
\hspace{2.8cm}}
\nn[0.5cm]
&=&\frac{2(d+1)\ln\lambda}{(d-1)(d+2)}
\mathfrak{V}^{\mu\,\rho}_{(1;4)}\rbr{\vv{x},\phi}
\mathcal{Q}^{\alpha\,\beta}_{\rho\,\sigma} x^{\sigma}
+O(\lambda^{0},\xi)
\label{structure:secondorder:v26}
\eeq
and
%%%%%%%%%%%%%%%%%%%%%%%%%%%%%%%%%%%%%%%%%%%%%%%%%%%%%%%%%%%%%%%%%%%%%%%%%%%%%%%
%% the commented diagram seems to have problems when compiled with pdflatex   %
%%%%%%%%%%%%%%%%%%%%%%%%%%%%%%%%%%%%%%%%%%%%%%%%%%%%%%%%%%%%%%%%%%%%%%%%%%%%%%%
%\beq
%\lefteqn{
%\lim_{\lambda \uparrow \infty} \lambda^{4}\,\sffv^{\alpha\,\beta\,;\mu\,\nu}_{(2;8)}
%\rbr{\frac{\vv{x}}{\lambda},\phi_{(1/\lambda)}}
%=\lim_{\lambda \uparrow \infty} \lambda^{4}\,\mathcal{I}_{\vv{x}}
%\setlength{\unitlength}{0.1cm}
%\begin{picture}(0,0) 
%\put(36.0,6.2){\mbox{$\matgrap{-1.4cm}{1.0cm}{EPSFIGURES/vertex14.eps}$}}
%\put(16.0,5.0){\mbox{\small$\alpha$}} 
%\put(16.0,-4.0){\mbox{\small$\beta$}} 
%\put(0.0,0.0){\mbox{$\matgrap{1.4cm}{1.0cm}{EPSFIGURES/vertex14.eps}$}}
%\put(20.0,5.0){\mbox{\small$\mu$}} 
%\put(20.0,-4.0){\mbox{\small$\nu$}} 
%\put(4.0,4.6){\mbox{$\bigstar$}} 
%\put(8.5,0.0){\mbox{$\bigstar$}} 
%\put(4.0,-4.2){\mbox{$\bigstar$}} 
%\end{picture}}
%\nn[0.5cm]
%&=&\frac{4\,(d+1)\ln\lambda}{(d-1)(d+2)}
%\mathfrak{V}^{\alpha\,\beta}_{(1;4)}\rbr{\vv{x},\phi}
%\mathcal{Q}^{\mu\,\nu}_{\rho\,\sigma}x^{\rho} x^{\sigma}
%+O(\lambda^{0},\xi)
%\label{structure:secondorder:v28}
%\eeq
\beq
\lefteqn{
\lim_{\lambda \uparrow \infty} \lambda^{4}\,\sffv^{\alpha\,\beta\,;\mu\,\nu}_{(2;8)}
\rbr{\frac{\vv{x}}{\lambda},\phi_{(1/\lambda)}}
=\lim_{\lambda \uparrow \infty} \lambda^{4}\,\mathcal{I}_{\vv{x}}
\setlength{\unitlength}{0.1cm}
\begin{picture}(0,0) 
\put(37.3,6.2){\mbox{$\matgrap{-3.8cm}{1.0cm}{EPSFIGURES/vertex28.eps}$}}
\put(16.0,5.0){\mbox{\small$\alpha$}} 
\put(16.0,-4.0){\mbox{\small$\beta$}} 
\put(20.0,5.0){\mbox{\small$\mu$}} 
\put(20.0,-4.0){\mbox{\small$\nu$}} 
\put(4.0,4.6){\mbox{$\bigstar$}} 
\put(8.5,0.0){\mbox{$\bigstar$}} 
\put(4.0,-4.2){\mbox{$\bigstar$}} 
\end{picture}}
\nn[0.5cm]
&=&\frac{4\,(d+1)\ln\lambda}{(d-1)(d+2)}
\mathfrak{V}^{\alpha\,\beta}_{(1;4)}\rbr{\vv{x},\phi}
\mathcal{Q}^{\mu\,\nu}_{\rho\,\sigma}x^{\rho} x^{\sigma}
+O(\lambda^{0},\xi)
\label{structure:secondorder:v28}
\eeq
%%%%%%%%%%%%%%%%%%%%%%%%%%%%%%%%%%%%%%%%%%%%%%%%%%%%%%%%%%%%%%%%%%%%%%%%%%%%
%%%%%%%%%%%%%%%%%%%%%%%%%%%%%%%%%%%%%%%%%%%%%%%%%%%%%%%%%%%%%%%%%%%%%%%%%%%%
The tensor structure of the diagrams is specified by 
\beq
\mathcal{Q}^{\alpha\,\beta}_{\mu\,\nu}:=\frac{1}{2}\de_{\mu}\de_{\nu}x^2
\ivec^{\alpha\,\beta}(\vv{x},2)=\delta^{\alpha\,\beta}\delta_{\mu\,\nu}-
\frac{\delta^{\alpha}_{\,\mu}\delta^{\beta}_{\,\nu}
+\delta^{\alpha}_{\,\nu}\delta^{\beta}_{\,\mu}}{d+1}
\eeq
Only the index symmetric part of the above diagrams contributes to
\eq{ir:structure:secondorder} owing to the contraction with fully 
symmetric quantities. 
As in sections~\eq{sec:smallxi} and  sections~\eq{sec:gradients} this
fact is emphasized by omitting semicolons between non-symmetric indices.

In section~\eq{sec:smallxi} it was shown that up to second order in $\xi$ 
the perturbative expression of any structure function in the inertial range
is compatible with that of a  homogeneous function. This information together with 
the scaling limits \eq{structure:secondorder:v24}, \eq{structure:secondorder:v26}
\eq{structure:secondorder:v28} permit to verify after some straightforward algebra 
that  the structure function operator has a finite 
scaling limit
\beq
\lefteqn{\lim_{\lambda \uparrow \infty}\lambda^{\zeta_{2n,j}}
\int d \Omega_{d} Y_{j0}^{*}(\hat{\vv{x}})
(\relin{\lambda}S_{2n})\rbr{\frac{\vv{x}}{\lambda};\phi}=
\rbr{\sum_{r=0}^{2}\frac{\xi^{r}}{r!}\,\der{^r\lambda^{\zeta_{2n,j}}}{\xi^r}}
\,\int d \Omega_{d} 
Y_{j0}^{*}(\hat{\vv{x}})\,\stf{2n}^{(0)}(\vv{x})}
\nn&&
+\xi\,\rbr{\sum_{r=0}^{1}\frac{\xi^{r}}{r!}\,\der{^r\lambda^{\zeta_{2n,j}}}{\xi^r}}
\int d \Omega_{d} Y_{j0}^{*}(\hat{\vv{x}})
\cbr{S_{2n}^{(1)}(\vv{x};\phi)
-\,\stf{2n}^{(0)}(\vv{x})\zeta^{(1)}_{2n}\ln \lambda}
\nn
&&+\frac{\xi^2}{2}
\int d \Omega_{d} Y_{j0}^{*}(\hat{\vv{x}})
\cbr{S_{2n}^{(2)}(\vv{x};\phi)
-2\,S_{2n}^{(1)}(\vv{x};\phi)\zeta^{(1)}_{2n;j}\ln \lambda
+\stf{2n}^{(0)}(\vv{x})\sbr{(\zeta^{(1)}_{2n;j}\ln \lambda)^2
-\zeta^{(2)}_{2n;j}\ln \lambda}}
\nn
&&+O(\xi^3)=S_{2\,n;j}^{\star}(\vv{x};\phi)
\eeq
with $S_{2n}^{(i)}(\vv{x};\phi)$, $i=1,2$ obtained by gathering the field dependence
in \eq{ir:structure:secondorder} according to its asymptotic behavior in 
$\lambda$ and provided the scaling dimension $\zeta_{2n,j}$ is specified by 
\eq{smallxi:secondorder:exponent}.

The conclusion of the above analysis is that the infra-red renormalization
of the structure function operator produces a well defined scaling field.
It is worth noting that this conclusion holds also had we averaged out the
velocity field from the outset. The resulting scaling field will then depend 
only on  the ultra-violet degrees of freedom of the scalar and ghost fields
but it has the same scaling exponent.

\section{Ultra-violet renormalization group}
\label{sec:uv}

Ultra-violet renormalization group addresses the question
of removing the ultraviolet cutoff in the theory stemming
from the one ($M$)  in the velocity covariance. Although
the correlation functions of the $\theta$ fields have a well
defined $M\to\infty$ limit the gradients will not have as
was discussed in Section 9. Ultra-violet renormalization will study that
divergence by finding the appropriate scaling fields.

Ultra-violet renormalization was applied to the Kraichnan model 
in ref's~\cite{AAV98,AABKV01a,AABKV01b} in the framework of the 
minimal subtraction scheme \cite{Co,LB,Va,Zi}. 
The minimal subtraction scheme has the merit to provide probably the 
most computationally efficient setting for the determination of the 
scaling exponents $\zeta_{2 n,j}$ which in \cite{AABKV01a,AABKV01b} 
were determined up to third order in $\xi$. 

The purpose of the present section is to reproduce to leading order in $\xi$ 
the same calculation using Wilson's original scheme in order to 
render the comparison with infra-red renormalization more transparent. 

\subsection{The effective action}
\label{sec:uv:action}

In order to inquire the limit $M$ tending to infinity, it is more convenient 
not to use rescalings in the renormalization group, i.e. to proceed as in 
eq. \eq{rg:general:uv:decomposition1}. The fluctuation covariances have momenta
on $[\scale,M]$ and they become in the limit of infinite ultra-violet cutoff
\beq
\lim_{M\uparrow \infty}\flu{R}(\vv{x},t)= \imath H_{0}\rbr{t}
\int\fvol{p}\,e^{\imath \vv{p}\cdot \vv{x} -
\eddy\,p^2\,t}\fset{[\scale,\infty]}{p}
\label{uv:action:response}
\eeq
and
\beq
\lim_{M\uparrow \infty}\flu{D}^{\alpha\,\beta}(\vv{x})=
\xi D_0 \int
\frac{d^dp}{(2\pi)^d}\frac{e^{\imath p\cdot \vv{x}}}{p^{d+\xi}}
\Pi^{\alpha\,\beta}(p)\fset{[\scale,\infty]}{p}
\label{uv:action:velocity}
\eeq
Since the main interest is to determine the statistical properties
of the scalar and the ghost fields, the renormalization group 
transformation will be applied directly to the interaction term
obtained by averaging out all the degrees of freedom of the velocity field.
The effective action at scale $\scale$ stabilizes trivially as $M\to\infty$ 
at all orders in perturbation theory. At first order it is simply given by 
\begin{eqnarray}
\ac_{\scale}= \frac{\xi}{2}\spr{\btheta \de_{\alpha} \theta}
{D^{(0)\,\alpha\,\beta}\btheta\de_{\beta} \theta}
+o(\xi^2)
\label{uv:action:rescaled}
\end{eqnarray}
with
\beq
D^{(r)\,\alpha\,\beta}=\eval{\der{^r}{\xi^r}}{\xi=0}D^{\alpha\,\beta}
\eeq
%\begin{eqnarray}
%\eval{\ac}{\scale}(\phi) = -\imath \,\spr{\btheta}{v^{\alpha}\de_{\alpha}\theta}
%+\frac{1}{2}\spr{\btheta \de_{\alpha} \theta}
%{D^{\alpha\,\beta}\btheta\de_{\beta} \theta}
%+\spr{\btheta}{v^{\alpha}\de_{\alpha} \flu{\theta}}
%\spr{\flu{\btheta}}{v^{\beta}\de_{\beta} \theta}
%\setlength{\unitlength}{0.1mm}
%\begin{picture}(5,0.6)(0,0)
%\put(-220,-20){\line(1,0){60}}
%\put(-220,-20){\line(0,1){10}}
%\put(-160,-20){\line(0,1){10}}
%\end{picture}
%+o(\xi)
%\label{uv:action:rescaled}
%\end{eqnarray}

\subsection{Ultra-violet renormalization group analysis of radial gradients}
\label{sec:uv:gradients}

More interesting flow is found once we look at the scaling fields involving 
gradients of the scalar. Let $\relin{\scale}$ be the linearization 
of the above renormalization group in the limit as $M$ tends to infinity. 
Acting on radial gradients ultra-violet renormalization gives
\beq
\lefteqn{\relin{\scale}\sbr{\hat{\vv{x}}\cdot \de \theta(\vv{y},t)}^{2n}=
\sbr{\hat{\vv{x}}\cdot \de \theta(\vv{y},t)}^{2n}}
\nn&&
-\imath\, \xi\,\rbr{\begin{array}{c} 2n \\ 1 \end{array}}
\sbr{\hat{\vv{x}}\cdot \de \theta(\vv{y},t)}^{2\,n-1}
\sbr{\hat{\vv{x}}\cdot \de \flu{\theta}(\vv{y},t)}
\spr{\flu{\btheta}\,\de_{\alpha} \theta}
{D^{(0)\,\alpha\,\beta}\btheta\de_{\beta} \theta}
\setlength{\unitlength}{0.1mm}
\begin{picture}(5,0.6)(0,0)
\put(-480,-20){\line(1,0){160}}
\put(-480,-20){\line(0,1){10}}
\put(-320,-20){\line(0,1){10}}
\end{picture}
\nn&&
+\xi\,\rbr{\begin{array}{c} 2n \\ 2 \end{array}}
\sbr{\hat{\vv{x}}\cdot \de \theta(\vv{y},t)}^{2\,n-2}
\sbr{\hat{\vv{x}}\cdot \de \flu{\theta}(\vv{y},t)}^{2}
\spr{\flu{\btheta} \de_{\alpha} \theta}
{D^{(0)\,\alpha\,\beta}\flu{\btheta}\de_{\beta} \theta}
\setlength{\unitlength}{0.1mm}
\begin{picture}(5,0.6)(0,0)
\put(-510,50){\line(1,0){180}}
\put(-510,40){\line(0,1){10}}
\put(-330,40){\line(0,1){10}}
\put(-510,-20){\line(1,0){410}}
\put(-510,-20){\line(0,1){10}}
\put(-100,-20){\line(0,1){10}}
\end{picture}
\label{uv:gradients:flow}
\nn&&
+O(\xi^2)
\eeq
In order to streamline the notation, on the right hand side of 
\eq{uv:gradients:flow} coarse grained fields with ultra-violet 
cutoff $\scale$ are represented by the letters $\theta$, $\btheta$.
Leading order corrections are specified by the couplings
\beq
\lefteqn{\sbr{\hat{\vv{x}}\cdot \de \flu{\theta}(\vv{y},t)}
\spr{\flu{\btheta}\,\de_{\alpha} \theta}
{D^{(0)\,\alpha\,\beta}\btheta\de_{\beta} \theta}
\setlength{\unitlength}{0.1mm}
\begin{picture}(5,0.6)(0,0)
\put(-480,-20){\line(1,0){160}}
\put(-480,-20){\line(0,1){10}}
\put(-320,-20){\line(0,1){10}}
\end{picture}=}
\nn&&
\int_{-\infty}^{t} ds \prod_{i=1}^{2} d^{d}y_{i}\,
\sbr{\hat{\vv{x}}\cdot(\de_{\vv{y}}\flu{R})(\vv{y}-\vv{y}_1,t-s)}\,
D^{\alpha_{1}\,\alpha_{2}}(\vv{y}_{1}-\vv{y}_{2})
\btheta(\vv{y}_{j},s)
\prod_{j=1}^{2}\de_{\alpha_{j}}\theta(\vv{y}_{j},s)
\label{uv:gradients:irrelevant}
\eeq
and
\beq
\lefteqn{\sbr{\hat{\vv{x}}\cdot \de \flu{\theta}(\vv{y},t)}^{2}
\spr{\flu{\btheta} \de_{\alpha} \theta}
{D^{(0)\,\alpha\,\beta}\flu{\btheta}\de_{\beta} \theta}
\setlength{\unitlength}{0.1mm}
\begin{picture}(5,0.6)(0,0)
\put(-510,50){\line(1,0){180}}
\put(-510,40){\line(0,1){10}}
\put(-330,40){\line(0,1){10}}
\put(-510,-20){\line(1,0){410}}
\put(-510,-20){\line(0,1){10}}
\put(-100,-20){\line(0,1){10}}
\end{picture}=}
\nn&&
\int_{-\infty}^{t} ds \prod_{i=1}^{2} d^{d}y_{i}\,
D^{(0)\,\alpha_{1}\,\alpha_{2}}(\vv{y}_{1}-\vv{y}_{2})
\prod_{j=1}^{2}\sbr{\hat{\vv{x}}\cdot(\de_{\vv{y}}\flu{R})(\vv{y}-\vv{y}_{j},t-s)}
\de_{\alpha_{j}}\theta(\vv{y}_{j},s)
\label{uv:gradients:relevant}
\eeq
Let us shift the variables $\vv{y}_{i}\rightarrow \vv{y}_{i}+\vv{y}$
and Taylor expand
\beq
\theta(\vv{y}+\vv{y}_{i},t+s)=
\sum_{n,k=0}^{\infty}\frac{1}{n!k!}
[(\vv{y}_{i}\cdot \de_{\vv{y}})^{n}(s\de_{t})^{k}\theta](\vv{y},t)
\label{uv:gradients:gradientexpansion}
\eeq
for $i=1,2$. Then
the leading order of this expansion gives the most singular part of the
integral as the ultra-violet cut-off tends to infinity. Higher orders
improve the ultra-violet behavior of the integral. Thus it is found
\beq
\lefteqn{\sbr{\hat{\vv{x}}\cdot \de \flu{\theta}(\vv{y},t)}
\spr{\flu{\btheta}\,\de_{\alpha} \theta}
{D^{(0)\,\alpha\,\beta}\btheta\de_{\beta} \theta}
\setlength{\unitlength}{0.1mm}
\begin{picture}(5,0.6)(0,0)
\put(-480,-20){\line(1,0){160}}
\put(-480,-20){\line(0,1){10}}
\put(-320,-20){\line(0,1){10}}
\end{picture}=}
\nn&&
\int_{0}^{\infty} ds \prod_{i=1}^{2} d^{d}y_{i}\,
\hat{\vv{x}}\cdot(\de_{\vv{y}}\flu{R})(\vv{y}_1,s)\,
D^{\alpha_{1}\,\alpha_{2}}(\vv{y}_{1}-\vv{y}_{2})
+O\rbr{\frac{1}{M^2},\frac{1}{\scale^2}}=O\rbr{\frac{1}{M^2},\frac{1}{\scale^2}}
\eeq
by parity and
\beq
\lefteqn{\sbr{\hat{\vv{x}}\cdot \de \flu{\theta}(\vv{y},t)}^{2}
\spr{\flu{\btheta} \de_{\alpha} \theta}
{D^{(0)\,\alpha\,\beta}\flu{\btheta}\de_{\beta} \theta}
\setlength{\unitlength}{0.1mm}
\begin{picture}(5,0.6)(0,0)
\put(-510,50){\line(1,0){180}}
\put(-510,40){\line(0,1){10}}
\put(-330,40){\line(0,1){10}}
\put(-510,-20){\line(1,0){410}}
\put(-510,-20){\line(0,1){10}}
\put(-100,-20){\line(0,1){10}}
\end{picture}=}
\nn&&
\gfv^{(0)\,\alpha\,\beta}_{(1:4)}(\scale,M)\sbr{\de_{\alpha} \theta
\rbr{\vv{y},t}}\sbr{\de_{\beta} \theta \rbr{\vv{y},t}}
+O\rbr{\frac{1}{M^2},\frac{1}{\scale^2}}
\eeq
where $\gfv^{\alpha\,\beta}_{(1:4)}$ was given in 
\eq{gradients:firstorder:v14} and it is here evaluated in the momentum
range $\sbr{\scale,M}$. The terms neglected are irrelevant for
the determination of the scaling dimension. This latter is exhibited
in the scaling limit which in the present context can be taken by fixing
the ratio
\beq
0\,<\,\lambda =\frac{\scale}{M}\,<\,1
\eeq
and by considering 
\beq
\lefteqn{
\lim_{M \uparrow \infty}\lambda^{\dd{\gf_{2n,j}}} 
\int d\Omega\, Y_{j0}(\hat{\vv{x}})\relin{\lambda M}
\sbr{\hat{\vv{x}}\cdot \de \theta}^{2n}
=}
\nn&&
A_{n}
\mmo{\de \theta}^{2\,n}
\cbr{1+\xi\sbr{\dd{\gf_{2n,j}}^{(1)}-
\frac{n\,(d+2\,n)}{d+2}+\frac{(d+1)\,j(d+j-2)}{2(d-1)(d+2)}}\ln\lambda} +O(\xi^2)
\label{uv:gradients:limit}
\eeq
where
\beq
\dd{\gf_{2n,j}}^{(1)}:=\eval{\der{}{\xi}}{\xi=0}\dd{(\de\theta)^{2n}}
\eeq
The derivation of \eq{uv:gradients:limit} exploits the integral identity 
\beq
\int d\Omega\, Y_{j,0}(\hat{\vv{x}})\sbr{\hat{\vv{x}}\cdot \de 
\theta(\vv{y},t)}^{2n}=A_{n,j}\,\sbr{\de \theta\cdot \de \theta}^{n}(\vv{y},t)
\eeq
with $A_{n}$ satisfying (see appendix~\ref{ap:angular} for the proof)
\beq
(d+1) \frac{A_{n-1,j}}{A_{n,j}}-2=\frac{d-1}{2\,n-1}
\sbr{d+2\,n-\frac{(d+1)\,j\,(d+j-2)}{2\,n\,(d-1)}}.
\label{uv:gradients:angular}
\eeq
Existence of scaling requires the cancellation of the $\lambda$ dependence in 
\eq{uv:gradients:limit} whence it follows
\beq
\dd{\gf_{2n,j}}^{(1)}=\frac{n\,(d+2\,n)}{d+2}-\frac{(d+1)\,j(d+j-2)}{2(d-1)(d+2)}
\eeq
Thus, $\mmo{\de \theta}^{2\,n}+O(\xi^2)$ is a scaling 
field and the isotropic component of the radial gradients scales with 
the ultraviolet cut-off
\beq
\int d\Omega\, Y_{j0}(\hat{\vv{x}})
\av{\rbr{\hat{\vv{x}}\cdot \de \theta}^{2n}}\sim 
M^{n\xi}\rbr{\frac{M}{m}}^{\rho_{2n,j}}
\eeq
with $\rho_{2n,j}$ given \eq{smallxi:firstorder:anomalous}.

Let us finish this section by comparing our treatment of the
UV problem with that of
ref's~\cite{AAV98,AABKV01a,AABKV01b}. 
In these papers 
the anomalous exponents of the scalar gradients are
computed using the field theoretic RG derived from
dimensionally regularized perturbation expansion.
Their starting point for the MSR theory differs from
our one in two ways. First
 their velocity covariance
\eq{ps:velcor} has $D_0\,\xi$ replaced by a constant 
with no explicit $\xi$ dependence. Second, perturbation theory
is done to the Stratonovich representation of the model \eq{ps:ps}.

The first difference means that the "bare" correlation functions
${\cal C}_{2n}$ need to be multiplied by a "renormalization constant"
proportional to $\xi^{-2n}$ to get a nontrivial limit as $\xi\to 0$.

The second difference leads to a logarithmic 
UV divergence at $\xi=0$ which can 
be traced to the $M$ dependence of the effective diffusivity $\kappa$ 
in  \eq{hopf:viscosity}. Since the authors work in dimensional 
regularization an UV cutoff  doesn't enter but its role 
is played by $\xi$ that plays the same role as $d-4$ in dimensional 
regularization. Using the Ito representation in the perturbation as we 
do the logarithmically UV divergent tadpole diagram doesn't enter and 
the only divergences to be dealt with are infrared. In both approaches  
the UV problem for the action is trivial, indeed, the authors find a simple  
fixed point $\xi$ for a running coupling constant describing 
the strength of the nonlinearity and their perturbation expansion becomes 
an expansion in powers of $\xi$.

\section{Conclusions}

The Kraichnan model has the rare feature to allow for detailed analytical 
study of a turbulent system. In particular it allows to address the question 
of what kind of renormalization group if any is appropriate
for turbulence, the traditional ultraviolet one or the more
exotic inverse or infrared one.
 
In the context of the Kraichnan model both direct and inverse renormalization
can be successfully applied. Direct renormalization is natural
when studying the short distance singularities that appear
when the dissipative scale is taken to zero.  The scaling fields
are local operators in the derivatives of the scalar and the
exponents may be computed using various versions of the
UV RG, in dimensional regularization as in 
ref's~\cite{AAV98,Wi00,AABKV01a,AABKV01b} or in the
Wilsonian framework as in the present paper.

The inverse RG appears more natural for dealing with
inertial range quantities such as
the nonlocal operators that enter the study of the structure functions.
The scaling fields are now very nonlocal in position space.
We implemented the inverse RG in the Wilsonian framework.
Although cumbersome it has the advantage of being  conceptually clear. 
However other more computationally effective inverse schemes should be possible too.

A natural question is whether infra-red renormalization may be of use in 
the analysis of physical models other than Kraichnan's. 
It is important here to stress that the possibility of successfully applying 
infra-red renormalization ultimately relies on the physics of the
system. It remains a challenge for the future to establish whether such tool 
may prove useful to inquire scaling properties of systems for which direct 
renormalization cannot be applied.

\section{Acknowledgments}

The authors are pleased to acknowledge many discussions with 
N.~Antonov, D.Bernard and K.~Gaw\c{e}dzki. 

P. M.-G. was supported
by the {\em``Fluid Mechanical Stirring and Mixing''}
research training network funded by the European Commission's 5th Framework 
programme (HPRN-CT-2002-00300) and by the centre of excellence 
{\em ``Geometric Analysis and Mathematical Physics''} funded by the 
Academy of Finland.

\appendix
\section*{Appendices}

\section{Mellin transform}
\label{ap:Mellin}
\setcounter{equation}{1}
\renewcommand{\theequation}{\ref{ap:Mellin}-\arabic{equation}}

We use the following definition of the Mellin transform of 
a  function $f:[0,\infty]\to \bf R$:
\begin{eqnarray}
\tilde{f}(x,z)=\int_{0}^{\infty}\frac{dw}{w}\frac{f(w\,x)}{w^z}=
x^z \int_{0}^{\infty}\frac{dw}{w}\frac{f(w)}{w^z}:=x^z\,J(z).
\end{eqnarray}
Suppose $f$ is given by
\begin{eqnarray}
f({x})=x^{a}\,g({x})
\end{eqnarray}
with $g$ decaying faster than any power at infinity. Then
the Mellin transform is defined and analytic for ${\rm Re}{z}<a$. It 
has a pole at $a$. Indeed,  write
\begin{eqnarray}
\tilde{f}(x;z)=x^z \int_{0}^{\infty}\frac{dw}{w}\frac{g(w)}{w^{z-a}}
=-\,\frac{x^{z}}{z-a}
\int_{-\infty}^{\infty}du\,e^{-u}g(e^{\frac{u}{z-a}})
\end{eqnarray}
so the residue is given by
\begin{eqnarray}
Res\tilde{f}(x;a)=-x^{z}\,g(0)
\end{eqnarray}
If $g$ is smooth at origin further {\em ``ultra-violet''} poles will occur in $a+n$ for any 
integer $n$, in correspondence with the Taylor expansion of $g$.
If  $g$ has only power law decay at infinity, say 
\begin{eqnarray}
g(x)\sim x^{-b}\,,\qquad\qquad b>0
\end{eqnarray}
the Mellin transform will exhibit {\em ``infra-red''} poles also 
in $-b-n$, $n=0,...$.  

The inverse Mellin transform is
\begin{eqnarray}
f(x)=\int_{c-\imath\,\infty}^{c+\imath\,\infty}\frac{dz}{(2\pi\imath)} 
x^z J(z)
\label{ap:Mellin:inverse}
\end{eqnarray}
with $c$ less than $a$. The inverse is usually computed by 
Cauchy's residue theorem. 

Finally for a function $f(\vv{x})$ on ${\bf R}^n$ we define
\begin{eqnarray}
\tilde{f}(\vv{x},z)=\int_{0}^{\infty}\frac{dw}{w}\frac{f(w\,\vv{x})}{w^z}=
x^z \int_{0}^{\infty}\frac{dw}{w}\frac{f(w\hat{\vv{x}})}{w^z}
\end{eqnarray}
with $x=|\vv{x}|$.

More details can be found for example in 
the textbook \cite{Po}. Useful examples of application of the Mellin
transform to diagrammatic expansions in field theory are given also
in \cite{OO00}.

\section{Evaluation of the inertial range asymptotics of the velocity field}
\label{ap:inertial}
\setcounter{equation}{1}
\renewcommand{\theequation}{\ref{ap:inertial}-\arabic{equation}}

An algorithmically efficient evaluation of (\ref{velasympt:Mellin}) is 
achieved using the integral representation of a power law:
\begin{eqnarray}
\frac{1}{x^{z}}=\int_{0}^{\infty}
\frac{du}{u} \frac{u^{\frac{z}{2}}\,e^{-u\,x^2}}
{\Gamma\left(\frac{z}{2}\right)}\,,\qquad z\,>\,0
\label{ap:inertial:gamma}
\end{eqnarray}
Namely, for negative values of $z$ the order of integration in 
(\ref{velasympt:Mellin}) can be inverted: 
\begin{eqnarray}
\tilde{D}^{\alpha\,\beta}(\vv{x},m;z)=-\,
\frac{D_{0}\,\xi\,m^{z-\xi}}{z-\xi}\int\fvol{q}  
\frac{e^{\imath \vv{q}\cdot \vv{x}}}{q^{d+z}}\,\Pi^{\alpha\,\beta}(\hat{\vv{q}})
\label{ap:inertial:Mellin}
\end{eqnarray}
The integral is absolutely convergent in the infra-red and is
convergent in the ultra-violet because of the oscillatory exponential. 
Furthermore incompressibility constrains the tensorial structure of
the integral to the form
\begin{eqnarray} 
\tilde{D}^{\alpha\,\beta}(\vv{x},m;z)=
\frac{\tilde{D}^{\gamma}_{\hspace{0.18cm}\gamma}(\vv{x};m)}
{\ivec^{\mu}_{\hspace{0.18cm}\mu}(\hat{\vv{x}},z)}\,
\ivec^{\alpha\,\beta}(\hat{\vv{x}},z)
\label{ap:inertial:general}
\end{eqnarray}
with $\ivec^{\alpha\,\beta}$ defined by (\ref{velasympt:realranktwo}).
Thus, the knowledge of the trace allows to reconstruct \eq{velasympt:real}.
The trace can be computed using the identity
\begin{eqnarray}
\int \fvol{q} \frac{e^{\imath q \cdot x}}{q^{d+z}}=
\frac{x^{z}}{2^{z}\,(4\,\pi)^{\frac{d}{2}}}
\frac{\Gamma\left(-\frac{z}{2}\right)}{\Gamma\left(\frac{d+z}{2}\right)}
\label{ap:inertial:dimensional}
\end{eqnarray}
which is derived using \eq{ap:inertial:gamma}. The result is
\begin{eqnarray}
\frac{\tilde{D}^{\gamma}_{\hspace{0.18cm}\gamma}(\vv{x},m;z)}
{\ivec^{\mu}_{\hspace{0.18cm}\mu}(\hat{\vv{x}},z)}
=\frac{D\,\xi\,m^{z-\xi}\,x^z}{z\,(z-\xi)}\,\frac{(z+d-1)\,d}{(d-1)(d+z)}
\frac{\Gamma\left(\frac{d}{2}\right)\Gamma\left(1-\frac{z}{2}\right)}
{2^{z}\,\Gamma\left(\frac{d+z}{2}\right)}
\label{ap:inertial:prefactor}
\end{eqnarray}
with $D$ now given by (\ref{ps:eddy}). Inserting (\ref{ap:inertial:prefactor}) 
in (\ref{ap:inertial:general}) recovers (\ref{velasympt:real}). 

An alternative derivation of \eq{ap:inertial:prefactor} which does not
use oscillatory integrals is available from \cite{website}.

\section{First order approximation}
\label{ap:firstorder}
\setcounter{equation}{1}
\renewcommand{\theequation}{\ref{ap:firstorder}-\arabic{equation}}

Rewriting (\ref{smallxi:firstorder:v14}) as the difference
\begin{eqnarray}
\stfv_{(1;4)}^{\alpha\,\beta}(\vv{x},m)
=2\,\frac{D_{0}\,m^{\xi}}{D}\underset{q\geq m}{\int}
\fvol{q}\,\frac{\Pi^{\alpha\,\beta}(\hat{\vv{q}})}{q^{d+2+\xi}}
-2\,\frac{D_{0}\,m^{\xi}}{D}\,
\underset{q\geq m}{\int}\fvol{q} \frac{e^{\imath q \cdot x\,w}}{q^2}
\frac{\Pi^{\alpha\,\beta}(\hat{\vv{q}})}{q^{d+\xi}}
\label{ap:firstorder:mv14}
\end{eqnarray}
permits to evaluate it using the same method expounded in 
appendix~\ref{ap:inertial}. The Mellin transform of the second interval 
on the right hand side of \eq{ap:firstorder:mv14} is well defined 
for $\Re z\,<\,-2$. 
However, the residue of the pole in $z=-2$ cancels exactly with the
first integral on the right hand side of \eq{ap:firstorder:mv14}.
Observing that \eq{ap:firstorder:mv14} has vanishing divergence 
its Mellin transform can be finally written as
\beq
\tilde{\stfv}^{\alpha\,\beta}(\vv{x},m;z+2)=
\frac{\tilde{\stfv}^{\gamma}_{(1;4)\,\gamma}(\vv{x};m,z+2)}
{\ivec^{\mu}_{\hspace{0.18cm}\mu}(\hat{\vv{x}},z+2)}\,
\ivec^{\alpha\,\beta}(\hat{\vv{x}},z+2)\,,\hspace{2.0cm}-2\,<\Re z\,<\,0
\label{ap:firstorder:generalz}
\eeq
with
\beq
\frac{\tilde{\stfv}_{(1;4)\,\gamma}^{\gamma}(\vv{x},m;z+2)}
{\ivec^{\mu}_{\hspace{0.2cm}\mu}(\hat{\vv{x}};z+2)}
\!\!&=&\!\!\frac{2\,(d+z+1)\,c(z)}{(z+2)(d+z-1)(d+2+z)}
\frac{m^{z}\,x^{2+z}}{z\,(z-\xi)}
\eeq
and $c(z)$ defined by (\ref{velasympt:g}). 
Setting $\xi$ to zero, (\ref{ap:firstorder:generalz}) has a double
pole at $z$ equal zero. The amplitude of such pole specifies is equal
to minus the prefactor of the logarithm in the short distance 
asymptotics of the diagram
\begin{eqnarray}
\lefteqn{\stfv_{(1;4)}^{(0)\,\alpha\,\beta}(\vv{x};m)\,=}
\nn&&
-\,\frac{(d+1)\,x^{2}\,\ivec^{\alpha\,\beta}(\hat{\vv{x}},2)}{(d-1)(d+2)}\,
\sbr{\ln\rbr{\frac{m\,x}{2}}-\frac{\psi\rbr{\frac{d+4}{2}}+\psi\rbr{1}}{2}}
+\frac{x^{2}\,\delta^{\alpha\,\beta}}{2\,(d+2)}+O(m^2 x^2)
\label{ap:firstorder:logarithm}
\end{eqnarray}
with
\begin{eqnarray}
\psi(x)=\eval{\frac{d\,}{du}}{u=0}\ln \Gamma(u+x)
\label{ap:firstorder:polygamma}
\end{eqnarray}
Differentiating the Mellin transform with respect to $\xi$ higher
orders contribution are found. In particular
\beq
\lefteqn{\stfv_{(1;4)}^{(1)\,\alpha\,\beta}(\vv{x};m)\,=
-\,\frac{(d+1)\,x^{2}\,\ivec^{\alpha\,\beta}(\hat{\vv{x}},2)}{2\,(d-1)(d+2)}
\sbr{\ln\rbr{\frac{m\,x}{2}}}^2}
\nn&&
+\cbr{\frac{(d+1)\,x^{2}\,\ivec^{\alpha\,\beta}(\hat{\vv{x}},2)}{(d-1)(d+2)}\,
\frac{\psi\rbr{\frac{d+4}{2}}+\psi\rbr{1}}{2}
+\frac{x^{2}\,\delta^{\alpha\,\beta}}{2\,(d+2)}}\ln\rbr{\frac{m\,x}{2}}
+O(m^2 x^2)
\eeq
For more details the reader is referred to \cite{website}

\section{Second order approximation}
\label{ap:secondorder}
\setcounter{equation}{1}
\renewcommand{\theequation}{\ref{ap:secondorder}-\arabic{equation}}

The scope of this appendix is to expound the computational strategy
followed in the evaluation of the second order diagrams and to give
the final results used in the determination of the scaling exponents.
Further details together with the computer packages used in the practical
evaluation are available from \cite{website}.

The evaluation of higher order integrals is hampered by the fact that
the Mellin transform with respect of the spatial argument of the structure
functions does not remove uniformly the mass cutoff from all the momentum
integrations. The problem is obviated by taking the convolution of as many
Mellin transforms as the number of momentum integration affected by the cutoff. 
The procedure can be illustrated by considering the general form of second
order diagrams: 
\beq
\tilde{\stfv}(\vv{x},z+n)=\int_{0}^{\infty}\frac{dw}{w}\frac{1}{w^{z+n}}
\underset{p\,,q\,>\,m}{\int} 
\frac{d^dpd^dq}{(2\pi)^{2d}}\frac{f(\vv{p},\vv{q},w\,\vv{x})}{q^{d}p^{d}}
\eeq
The integer $n$ is fixed by the degree of homogeneity of the function $f$:
\beq
f(w\,\vv{p},w\,\vv{q},w\,\vv{x})=w^{n}\,f(\vv{p},\vv{q},\vv{x})
\eeq
Thus the translation of the origin in the complex $z$-plane 
makes sure that the integration contour of the Mellin anti-transform poles  
lies to the left of the pole in $z$ equal zero. 
By rescaling the integral becomes
\beq
\tilde{\stfv}(\vv{x},z+n)=
\int_{0}^{\infty}\frac{dw}{w}\frac{m^{z}}{w^{z}}
\underset{p\,,q\,>\,w}{\int} 
\frac{d^dpd^dq}{(2\pi)^{2d}}\frac{f(\vv{p},\vv{q},\vv{x})}{q^{d}p^{d}}\,,
\qquad\qquad z<0
\label{ap:secondorder:mellin}
\end{eqnarray}
Taking the convolution with a second Mellin transform allows to deal with
unbounded momentum integrations:
\begin{eqnarray}
\tilde{\stfv}(\vv{x},z+n)\!\!&=&\!\!\int_{0}^{\infty}\frac{dw}{w}\frac{m^{z}}{w^{z}}
\underset{\substack{z<\Re \zeta<0\\-\infty<\Im \zeta <\infty}}{\int}
\!\!\!\!\!
\frac{d\zeta}{(2\,\pi\imath)}\int_{0}^{\infty}
\frac{du}{u}\frac{1}{u^{\zeta}}\underset{p\,,q\,>\,w}{\int} 
\frac{d^dpd^dq}{(2\pi)^{2d}}\frac{f(\vv{p},\vv{q},\vv{x})}{q^{d}p^{d}}
\nonumber\\
&=&\!\!\underset{\substack{z<\Re \zeta<0\\-\infty<\Im \zeta <\infty}}{\int}
\!\!\!\!\!
\frac{d\zeta}{(2\,\pi\imath)} \frac{m^{z}}{\zeta\,(z-\zeta)}\int 
\frac{d^dpd^dq}{(2\pi)^{2d}}\frac{f(\vv{p},\vv{q},\vv{x})}{q^{d+\zeta}p^{d+z-\zeta}}
\end{eqnarray}
Performing, eventually with the help of the representation 
of a power-law (\ref{ap:inertial:gamma}), the integration 
over momenta leaves with an integral over the Mellin variable $\zeta$.
In order to determine the scaling exponents up to second order, it is necessary to
compute the Mellin transform of each diagram up to $O(z^{-1})$ accuracy. 
The observation allows for some simplifications. Namely dimensional considerations
impose
\begin{eqnarray}
\tilde{f}(\vv{x},\zeta,z-\zeta):=\int 
\frac{d^dpd^dq}{(2\pi)^{2d}}\frac{f(\vv{p},\vv{q},\vv{x})}{q^{d+\zeta}p^{d+z-\zeta}}
=x^{z+n}\,\tilde{f}(\hat{\vv{x}},\zeta,z-\zeta)
\end{eqnarray}
with $\tilde{f}$ not vanishing for $z$ equal zero. Thus, it is possible to 
infer the form of the Laurent expansion for $z$ and $\zeta$ in the neighborhood
of zero
\begin{eqnarray}
\tilde{f}(\vv{x},\zeta,z-\zeta)=\frac{x^{z+n}}{z}
\left[\frac{\tilde{f}_{(-3;1)}(\hat{\vv{x}})}{\zeta}+
\frac{\tilde{f}_{(-3;2)}(\hat{\vv{x}})}
{z-\zeta}+\tilde{f}_{(-2)}+O(\zeta,z-\zeta)\right]
\end{eqnarray}
The knowledge of the first two poles of $\tilde{f}$ around $\zeta$ equal zero 
suffices to reconstruct the first two terms of the Laurent series around $z$ equal
zero of $\tilde{\stfv}$:
\begin{eqnarray}
\tilde{\stfv}(\vv{x},z+n)&=&x^{n}\,(m x)^z\!
\underset{\substack{z<\Re \zeta<0\\-\infty<\Im \zeta <\infty}}{\int}
\!\!
\frac{d\zeta}{2\,\pi\,\imath}\frac{m^{-z}}{z\,\zeta\,(z-\zeta)}
\left[\frac{\tilde{f}_{(-3;1)}(\hat{\vv{x}})}{\zeta}+
\frac{\tilde{f}_{(-3;2)}(\hat{\vv{x}})}
{z-\zeta}+\tilde{f}_{(-2)}+O(\zeta,z-\zeta)\right]
\nonumber\\
&=&x^{n}(m x)^z\left[\frac{\tilde{f}_{(-3;1)}(\hat{\vv{x}})
+\tilde{f}_{(-3;2)}(\hat{\vv{x}})}{z^3}+
\frac{\tilde{f}_{(-2)}(\hat{\vv{x}})}{z^2}\right]+O(z^{-1})
\end{eqnarray}
In essence, inner Mellin transforms work as regularized Taylor expansions for 
the argument of multi-loop integrals. The method was applied to the evaluation
of $\stfv_{(2;4)}$ and $\stfv_{(2;6)}$. The integral $\stfv_{(2;8)}$ in the limit
of infinite integral scale of the forcing reduces to the product of two first 
order integrals. The evaluation of its Mellin transform through convolutions can be
used to check of the procedure.
  
\subsection{Evaluation of $\stfv_{(2;4)}$}

The diagram $\stfv_{(2;4)}^{\alpha\,\beta}$ is most conveniently
evaluated if its real space representation
\begin{eqnarray}
\stfv_{(2;4)}^{\alpha\,\beta}= \frac{2}{\xi}\int_{0}^{\infty}ds\,
\int d^dy\,\frac{e^{-\frac{y^2}{4\,\eddy\,s}}-
e^{-\frac{(\vv{x}-\vv{y})^2}{4\,\eddy\,s}}}
{(4\,\pi\,\eddy\,s)^{\frac{d}{2}}}\,D^{\mu\,\nu}(\vv{y};m)
(\de_{\mu}\de_{\nu}\stfv_{(1;4)}^{\alpha\,\beta})(\vv{y};m)
\label{ap:secondorder:v24}
\end{eqnarray}
is adopted as a starting point for its evaluation. The choice of the 
prefactor guarantees that at zero molecular viscosity the interaction
term is order is of the order $O(\xi^{0})$. 
The reason is that in real space space (\ref{ap:secondorder:v24}) has the 
same structure of the first order vertex $\stfv_{(1;4)}^{\alpha\,\beta}$.
Once the Mellin transform of this latter is known, the use of the Mellin 
convolution technique (appendix~\ref{ap:Mellin}) reduces the evaluation
(\ref{ap:secondorder:v24}) to that of an integral similar of the same type
of (\ref{smallxi:firstorder:v14}). More explicitly the Mellin transform is
\begin{eqnarray}
\lefteqn{\tilde{\stfv}_{(2;4)}^{\alpha\,\beta}(z+2)=}
\nonumber\\
&&\frac{2}{\xi}\hspace{-0.4cm}
\underset{\substack{z<\Re \zeta<0\\-\infty<\Im \zeta <\infty}}{\int}
\hspace{-0.4cm}\frac{d\zeta}{2\,\pi\,\imath}\int_{0}^{\infty}ds\,
\int d^dy\,\frac{e^{-\frac{y^2}{4\,\eddy s}}
-e^{-\frac{(\vv{x}-\vv{y})^2}{4\,\eddy\,s}}}
{(4\,\pi\,\eddy\,s)^{\frac{d}{2}}}\,\tilde{D}^{\mu\,\nu}(\vv{y},m,\zeta)
(\de_{\mu}\de_{\nu}
\tilde{\stfv}_{(1;4)}^{\alpha\,\beta})(\vv{y},m,z-\zeta)
\label{ap:secondorder:v24bis}
\end{eqnarray} 
The integrals over space-time variables can be performed exactly: 
\begin{eqnarray}
\lefteqn{\!\!\!\!\!\!\!\tilde{\stfv}_{(2;4)}^{\alpha\,\beta}(z+2)=}
\nonumber\\
&&\!\!\!\!\!\!\!\!\!\!\!\!\!\!\!\!
\underset{\substack{z<\Re \zeta<0\\-\infty<\Im \zeta <\infty}}{\int}
\!\!\!\!\!\frac{d\zeta}{2\,\pi\,\imath}\,
\frac{2^{-z} m^{z-2\xi} x^z \Gamma\left(\frac{2 + d}{2}\right)^2 
\Gamma\rbr{1 - \frac{\zeta}{2}} \Gamma\rbr{\frac{2 + \zeta - z}{2}}
\sbr{P_{0}(z,\zeta,d)\,x^2\,\delta^{\alpha\,\beta}-
P_{1}(z,\zeta,d)\,x^{\alpha}x^{\beta}}}
{(d-1)^2\,\zeta\,(\zeta - z)\,z\,(2 + z)\,(2 + d + z)\,(\zeta-\xi)\,(\zeta - z+\xi)
\Gamma\left(\frac{4 + d + \zeta}{2}\right)\Gamma\left(\frac{2 + d - \zeta + z}{2}
\right)}
\label{ap:secondorder:mv24}
\end{eqnarray}
with
\sbeq
P_{0}(z,\zeta,d)=
z\,[d^2\,(2 + z)+\,d^3 - 3\,z-4 - 3\,d] + \zeta\,[2 + d^2\,z - z^2 + 
d\,(2 + z + z^2)]
\smeq
P_{1}(z,\zeta,d)=(2 + z)\,[d^2\, \zeta - (2 + \zeta)\, z + d\, \zeta\, (1 + z)]
\seeq
In order to determine scaling exponents with second order accuracy it is enough 
to evaluate \eq{ap:secondorder:mv24} at $\xi$ equal zero. To that goal 
the integral over $\zeta$ can be performed by applying Cauchy theorem in the
complex $\zeta$ plane. Since the contour is has clockwise orientation residues 
must multiplied by a minus sign. Logarithmic contributions to 
\eq{ap:secondorder:v24} are associated to the triple and double pole in $z$ 
of the Mellin transform. These latter ones are fully specified if 
\eq{ap:secondorder:mv24} is approximated by its first residue for $\zeta$ equal
zero. The result can be couched into the form
\begin{eqnarray}
\lefteqn{\tilde{\stfv}_{(2;4)}^{(0)\,\alpha\,\beta}(\vv{x};z+2)=}
\nn&&- \rbr{\frac{m\,x}{2}}^{z}\cbr{
\left[\frac{1}{z^3}-\frac{\psi\rbr{\frac{d+2}{2}}
+\psi(1)}{2\, z^{2}}\right]V_{(2;4;1)}^{\alpha\,\beta}(\vv{x})+
\frac{V_{(2;4;2)}^{\alpha\,\beta}(\vv{x})}{z^2}}
+\,O(z^{-1})
\label{ap:secondorder:v24final}
\end{eqnarray}
with the function $\psi$ defined by \eq{ap:firstorder:polygamma}. 
The tensor coefficients appearing in \eq{ap:secondorder:v24final} are
\begin{eqnarray}
V_{(2;4;1)}^{\alpha\,\beta}(\vv{x}):= 2 
\frac{(d+1)(d^2+d-3)\,x^2\,\delta^{\alpha\,\beta}-(d^2+d-4)\,
\vv{x}^{\alpha}\vv{x}^{\beta}}{(d-1)^{2}\,(d+2)^2}
\label{ap:secondorder:tensorv241}
\end{eqnarray}
and
\begin{eqnarray}
V_{(2;4;2)}^{\alpha\,\beta}(\vv{x}):=
\frac{[4 - d\,(d^3+4\,d^2+d-10)]\, x^2 \delta^{\alpha\,\beta}-8\,
\vv{x}^{\alpha} \vv{x}^{\beta}}{(d-1)^2 (d+2)^3}
\end{eqnarray}
The residue in $z$ equal zero of \eq{ap:secondorder:v24} specifies
the inertial range asymptotics of the diagram at leading order in $\xi$.

\subsection{Evaluation of $\stfv_{(2;6)}$}

The Mellin transform of $\stfv_{(2;6)}$ can be written as the sum of 
three terms
\begin{eqnarray}
\tilde{\stfv}_{(2;6)}^{\alpha\,\beta\,;\,\mu}(z+3)
=\sum_{i=1}^{3}\tilde{\stfv}_{(2;6;i)}^{\alpha\,\beta\,;\,\mu}(z+3)
\end{eqnarray}
with
\begin{eqnarray}
&&\tilde{\stfv}_{(2;6;1)}^{\alpha\,\beta\,;\mu}(z+3)=
\int_{0}^{\infty}\frac{dw}{w}\frac{D_{0}^{2}\,m^{z-2\,\xi}}{D^{2}\,w^{z-2\,\xi}}
\underset{\substack{q\geq w\\ p \geq w}}{\int} 
\frac{d^dq\,d^dp}{(2\,\pi)^{2\,d}}\,
\frac{2\,\sin(\vv{p}\cdot\vv{x})}{(q^2+\vv{q}\cdot\vv{p}+p^2)\,q^2}
\frac{\vv{q}_{\nu}\,\Pi^{\mu\,\nu}(\hat{\vv{p}})\, 
\Pi^{\alpha\,\beta}(\hat{\vv{q}})}{p^{d+\xi}\,q^{d+\xi}}
\nonumber\\
&&\tilde{\stfv}_{(2;6;2)}^{\alpha\,\beta\,;\mu}(z+3)=
\int_{0}^{\infty}\frac{dw}{w}\frac{D_{0}^{2}\,m^{z-2\,\xi}}{D^{2}\,w^{z-2\,\xi}}
\underset{\substack{q\geq w\\ p \geq w}}{\int} 
\frac{d^dq\,d^dp}{(2\,\pi)^{2\,d}}\,
\frac{2\,\sin(\vv{q}\cdot\vv{x})}{(q^2+\vv{q}\cdot\vv{p}+p^2)\,q^2}
\frac{\vv{q}_{\nu}\,\Pi^{\mu\,\nu}(\hat{\vv{p}})\, 
\Pi^{\alpha\,\beta}(\hat{\vv{q}})}{p^{d+\xi}\,q^{d+\xi}}
\nonumber\\
&&\tilde{\stfv}_{(2;6;3)}^{\alpha\,\beta\,;\mu}(z+3)=\,-
\int_{0}^{\infty}\frac{dw}{w}\frac{D_{0}^{2}\,m^{z-2\,\xi}}{D^{2}\,w^{z-2\,\xi}}
\underset{\substack{q\geq w\\ p \geq w}}{\int} 
\frac{d^dq\,d^dp}{(2\,\pi)^{2\,d}}\,
\frac{2\,\sin[(\vv{q}+\vv{p})\cdot\vv{x}]}{(q^2+\vv{q}\cdot\vv{p}+p^2)\,q^2}
\frac{\vv{q}_{\nu}\,\Pi^{\mu\,\nu}(\hat{\vv{p}})\, 
\Pi^{\alpha\,\beta}(\hat{\vv{q}})}{p^{d+\xi}\,q^{d+\xi}}
\end{eqnarray}
In order to determine the scaling exponent within second order, 
$\xi$ can be set to zero in the integrands.
The three Mellin integrals exist separately in the complex $z$-plane 
for values of $z$ such that $\Re z<-2$.  The sum of the three integrals 
brings about the cancellations restoring the original domain of convergence 
of the Mellin transform of $\stfv_{(2;6)}$.

The explicit evaluation of the integrals is cumbersome but can be performed
using some software for symbolic manipulations. The packages used for the 
evaluation are available for free download from \cite{website}.
The final result is 
\begin{eqnarray}
\lefteqn{\tilde{\stfv}_{(2;6)}^{(0)\,\alpha\,\beta\,;\,\mu}(\vv{x};z+3)=}
\nonumber\\
&&- \rbr{\frac{m\,x}{2}}^{z}\left\{\left[\frac{1}{z^3}
-\frac{\psi\rbr{\frac{2+d}{2}}+\psi\rbr{1}}{2\, z^{2}}\right]
V_{(2;6;1)}^{\alpha\,\beta\,;\,\mu}(\vv{x})+
\frac{V_{(2;6;2)}^{\alpha\,\beta\,;\,\mu}(\vv{x})}{z^2}+O(z^{-1})\right\}
\end{eqnarray}
The tensor coefficients are
\begin{eqnarray}
&&V_{(2;6;1)}^{\alpha\,\beta\,;\,\mu}(\vv{x}):=
\frac{4\,x^{\alpha}\,x^{\beta}\,x^{\mu}  + (d+1)\,x^2\, 
[\,(d-1)\, \delta^{\alpha\,\beta}\, x^{\mu}-\delta^{\alpha\,\mu}\,x^{\beta}
-\delta^{\mu\,\beta}\,x^{\alpha}]}{(d-1)^2\,(d+2)^2}
\label{ap:secondorder:tensorv261}
\\
&&V_{(2;6;2)}^{\alpha\,\beta\,;\,\mu}(\vv{x}):=
\upsilon_{(1)}\,x^2\, x^{\mu}\,\delta^{\alpha\,\beta}+
\upsilon_{(2)}\,x^2\,( x^{\beta}\,\delta^{\alpha\,\mu}+x^{\alpha}\,
\delta^{\mu\,\beta})+
\upsilon_{(3)} x^{\alpha}\,x^{\beta}\, x^{\mu}
\end{eqnarray}
where the scalar coefficients $\{\upsilon_{(i)}\}_{i=1}^{3}$ are
\begin{eqnarray}
&&\upsilon_{(1)}=\frac{2\,(28 + 20\,d - 15\,d^2 - 8\,d^3 - d^4) 
- 3\,(12 + 5\,d - 4\,d^2 - d^3)\, 
\mathrm{Hyp}_{21}\rbr{1,1,2+\frac{d}{2},\frac{1}{4}}}
{4\,(d-1)^2\,(d+2)^3\,(d+4)}
\nonumber\\
&&\upsilon_{(2)}=\frac{2\,(-4 + 4\,d + 3\,d^2 + d^3) + 3\,(4 + 3\,d 
+ d^2)\,\mathrm{Hyp}_{21}\rbr{1,1,2+\frac{d}{2},\frac{1}{4}}}
{4\,(d-1)^2\,(d+2)^3\,(d+4)}
\nonumber\\
&&\upsilon_{(3)}=-\frac{2\,(6+d+d^2)+3\,d \,
\mathrm{Hyp}_{21}\rbr{1,1,2+\frac{d}{2},\frac{1}{4}}}
{(d-1)^2\,(d+2)^3\,(d+4)}
\end{eqnarray}

\subsection{Evaluation $\stfv_{(2,8)}$}

As shown in the main text, when the integral scale of the forcing
tends to infinity, this integral factorizes to
\begin{eqnarray}
\stfv_{(2,8)}^{\alpha\,\beta\,;\,\mu\,\nu}=
\stfv_{(1,4)}^{\alpha\,\beta}\,
\stfv_{(1,4)}^{\mu\,\nu}
\label{ap:secondorder:v28}
\end{eqnarray}
Hence, the knowledge of the small scale asymptotics of the first order vertex 
$\stfv_{(1,4)}$ suffices to determine the one of $\stfv_{(2,8)}$.  
Scaling exponents are conveniently evaluated using the Mellin 
transform of diagrams. Noting that $\stfv_{(1,4)}$ has the form 
\begin{eqnarray}
\mathcal{V}=A(x) \ln x +B(x) 
\end{eqnarray}
with $A$, $B$ some cut-off functions having finite value in zero and 
vanishing at infinity, the Mellin transform of \eq{ap:secondorder:v28}
can be represented as
\begin{eqnarray}
\tilde{\mathcal{V}^{2}}(x,z)
=x^z\,\int_{-\infty}^{\infty}du\, e^{-u}
\sbr{\,A^{2}(e^{\frac{u}{z}})\frac{u^{2}}{z^3} 
+2\,B(e^{\frac{u}{z}})\,A(e^{\frac{u}{z}}) \frac{u}{z^2}
+\frac{B^{2}(e^{\frac{u}{z}})}{z}}
\end{eqnarray}
For $z$ tending to zero from below one finds
\beq
\lim_{z\uparrow 0}\tilde{\mathcal{V}^{2}}(x,z)
=\int_{0}^{\infty}du\, e^{-u}
\sbr{\,A^{2}(0)\frac{u^{2}}{z^3} +2\,B(0)\,A(0) \frac{u}{z^2}
+\frac{B^{2}(0)}{z}}
+O(z^0)
\eeq
The equality entails that
\begin{eqnarray}
\lefteqn{\tilde{\stfv}_{(2;8)}^{(0)\alpha\,\beta\,;\,\mu\,\nu}(\vv{x};z+4)=}
\nonumber\\
&&- \rbr{\frac{m\,x}{2}}^{z}\left\{\left[\frac{1}{z^3}-\frac{
\psi\rbr{\frac{2+d}{2}}+\psi\rbr{1}}{2\, z^{2}}\right]
V_{(2;8;1)}^{\alpha\,\beta\,\mu\,\nu}(\vv{x})
+\frac{V_{(2;8;2)}^{\alpha\,\beta\,\mu\,\nu}(\vv{x})}{z^2}+O(z^{-1})\right\}
\end{eqnarray}
with
\sbeq
V_{(2;8;1)}^{\alpha\,\beta\,;\,\mu\,\nu}(\vv{x}):=
\frac{2\,(d+1)^2\,x^{4}\,\ivec^{\alpha\,\beta}(\hat{\vv{x}},2)
\ivec^{\mu\,\nu}(\hat{\vv{x}},2)}{(d-1)^2(d+2)^2}
\smeq
V_{(2;8;2)}^{\alpha\,\beta\,;\,\mu\,\nu}(\vv{x}):=
-\,
\frac{d\,(3+4\,d+d^2)\,x^4\,\delta^{\alpha\,\beta} \delta^{\mu\,\nu}+
8\,x^{\alpha}\,x^{\beta}\,x^{\mu}\,x^{\nu}- (2 + 5\,d + d^2)\,x^2\,
(\delta^{\mu\,\nu}\,x^{\alpha}\,x^{\beta} + \delta^{\alpha\,\beta}\,x^{\mu}\,
x^{\nu})}{(d-1)^2 (d+2)^3}
\seeq

\section{Angular integrals}
\label{ap:angular}
\setcounter{equation}{1}
\renewcommand{\theequation}{\ref{ap:angular}-\arabic{equation}}

Let $\omega$ denote the azimuthal angle in a given reference frame.
The projection of powers of $\cos\omega$ onto hyperspherical 
harmonics with zero magnetic numbers can be evaluated by considering
the generating function 
\beq
\sum_{n=0}^{\infty}\frac{(\imath z)^{n}}{n!}
\int d\Omega\,Y_{j;0}^{*}(\Omega)\,\cos^{n}(\omega)=
\int d\Omega\,Y_{j;0}(\Omega)\,e^{\imath\,z\,\cos(\omega)}
\eeq
The exponential can be expanded in hyperspherical harmonics
\beq
e^{\imath\,z\,\cos_{\angle}(\omega)}=
\sum_{j=0}^{\infty}\,\frac{\imath^{j}N_{j,d}}{z^{\frac{d-2}{2}}}\mathrm{BesJ}
\rbr{j+\frac{d-2}{2};z}Y_{j;0}(\Omega)
\eeq
with $N_{j,d}$ a normalization factor irrelevant for the present considerations.
Since
\beq
\mathrm{BesJ}\rbr{j+\frac{d-2}{2};z}=\rbr{\frac{z}{2}}^{j + \frac{d - 2}{2}} 
\sum_{k=0}^{\infty}\frac{\rbr{-\frac{z^2}{4}}^k}
{\Gamma\rbr{k + 1} \Gamma\rbr{j + \frac{d - 2}{2} + k + 1}}
\eeq
angular integrals are just the $n$-th coefficient of the Taylor expansion 
\beq
\int d\Omega\,Y_{j;0}^{*}(\Omega)\,\cos^{n}(\omega)=
\frac{2^{1 - \frac{d}{2} - n}\,N_{j,d} \Gamma(1 + n)}
{\Gamma\rbr{\frac{2 - j + n}{2}}\Gamma\rbr{\frac{d + j + n}{2}}}
\eeq
whence \eq{uv:gradients:angular} follows.

\section{Gradient expansion integrals}
\label{ap:gradients}
\setcounter{equation}{1}
\renewcommand{\theequation}{\ref{ap:gradients}-\arabic{equation}}

The first order integral can be readily performed
\begin{eqnarray}
\tilde{\gfv}_{(1;4)}^{\alpha\,\beta}(z)&=&\frac{D_{0}\,m^{\xi}\,M^z}{z\,D}\,
\underset{\infty > p\geq m}{\int}
\frac{d^dp}{(2\pi)^d}\frac{(\vv{p}\cdot \vv{x})^2}
{p^2}\frac{\Pi^{\alpha\,\beta}(\hat{\vv{p}})}{p^{d+z+\xi}}
\nonumber\\
&=&\frac{(d+1)\,x^2}{z\,(z+\xi)\,(d-1)\,(d+2)}\left(\frac{M}{m}\right)^{z}
\ivec^{\alpha\,\beta}(\hat{\vv{x}},2)
\label{ap:gradients:firstorder:mv14}
\end{eqnarray}
Second order integrals can be performed by resorting to the Mellin-convolution
techniques expounded in the previous appendix~\ref{ap:secondorder}. 
Second order integrals can be always reduced to the scalar form
\begin{eqnarray}
\gfv(M,m;z)= 
\underset{\substack{\Re \zeta\,<\,z\\-\infty<\Im \zeta <\infty}}{\int}
\!\!\!\!\!
\frac{d\zeta}{2\,\pi\,\imath}\,\frac{M^{z}\,G(m,\zeta,z-\zeta)}{z\,\zeta\,(z-\zeta)}
\label{ap:gradients:mtwoloops}
\end{eqnarray}
with
\begin{eqnarray}
G(m,\zeta,z-\zeta)=\int_{m}^{\infty}\int_{m}^{\infty}\frac{dp dq}{p\,q}\,
\frac{\varphi(p,q)}{p^{z-\zeta}\,q^{\zeta}}=\frac{m^{z}}{z}
\int_{1}^{\infty}\frac{dq}{q}\,\left[
\frac{\varphi(1,q)}{q^{\zeta}}+\frac{\varphi(q,1)}{q^{z-\zeta}}\right]
\end{eqnarray}
and angular degrees of freedom re-absorbed into the definition of $\varphi$.
 
The integrals to perform are essentially those calculated in \cite{AABKV01b}
when direct renormalization and dimensional regularization were applied to
compute the scaling exponents of the Kraichnan model. The reader interested
to the details of the calculations is therefore referred to \cite{AABKV01b}.
Here the results are presented in the 
\begin{eqnarray}
&&\gfv_{(2;4)}^{(0)\,\alpha\,\beta}=\left(\frac{M}{m}\right)^{z}\sbr{
\frac{V_{(2;4;1)}^{\alpha\,\beta}(\vv{x})}{z^3}+
\frac{U_{(2;4;1)}^{\alpha\,\beta}(\vv{x})}{z^2}+O(z^{-1})}
\nonumber\\
&&\gfv_{(2;6)}^{(0)\,\alpha\,\beta\,;\mu}=\left(\frac{M}{m}\right)^{z}\sbr{
\frac{V_{(2;6;1)}^{\alpha\,\beta\,;\mu}(\vv{x})}{z^3}+
\frac{U_{(2;6;1)}^{\alpha\,\beta\,;\mu}(\vv{x})}{z^2}+O(z^{-1})}
\end{eqnarray}
where $V_{(2;4;1)}$ and $V_{(2;6;1)}$ were respectively defined in
(\ref{ap:secondorder:tensorv241}), (\ref{ap:secondorder:tensorv261})
while 
\begin{eqnarray}
U_{(2;4;1)}^{\alpha\,\beta}(\vv{x}):=-\,2\,(d+1)\,\,
\frac{x^2\,\delta^{\alpha\,\beta} - d\, x^{\alpha} x^{\beta}}
{(d-1)^2\ (d+2)^3}
\end{eqnarray}
and
\begin{eqnarray}
&&U_{(2;6;2)}^{\alpha\,\beta\,;\mu}(\vv{x}):=
u_{(1)} x^{\mu}\,x^2\,\delta^{\alpha\,\beta}+
u_{(2)}\,x^2\,(x^{\beta}\delta^{\alpha\,\mu}+x^{\alpha}\delta^{\mu\,\beta})+
u_{(3)} x^{\alpha} x^{\beta} x^{\mu}
\end{eqnarray}
where the scalar coefficients $\{u_{(i)}\}_{i=1}^{3}$ are
\begin{eqnarray}
&&u_{(1)}=-\,3\,\,\frac{8\,(1 + d) + (-12 - 5\,d + 4\,d^2 + d^3)\,
\mathrm{Hyp}_{21}\rbr{1, 1, 2 + \frac{d}{2},\frac{1}{4}}}
{4\,(d-1)^2\,(2 + d)^3\,(4 + d)}
\nonumber\\
&&u_{(2)}= \frac{8\,(1 + d)^2 - 3\,(4 + 3\,d + d^2)\,
\mathrm{Hyp}_{21}\rbr{1, 1, 2 + \frac{d}{2}, \frac{1}{4}}}
{4\,(d-1)^2\,(2 + d)^3\,(4 + d)}
\nonumber\\
&&u_{(3)}=\frac{-4 - 2\,d + 2\,d^2 + 3\,d\,\mathrm{Hyp}_{21}\rbr{1, 1, 2 + 
\frac{d}{2}, \frac{1}{4}}}{(d-1)^2\,(2 + d)^3\,(4 + d)}
\end{eqnarray}

\end{document}